\documentclass[11pt,a4paper]{article}
\pdfoutput=1
\usepackage{jheppub}
\usepackage{lmodern}
\usepackage{amsmath,amssymb,mathtools,empheq}
\usepackage{tikz}
\usepackage{mathrsfs}
\usepackage{caption}
\usepackage{stmaryrd}
\usepackage{mdframed}
\usepackage{amscd}
\usepackage{amsfonts}
\usepackage{graphicx}
\usepackage{color}
\usepackage[normalem]{ulem}
\usepackage{hyperref}
\usepackage{enumerate}
\usepackage{slashed}
\usepackage{dsfont}
\usepackage{comment}
\def\calg{{\cal G}}
\def\calh{{\cal H}}

\def\calm{{\cal M}}
\def\caln{{\cal N}}

\global\mdfdefinestyle{exampledefault}{%
	leftmargin=0.1cm,rightmargin=0.1cm
}

\title{Multi-Centered Black Hole Index from a Superconformal Quiver Index}

\author[a]{Canberk \c{S}anl\i}

\affiliation[a]{CEICO, Institute of Physics of the Czech Academy of Sciences, Na Slovance 2, 182 00 Prague 8, Czech Republic}

\emailAdd{sanli@fzu.cz}

\abstract{We formulate a $D(2,1;0)$ superconformal index for scaling quiver quantum mechanics in its gauged sigma model description which preserves the hyperK\"ahler with torsion geometry. These quivers describe the Coulomb-branch dynamics of multi-centered D-brane configurations in the $AdS_2$ scaling limit of type II Calabi--Yau compactifications. Using supersymmetric localization, we derive a fixed-point formula for this superconformal quiver index and identify the geometric origin of its saddle equations and fixed-point contributions. We compare the result with the configurational index of multi-centered BPS black holes, finding the same universal denominator and angular momentum weights but different saddle equations. We then examine the additional scaling contributions to the microscopic quiver index and show that, for the symmetric three-center cyclic family, the unequal-sign scaling term is exactly reproduced by the superconformal index.}

\begin{document}
	
	\maketitle 
	
	\section{Introduction}
	\label{sec: intro}
	
	Scaling quiver quantum mechanics \cite{Denef:2002ru, Anninos:2013nra, Mirfendereski:2020rrk} describes the low-energy dynamics of multi-centered D-brane configurations in type II string theory compactifications on Calabi--Yau threefolds in a scaling regime, where an $AdS_2$ throat emerges in the near-horizon geometry. These configurations correspond to multi-centered BPS black-hole bound states in the four-dimensional $\caln=2$ supergravity picture \cite{Denef:2000nb, Denef:2007vg}. In addition to the equations of motion, supersymmetric multi-center configurations obey equilibrium equations whose solutions depend on the stability chamber determined by the asymptotic moduli. The quiver quantum mechanics provides a microscopic description of these BPS bound states, and its spectrum is subject to the same stability conditions. It therefore connects the stringy D-brane picture with the supergravity description of multi-centered BPS black holes, allowing supersymmetric states to be tracked in a regime where the black-hole lives. This differs from many standard $AdS/CFT$ constructions, where the microscopic counting is usually performed in a regime in which one does not have direct control over which classical BPS solutions are described by the counted states. Hence, it might help us understand ``which microscopic degrees of freedom can be described and distinguished by an observer having only gravitational probes at hand, and which ones cannot''  \cite{Bena:2012hf}.
	
	In the Coulomb branch $\caln=4$ quiver quantum mechanics description of $N$-centered D-brane configurations, each center carries a charge $\Gamma_a$, with $a=1,\cdots, N$, and interactions are described via the antisymmetric Dirac--Schwinger--Zwanziger (DSZ) product $\Gamma_{ab} = \langle \Gamma_a, \Gamma_b \rangle$. The system is encoded in a quiver \cite{Douglas:1996sw}, whose nodes represent centers and arrows are determined by $\Gamma_{ab}$. We assume that each center carries a primitive brane charge, so that each quiver node is $U(1)$. The effective Coulomb-branch description arises in the regime where the centers are separated so that the hypermultiplet modes are heavy and can be integrated out. The resulting description is given by $N$ abelian $d=1,~ \caln=4$ vector multiplets, each containing three bosonic, four fermionic and one auxiliary degree of freedom, commonly referred to as a $(3,4,1)$ multiplet. 
	
	This low-energy effective $\caln=4$ supersymmetric mechanics develops an additional conformal symmetry in the scaling limit \cite{Anninos:2013nra,Mirfendereski:2020rrk}. The resulting superconformal quantum mechanics of $N$ $(3,4,1)$ multiplets is invariant under the Lie superalgebra $D(2,1;0)$,\footnote{$D(2,1;0)$ also appears in the near-horizon moduli-space dynamics of the earlier superconformal multi-black-hole constructions studied in \cite{Maloney:1999dv,BrittoPacumio:1999ax}.} which is a special case of one-dimensional $\mathcal{N}=4$ superconformal symmetry. As the scaling limit removes the terms controlled by the asymptotic moduli, it describes the limiting Coulomb-branch dynamics in the deep scaling region, and its protected trace counts superconformal short multiplets. 
	
	Sigma models with $(3,4,1)$ multiplets can be reformulated in terms of root $(4,4,0)$ multiplets through a one-dimensional automorphic duality \cite{Delduc:2006yp,Fedoruk:2011aa}. In this reformulation, the auxiliary field is replaced by the time derivative of a new bosonic coordinate. Since the new coordinate enters only through its derivative, it has a shift symmetry, which is then gauged. This leads to a gauged superconformal mechanics \cite{Mirfendereski:2022omg}, which is defined on an extended target space parametrized by $(4,4,0)$ multiplets and constrained by a deformed version of the well known geometric constraints of the ungauged models \cite{Michelson:1999zf}.
	
	This gauged description of the scaling quiver mechanics is particularly well suited to study the supersymmetric ground states. In particular, the gauged superconformal mechanics formulation is defined on a larger target space with hyperk\"ahler with torsion (HKT) structure, and provides a geometric description in which supersymmetric states can be represented by elements of a twisted Dolbeault cohomology, with a standard $\mathfrak{su}(2)_L$ Lefschetz-like action realized by the $R$-symmetry generators.
	
	In this work, using the algebraic $D(2,1;0)$ index of Gaiotto--Simons--Strominger--Yin \cite{Gaiotto:2004pc}, we formulate the following $D(2,1;0)$ superconformal index for scaling quiver mechanics:
	\begin{equation}
		\mathcal{I}_\pm
		=
		\mathrm{Tr} \left[
		(-1)^{2J_L^3}
		e^{-\beta \{ \mathcal{G}_{\pm 1/2},
			\mathcal{G}^\dagger_{\pm 1/2} \}}
		y^{\pm (J_L^3 + J_R^3)}
		\right],
		\label{eq: intro defn index}
	\end{equation}
	where $\mathcal{G}_{\pm 1/2}$ are complex supercharges of the $D(2,1;0)$ algebra and $J_L^3,J_R^3$ are the Cartan generators of the $SU(2)_L\times SU(2)_R$ R-symmetry. The fugacity $y=e^{i\lambda}$ refines the index by the combination $J_L^3+J_R^3$. Algebraically, $\mathcal I_+$ and $\mathcal I_-$ count chiral and anti-chiral short multiplets of $D(2,1;0)$.
	
	Superconformal indices on quantum-mechanical targets are evaluated by passing from the conformal Hamiltonian to the supersymmetric Hamiltonian generated by the anticommutator of the chosen supercharge with its conjugate \cite{Dorey:2018klg,Barns-Graham:2018xdd,Dorey:2019kaf,Raeymaekers:2024usy,Raeymaekers:2024ics}. We use the same construction for gauged superconformal mechanics. In the $D(2,1;0)$ quiver model, the auxiliary Hamiltonian is
	\begin{equation}
		\mathcal H_\pm
		=
		\{\mathcal{G}_{\pm 1/2},\mathcal{G}^\dagger_{\pm 1/2}\}
		=
		L_0 \mp J_R^3 .
	\end{equation} The refinement by $J_L^3+J_R^3$ commutes with the localizing supercharge and is implemented as the corresponding equivariant deformation. The Euclidean path integral is then evaluated by supersymmetric localization.
	
	\paragraph{Main result.}
	Supersymmetric localization reduces the calculation to admissible collinear zeros of the section
	\begin{equation}
		E_{\pm,a}(z):=
		\sum_{b\neq a}
		\frac{\Gamma_{ab}}{2|z^a-z^b|}
		\mp G_{ab}(z)z^b .
	\end{equation}
	Here admissible means that the zero lies in the smooth collinear configuration space, with nonzero separation on every pair carrying a nonzero DSZ product, and has nonzero Jacobian on the relative subspace. Expanding around each such zero, the transverse modes give a universal equivariant oscillator factor, the classical first-order coupling gives the DSZ ordering phase, and the longitudinal constant modes produce the orientation sign. The resulting fixed-point formula is
	\begin{equation}
		\boxed{
			\mathcal I_\pm(y)
			=
			\sum_{p:E_\pm(p)=0}
			\sigma_\pm(p)
			\left(
			\frac{\pm y^{1/2}}{1-y}
			\right)^{N-1}
			y^{\mp\frac12
				\sum_{a<b}
				\Gamma_{ab}\operatorname{sign}_p(z^b-z^a)}
		}\, ,
		\label{eq:main-result-intro-boxed}
	\end{equation}
	where
	\begin{equation}
		\sigma_\pm(p)=\operatorname{sign}\det{}'
		[\partial_bv_a\pm2G_{ab}]_p ,
	\end{equation}
	where the prime denotes restriction to the relative configuration space. 
	
	We compare the result with the Manschot--Pioline--Sen configurational index $g_{\rm ref}$ \cite{Manschot:2010qz,Manschot:2011xc}. The universal denominator and DSZ ordering weights agree after matching fugacity conventions, while the saddle equations and orientation matrices differ. That our result need not coincide with the regular Coulomb index $g_{\rm ref}$ is indeed expected in the deep scaling regime. For quivers admitting scaling solutions, the microscopic index receives additional contributions involving single-centered (or pure-Higgs) invariants and minimal-modification terms \cite{Manschot:2012rx}. We then compare with the microscopic localization calculation of Beaujard, Mondal and Pioline \cite{Beaujard:2021fsk}, which separates the regular Coulomb contribution from scaling contributions. For the symmetric three-center cyclic family, our result reproduces their unequal-sign scaling contribution exactly\footnote{While this paper was under review, \cite{DeMarco:2026jje} appeared, deriving a fixed-point formula for a superconformal index in the $(3,4,1)$ formulation. After matching conventions, their formula agrees with the result obtained here in the $(4,4,0)$ formulation.}. 
	
	We start by reviewing the gauged superconformal mechanics framework and then specializing it to scaling quiver quantum mechanics in Section~\ref{sec:SCQM}. In Section~\ref{sect index general}, we define the $D(2,1;0)$ superconformal index, and derive the auxiliary localization problem. We then specialize the localization equations to the quiver model, derive the fixed-point formula, and compare it with the multi-centered black hole index and the microscopic scaling contribution. We comment on possible generalizations in Section~\ref{sect: discussion}. There are also six appendices collecting relevant technical details.
	
	\section{Gauged superconformal mechanics and quivers}
	\label{sec:SCQM}
	
	The gauged superconformal mechanics \cite{Mirfendereski:2022omg}, which lies at the core of our analysis, is formulated as a one-dimensional $\mathcal{N}=4B$ supersymmetric sigma model with $N$ $(4,4,0)$ multiplets\footnote{i.e. four bosonic, four fermionic, and zero auxiliary degrees of freedom per multiplet.} coupled to worldline gauge fields. The geometry of the target space $\mathcal{M}$, $\dim \calm =4N$, admits HKT structure, which is required for the off-shell realization of $\mathcal{N}=4B$ supersymmetry. This requires three integrable complex structures $(J^i)^A{}_{B}$ satisfying the quaternionic algebra
	\begin{equation}
		(J^i)^A{}_{B} (J^j)^B{}_{C} = -\delta^{ij} \delta^A_{C} + \epsilon^{ijk} (J^k)^A{}_{C}, \label{eq:quaternion}
	\end{equation}
	where $A=1,\cdots,4N$. They satisfy the covariant constancy condition with respect to the torsionful connection
	\begin{equation}
		\hat{\nabla}_A (J^i)^B{}_{C} = 0, \qquad\qquad \hat{\Gamma}^A{}_{BC} = \Gamma^A{}_{BC} + \frac{1}{2} C^A{}_{BC},
	\end{equation}
	which fixes the torsion $C_{ABC}$ to be of the Bismut type\footnote{See \cite{Gates:1984nk, Strominger:1986uh} for the original works on the appearance of Bismut connection in string theory, where it is also referred to as the $H$-connection.}:
	\begin{equation}
		C_{ABC} = -3 (J^i)^D{}_{A} (J^i)^E{}_{B} (J^i)^F{}_{C} \nabla_{[D} (J^i)_{EF]}, \qquad \text{for } i = 1,2,3,
	\end{equation}
	where $\Gamma^A{}_{BC}$ is the Levi-Civita connection of the metric $G_{AB}$, and $\nabla_A$ denotes the associated covariant derivative. 
	
	The gauge symmetry of the sigma model is generated by Killing vectors $k_I^A(x)$ that preserve the target space metric and the complex structures
	\begin{equation}
		\mathcal{L}_{k_I} G_{AB} = 0, \qquad \mathcal{L}_{k_I} (J^i)^A{}_{B} = 0,
	\end{equation}
	where $I = 1, \dots, N$. For most of the paper, we will focus on an abelian gauge group generated by a commuting set of Killing vectors. 
	
	In addition to the isometries, there exists a conformal Killing vector $\xi^A$ that generates one-dimensional conformal transformations, and satisfies
	\begin{equation}
		\mathcal{L}_\xi G_{AB} = -G_{AB}, \qquad \mathcal{L}_\xi (J^i)^A{}_{B} = 0, \qquad \mathcal{L}_\xi C_{ABC} = -C_{ABC}. \label{eq:CKV_conditions}
	\end{equation}
	
	The conformal Killing vector can be decomposed into a component orthogonal to the gauge directions:
	\begin{equation}
		\xi_\perp^A = \xi^A - h^I k_I^A,
	\end{equation}
	where the scalar functions $ h^I(x)$ are given by the projection
	\begin{equation}
		h^I = G^{IJ} k_J^A \xi_A, \label{eq: h potentials}
	\end{equation}
	with $G_{IJ} = k_I^A k_{J A}$ and $G^{IJ}$ its inverse. The orthogonality condition is then given by
	\begin{equation}
		G_{AB} \xi_\perp^A k_I^B = 0.
	\end{equation}
	
	A key geometric feature of gauged sigma models is that the gauge-invariant projection $G_{AB}\xi_\perp^B$ of the conformal Killing vector -- not $\xi_A$ itself\footnote{as in the ungauged case \cite{Michelson:1999zf}.} -- is exact with respect to the special conformal generator, satisfying
	\begin{equation}
		\xi_{\perp A} = -\frac{1}{2} \partial_A K. \label{eq: exact CKV}
	\end{equation}

	Finally, one $R$ symmetry factor is generated by the Reeb vector fields associated with the gauged conformal Killing vector as
	\begin{equation}
		(\omega^i)^A = -(\alpha + 1)(J^i)^A{}_{B} \, \xi_\perp^B, \label{eq: Reeb vectors general}
	\end{equation}
	where $\alpha$ is the parameter of the $D(2,1;\alpha)$ superalgebra, and the normalization is fixed by the closure condition. 
	
	These satisfy a useful identity \cite{Mirfendereski:2022omg}
	\begin{equation}
		\omega^{iA} C_{ABC} = (\alpha+1)(J^i)_{BC} + 2 \nabla_{[B} \omega^i_{C]} - 2 k_{I[B} V^{iI}_{C]}, \label{eq: identity omega torsion}
	\end{equation} where
	\begin{equation}
		V^{\rho I}_A = -(\alpha + 1) \, \partial_B h^I \, (J^\rho)^B{}_{A}. \label{eq: useful one form}
	\end{equation}
	
	The general gauged superconformal mechanical sigma model on such a target $\calm$ is described by the Lagrangian
	\begin{equation}\label{eq:gauged_model_general}
		\begin{aligned}
			L ={}& \frac{1}{2} G_{AB} D_t x^A D_t x^B + \frac{i}{2} G_{AB} \chi^A \hat{D}_t \chi^B + a^I v_I + A_A \dot{x}^A \\ &- \frac{i}{2} F_{AB} \chi^A \chi^B - \frac{1}{12} \partial_{[A} C_{BCD]} \chi^A \chi^B \chi^C \chi^D\, ,
	\end{aligned}
	\end{equation}
	where the coupling of the worldline gauge fields $a^I$ to the $(4,4,0)$ multiplets containing real bosonic and fermionic fields $x^A,\chi^A$ is realized through the gauge covariant derivatives
	\begin{align}
		D_t x^A &= \dot{x}^A - a^I k_I^A(x), \label{eq:Dt-bosonic} \\
		\hat{D}_t \chi^A &= \dot{\chi}^A + \left( \Gamma^A{}_{BC} + \frac{1}{2} C^A{}_{BC} \right) \dot{x}^B \chi^C + a^I \left( \nabla^A k_{IB} + \frac{1}{2} C^A{}_{BC} k_I^C \right) \chi^B,  \label{eq:Dt-fermionic}
	\end{align} and scalar functions $v_I(x)$, which satisfy 
	\begin{equation}
		i_{k_I} F = dv_I. \label{eq: vi definition}
	\end{equation}
	
	The local transformations realizing the $D(2,1;\alpha)$ superconformal symmetry are collected in Appendix~\ref{app:D21a-transformations}. We will only need a few structural facts here. Gauge transformations are generated by the Killing vectors $k_I$. The fermionic transformations involve the covariant velocity $D_t x^A$ and the projected conformal Killing vector $\xi_\perp^A$. The two $SU(2)$ $R$-symmetries act through $\omega^i$, with the accompanying gauge-field and fermion transformations, and the fermionic $SU(2)$ action. The corresponding Noether charges are:
	\begin{align}
		M_I &= k_I^A \left( \Pi_A - \frac{i}{4} C_{ABC} \chi^B \chi^C\right) - \frac{i}{2} \chi^A \chi^B \nabla_A k_{IB} - v_I, \label{eq:charge_MI} \\[4pt]
		H &= \frac{1}{2} \Pi_A G^{AB} \Pi_B + \frac{i}{2} F_{AB} \chi^A \chi^B + \frac{1}{12} \partial_{[A} C_{BCD]} \chi^A \chi^B \chi^C \chi^D + a^I M_I, \\
		Q^4 &= \chi^A \Pi_A - \frac{i}{6} C_{ABC} \chi^A \chi^B \chi^C, \\
		Q^i &= \chi^A (J^i)^B{}_A \Pi_B + \frac{i}{2} (J^i)_A{}^D C_{BCD} \chi^A \chi^B \chi^C, \\
		S^4 &= -2 \xi_{\perp A} \chi^A, \qquad
		S^i = 2 (J^i)_{AB} \xi_\perp^B \chi^A, \qquad
		K = 2 \xi_{\perp A} \xi_\perp^A, \\
		R^i &= \omega^{iA} \left( \Pi_A - \frac{i}{4} C_{ABC} \chi^B \chi^C \right) - \frac{i}{2} \left( \nabla_A \omega^i_B + k_{IA} V^{iI}_B \right) \chi^A \chi^B, \\
		\tilde{R}^i &= \frac{i}{4} (J^i)_{AB} \chi^A \chi^B, \label{eq:charge_Rtilde}
	\end{align}
	with the generalized momentum 
	\begin{equation}
		\Pi_A = {p}_A - A_A - \frac{i}{2} \left( \Gamma_{BCA} - \frac{1}{2} C_{BCA} \right) \chi^B \chi^C.
		\label{eq:PiA_def}
	\end{equation}
	Since the worldline gauge fields $a^I$ enter the action without time derivatives, their conjugate momenta vanish as primary constraints. In the Hamiltonian formulation they appear as multipliers for the Gauss-law generators $M_I$, which generate the gauge transformations. The charges $H,D,K$ generate the one-dimensional conformal algebra, with $H$ the original Hamiltonian, $D$ the dilatation generator, and $K$ the special conformal generator. The charges $Q^\mu$ and $S^\mu$ are the ordinary and conformal supercharges, respectively. Finally, $R^i$ and $\tilde R^i$ generate the two $SU(2)$ actions which are later combined into the commuting $SU(2)_R\times SU(2)_L$ generators used in the index.
	
	We also note that the gauged sigma model description on $\calm$ is equivalent to an ungauged description on the quotient $\calm/G$, which is realized in terms of $(3,4,1)$ multiplets\footnote{In general, the equivalent ungauged model can be more complicated, but the quiver models nicely fit into this particular class of gauged sigma models with an equivalent ungauged $(3,4,1)$ description.} and does not have the useful geometric properties of $\calm$. 
	
	\subsection{Quiver Mechanics} \label{sec: quivers}
	
	We now focus on the physically most interesting class of gauged superconformal quantum mechanical models, namely the ones originating from ungauged models with $(3,4,1)$ multiplet structure. This class includes a large family of models, with the quiver superconformal mechanics corresponding to a special member defined by specific geometric data. 
	
	In these models, the bosonic coordinates of the target space $\calm$ naturally split as 
	\begin{equation}
		x^A := x^{\mu a} = (x^{i a}, x^{4 a}), \qquad a=1,\cdots, N, \qquad  i = 1,2,3, \quad \mu = 1,2,3,4,
	\end{equation}
	where $x^{i a}$ parametrize the spatial position of the $a$-th center in $\mathbb{R}^3$, and $x^{4 a}\in\mathbb R$ is an auxiliary affine coordinate whose translations are gauged, generated by 
	\begin{equation}
		k_a = \partial_{4 a}. \label{eq: Killing vectors 3,4,1}
	\end{equation}
	
	The target space metric, torsion, and complex structures take the form
	\begin{align}
		G_{\mu a\, \nu b} &= \delta_{\mu\nu} G_{ab}, \qquad \qquad \; G_{ab} = \frac{1}{2} \partial_{i a} \partial_{i b} \mathcal{H}, \label{eq: Gmu a nu b metric decomposition} \\
		C_{\mu a\, \nu b\, \rho c} &= -\partial_{\lambda a} G_{b c} \, \epsilon_{\lambda \mu \nu \rho}, \qquad
		(J^i)^{\mu a}{}_{\nu b} = (j^i_+)_{\mu \nu} \delta^a_b,
	\end{align}
	where $\calh$ is a scalar Hesse potential, $\epsilon_{1234}=+1$, and $j^i_+$ are the standard 't Hooft symbols with components
	\begin{equation}
		(j^i_+)_{jk} = - \epsilon_{ijk4}, \qquad (j^i_+)_{4\nu} = \delta_{i\nu}.
	\end{equation}
	
	A general property of such gauged models described by $N$ $(4,4,0)$ multiplets is that via one-dimensional automorphic duality  \cite{Delduc:2006yp,Fedoruk:2011aa} they can be reformulated as ungauged sigma models in terms of $N$ $(3,4,1)$ multiplets on $\calm/G$. This is realized by choosing a gauge for the coordinates $x^{4a}$, so that the covariant velocities $D_t x^{4a}$ reduce to the gauge-invariant variables $-B^a$. These are then identified with the auxiliary scalars $D^a$ of the $(3,4,1)$ description. The resulting ungauged model is parameterized by the $3N$ coordinates $x^{ia}$. The role of duality is essentially to trade the gauge redundancy for auxiliary field dependence. More precisely, the D-term constraint of the $(3,4,1)$ theory is identified with the Gauss constraint associated with the gauged shift symmetry in the $(4,4,0)$ formulation.
	
	Such gauged sigma models with a $(3,4,1)$ origin appear in the description of $N$-centered D-brane systems with primitive brane charges, governed by $U(1)^N$ quiver quantum mechanics \cite{Denef:2002ru}, which captures the effective Coulomb branch dynamics. It is obtained after integrating out the sufficiently heavy hypermultiplets, which gives the low energy effective Lagrangian that is purely described in terms of $N$ interacting off-shell $d=1$, $\mathcal{N}=4$ vector multiplets, obtained via dimensional reduction of the $d=4$, $\mathcal{N}=1$ vector multiplet \cite{Smilga:1986rb}. Each such multiplet includes bosonic coordinates $x^i$ ($i=1,2,3$), an auxiliary scalar $D$, and fermions described by a two-component spinor $\lambda_\alpha$ and its conjugate $\bar{\lambda}_\alpha$, forming a $(3,4,1)$ multiplet.
	
	In the $AdS_2$ scaling limit of the Coulomb-branch quiver mechanics the moduli-dependent terms are removed and the remaining scaling theory has $D(2,1;0)$ superconformal symmetry \cite{Anninos:2013nra,Mirfendereski:2020rrk}. The corresponding sigma model can be described either in terms of $N$ off-shell $(3,4,1)$ multiplets in the ungauged formulation or equivalently by $N$ off-shell $(4,4,0)$ multiplets in the gauged formulation with the following fermion redefinitions 
	\begin{equation}
		\chi^{4a} = \frac{1}{\sqrt{2}} \left( \kappa \bar{\lambda}^a + \bar{\kappa} \lambda^a \right), \qquad
		\chi^{ia} = \frac{i \sigma^i}{\sqrt{2}} \left( \bar{\kappa} \lambda^a - \kappa \bar{\lambda}^a \right),
		\label{eq:fermion_redefinition}
	\end{equation}
	where $\kappa$ is a constant complex two-vector used to convert spinor indices to vector indices, normalized such that $\kappa \bar{\kappa} = 1$.
	
	The gauged sigma model target space metric of the superconformal quiver mechanics is determined by a scalar Hesse potential as
	\begin{equation} \label{eq: hesse potential scaling}
		G_{ab} = \frac{1}{2} \partial_{ia}\partial_{ib} \calh \qquad\qquad	\mathcal{H} = -\sum_{\substack{a,b\\ a \neq b}} \frac{|\Gamma_{ab}|}{4r_{ab}} \log  |x_a - x_b|,
	\end{equation} or,
	\begin{equation}
		G_{ab}(x) = \delta_{ab} \sum_{c \neq a} \frac{|\Gamma_{ac}|}{4 r_{ac}^3} - \frac{|\Gamma_{ab}|}{4 r_{ab}^3}, \qquad r_{ab} = |\vec{x}^a - \vec{x}^b|,  \label{eq: scaling Gab}
	\end{equation}
	where $|\Gamma_{ab}|$ are the absolute values of the DSZ intersection products
	\begin{equation}
		\Gamma_{ab} = \left<\Gamma_a, \Gamma_b\right> = -\left<\Gamma_b, \Gamma_a\right>,
	\end{equation}
	which also determine the electromagnetic vector and scalar potentials
	\begin{equation}
		v_a(x)
		=
		\sum_{b\neq a}
		\frac{\Gamma_{ab}}{2r_{ab}},
		\qquad
		A_{4a}=-v_a,
		\qquad
		A_{ia}
		=
		-\sum_{b\neq a}
		\Gamma_{ab}\,A_i^D(\tilde x_{ab}) ,
		\label{eq:quiver-potentials}
	\end{equation}
	where $A_i^D(x)$ is the Dirac monopole potential
	\begin{equation}
		A_i^D(x) = \frac{\epsilon_{ijk} n_j x_k}{2r(x_l n_l - r)},
	\end{equation}
	with $n_i$ a fixed unit vector specifying the direction of the Dirac string, and the relative position vector $\tilde{x}_{ab}$ defined by:
	\begin{equation}
		\tilde{x}_{ab} =
		\begin{cases}
			x_a - x_b & \text{if } a < b, \\
			x_b - x_a & \text{if } a > b.
		\end{cases}
	\end{equation} 
	
	The matrix $G_{ab}$ has the common translation direction as a null vector,
	\begin{equation}
		\sum_b G_{ab}=0 . \label{eq: null condition}
	\end{equation}
	Thus, for a connected quiver, the interacting Coulomb-branch metric is invertible only after removing the common translation direction, i.e. on the relative configuration space.
	
	A scaling configuration is a choice of DSZ charge data for which the homogeneous equilibrium equations have a solution:
	\begin{equation}
		\sum_{b\neq a}
		\frac{\Gamma_{ab}}{2r_{ab}} =0. \label{eq: scaling denef equations}
	\end{equation}
	From the quiver mechanics point of view the scaling limit is understood as the infrared limit of the Coulomb-branch theory near such solutions \cite{Anninos:2013nra}. Since these equations are invariant under the common rescaling $r_{ab}\mapsto \Lambda r_{ab}$, they fix only the ratios of inter-center distances. For example, for three centers with cyclic DSZ orientation, $\Gamma_{12},\Gamma_{23},\Gamma_{31}>0$, (\ref{eq: scaling denef equations}) implies
	\begin{equation}
		\frac{|\Gamma_{12}|}{r_{12}}
		=
		\frac{|\Gamma_{23}|}{r_{23}}
		=
		\frac{|\Gamma_{31}|}{r_{13}},
	\end{equation}
	or equivalently
	\begin{equation}
		r_{12}:r_{23}:r_{13}
		=
		|\Gamma_{12}|:|\Gamma_{23}|:|\Gamma_{31}| .
	\end{equation}
	Thus a three-center scaling configuration exists precisely when $|\Gamma_{12}|,|\Gamma_{23}|,|\Gamma_{31}|$ can form the sides of a possibly degenerate triangle,
	\begin{equation}
		|\Gamma_{12}|+|\Gamma_{23}| \ge |\Gamma_{31}|,
		\qquad
		|\Gamma_{23}|+|\Gamma_{31}| \ge |\Gamma_{12}|,
		\qquad
		|\Gamma_{31}|+|\Gamma_{12}| \ge |\Gamma_{23}| .
		\label{eq: triangle inequality}
	\end{equation}
	In the symmetric equal-distance case this reduces to
	\begin{equation}
		\sum_{b\neq a}\Gamma_{ab}=0,
		\qquad a=1,\ldots,N .
		\label{eq: sum gamma zero}
	\end{equation}
	For example, the equal-distance three-node configuration is scaling for the cyclic charge assignment
	\begin{equation}
		\Gamma_{12}=\Gamma,
		\qquad
		\Gamma_{23}=\Gamma,
		\qquad
		\Gamma_{13}=-\Gamma .
		\label{eq: 3node scaling DSZ assignment}
	\end{equation}
	
	More generally, for $N\geq 3$, a cyclic $N$-node quiver with only nearest-neighbor DSZ pairings of equal magnitude gives a simple symmetric scaling configuration. Taking
	\begin{equation}
		\Gamma_{a,a+1}=k,
		\qquad
		\Gamma_{a+1,a}=-k,
		\qquad
		k>0,
	\end{equation}
	with cyclic identification of the node labels, each node has one incoming and one outgoing pairing of equal magnitude. Hence \eqref{eq: sum gamma zero} is automatically satisfied.
	
	The above scaling quiver data satisfy the $D(2,1;0)$ invariance conditions of the gauged sigma model through the key identities:
	\begin{align}
		x^{ic} \partial_{ic} G_{ab} &= -3 G_{ab}, &
		G_{ab} x^{ib} &= -\partial_{ia} \left( G_{bc} x^{jb} x^{jc} \right), & \epsilon_{ijk} x^{ja} \partial_{ka} G_{bc} &= 0, \label{eq: quiver metric identities} \\
		x^{jb}\partial_{jb}A_{4a}
		&=
		-A_{4a},
		&
		x^{jb}\partial_{jb}A_{ia}
		&=
		x^{jb}\partial_{ia}A_{jb}.
	\end{align}
	
	The first identity in (\ref{eq: quiver metric identities}) implies that the conformal Killing vector for the quiver sigma model takes the simple form
	\begin{equation}
		\xi^{\mu a} = x^{\mu a} \qquad  \qquad \xi_\perp^{\mu a} = \delta^{\mu i} x^{ia}, \label{eq: xi quiver}
	\end{equation}
	which determines the potentials (\ref{eq: h potentials}) as 
	\begin{equation}
		h^a=x^{4a}, \label{eq: ha potentials}
	\end{equation}
	and the Reeb vector fields (\ref{eq: Reeb vectors general}) become
	\begin{equation}
		(\omega^i)^{4a} = -x^{ia}, \qquad (\omega^i)^{ja} = \epsilon_{ijk} x^{ka}. \label{eq: omega quiver}
	\end{equation}
	
	Indeed, substituting the quiver data into the general gauged sigma model Lagrangian~\eqref{eq:gauged_model_general} and using automorphic duality and the fermion redefinitions in (\ref{eq:fermion_redefinition}), we obtain 
	\begin{align}
		L &= D^a v_a + \frac{1}{2} G_{ab} \left( \dot{x}^{ia} \dot{x}^{ib} + D^a D^b \right) + A_{ia} \dot{x}^{ia} + \frac{i}{2} G_{ab} \left( \bar{\lambda}^a \dot{\lambda}^b - \dot{\bar{\lambda}}^a \lambda^b \right) \nonumber \\
		&\quad + \partial_{ib}v_a\,\bar{\lambda}^a\sigma^i\lambda^b
		+ \frac{1}{2} \partial_{ic} G_{ab} \left( \bar{\lambda}^a \sigma^i \lambda^b D^c - \epsilon_{ijk} \bar{\lambda}^a \sigma^j \lambda^b \dot{x}^{kc} \right) \nonumber \\
		&\quad + \frac{1}{4} \partial_{jd} \partial_{kc} G_{ab} \, \lambda^a \sigma^j \bar{\lambda}^b \lambda^c \sigma^k \bar{\lambda}^d,
		\label{eq: quiver 341 lagrangian}
	\end{align}
	which is the corresponding $(3,4,1)$ Lagrangian.
	
	These field identifications also apply to the superconformal generators and to the refinement defining the index below. Hence the localization calculation can equivalently be carried out in the ungauged $(3,4,1)$ formulation, with $D^a=-D_t x^{4a}$ and the fermions related by \eqref{eq:fermion_redefinition}. With the same index prescription and boundary conditions, this gives the same saddle equations and local fluctuation factors. The gauged formulation has the advantage that it preserves the HKT geometry which in particular organizes the supercharges and the two $SU(2)$ actions and allows us to trace the geometric origin of individual terms in the fixed point formula\footnote{We discuss further relevant advantages of the $(4,4,0)$ formalism in Section~\ref{sect: discussion}.}.
	
	The gauged formulation also admits a coordinate system adapted to the node-wise complex structure. We derive this in Appendix~\ref{app:metric-decomposition}. The result is a decomposition of the quiver metric into radial, angular, radial-angular mixing, and gauge-fiber sectors.
	
	\section{$D(2,1;0)$ superconformal index} \label{sect index general}
	
	The class of gauged quantum mechanical models we are interested in, namely the superconformal quiver quantum mechanics, realizes $D(2,1;0)$ superconformal algebra, which plays a foundational role in constructing the corresponding index \cite{Gaiotto:2004pc}. Hence, we first briefly review its main properties. 
	
	As a special case of $D(2,1;\alpha)$ algebra, labeled by a continuous real parameter $\alpha$,  it is equivalent to the classical Lie superalgebra (see e.g. \cite{VanDerJeugt:1985hq, Gunaydin:1986fe, Frappat:1996pb, deBoer:1999gea, Gaiotto:2004pc})
	\begin{equation}
		D(2,1;0) = \mathfrak{psu}(1,1|2) \ltimes \mathfrak{su}(2),
	\end{equation}
	which has the bosonic subalgebra 
	\begin{equation}
		\mathfrak{sl}(2,\mathbb{R}) \oplus \mathfrak{su}(2)_L \oplus \mathfrak{su}(2)_R,
	\end{equation}
	where $\mathfrak{sl}(2,\mathbb{R})$ is generated by $L_{-1}, L_0, L_{+1}$, and the two $\mathfrak{su}(2)$ factors correspond to left and right R-symmetries.
	
	The eight supercharges are denoted as $G^{\alpha\alpha'}_p$, where $p = \pm \frac{1}{2}$ denotes conformal weight and $\alpha, \alpha' = \pm$ label doublets under $\mathfrak{su}(2)_R$ and $\mathfrak{su}(2)_L$, respectively. For $D(2,1;0)$, the anticommutation relations take the form
	\begin{align}
		\{G^{\alpha\alpha'}_{+1/2}, G^{\beta\beta'}_{-1/2}\} &= \epsilon^{\alpha\beta} \epsilon^{\alpha'\beta'} L_0 + \epsilon^{\alpha'\beta'} T_R^{\alpha\beta}, \\
		\{G^{\alpha\alpha'}_{\pm 1/2}, G^{\beta\beta'}_{\pm 1/2}\} &= \epsilon^{\alpha\beta} \epsilon^{\alpha'\beta'} L_{\pm 1}.
	\end{align}
	The conjugation property is
	\begin{equation}
		(G^{\alpha\alpha'}_{\pm 1/2})^\dagger = \epsilon_{\alpha\beta} \epsilon_{\alpha'\beta'} G^{\beta\beta'}_{\mp 1/2}.
	\end{equation}
	
	The BPS bound\footnote{which in this special case ($\alpha=0$) is also a sufficient condition for the unitarity of all states contained in the multiplet.} 
	\begin{equation}
		h \geq |j_R|
	\end{equation}
	is saturated by chiral or anti-chiral primaries annihilated by a subset of the $G^{\alpha\alpha'}_{-1/2}$:
	\begin{align}
		G^{++}_{-1/2} |h, h, r_0\rangle = G^{+-}_{-1/2} |h, h, r_0\rangle = 0 &\quad \Rightarrow\quad \text{chiral primary}, \label{eq: D210 chiral} \\
		G^{--}_{-1/2} |h, -h, r_0\rangle = G^{-+}_{-1/2} |h, -h, r_0\rangle = 0 &\quad \Rightarrow\quad \text{anti-chiral primary}, \label{eq: D210 antichiral}
	\end{align}
	which give rise to unitary, lowest weight, infinite dimensional irreducible representations of $D(2,1;0)$ algebra.
	
	To connect the algebraic index with the sigma-model localization problem, we use the dictionary between the spinor-index generators $G^{\alpha\alpha'}_{\pm1/2}$ and the real sigma-model charges $Q^\mu,S^\mu$. The full dictionary is given in Appendix~\ref{app:D21a-dictionary}. The complex supercharge used in the index is
	\begin{equation}
		\mathcal G_{\pm1/2}
		=
		G^{-+}_{\pm1/2}
		=
		\frac{i}{\sqrt{2\omega}}
		\left(
		\mathcal Q_- \mp i\omega\mathcal S_-
		\right),
		\qquad
		\mathcal Q_-=\frac12(Q^3+iQ^4),
		\qquad
		\mathcal S_-=\frac12(S^3+iS^4).
		\label{eq: dictionary G-+}
	\end{equation}
	
	For the bosonic generators, the mapping to the standard conformal basis is given by
	\begin{align}
		L_0 &= \frac{1}{2\omega}(H + \omega^2 K), \qquad
		L_\pm = \frac{1}{2\omega}(H - \omega^2 K \pm 2i \omega D), \label{eq: dictionary L0 L-}
	\end{align}
	and the two commuting $\mathfrak{su}(2)$ R-symmetry factors are 
	\begin{equation}
		J^i_R = -(R^i + \tilde{R}^i), \qquad J^i_L = \tilde{R}^i. \label{eq: JL and JR}
	\end{equation}
	
	Here, $\omega$ is an arbitrary normalization needed to match the units of generators, that is inherent to conformal quantum mechanics \cite{deAlfaro:1976vlx} and will be fixed later on as $2\omega =1$ for convenience.
	
	This completes the brief review of the $D(2,1;0)$ superalgebra needed for the index construction. We relegate the more detailed and self-contained summary of the $D(2,1;\alpha)$ algebra with arbitrary $\alpha$ -- including its BPS shortening conditions and short representations, and a complete dictionary of generators in the two different conventions -- to Appendix~\ref{app:D21a-algebra}.
	
	We define the superconformal index associated with the $D(2,1;0)$ gauged superconformal sigma models by the algebraic index \cite{Gaiotto:2004pc},
	\begin{equation}
		\mathcal{I}_\pm(y) =  \mathrm{Tr}\left[(-1)^{2J_L^3} \, e^{-\beta \{\mathcal{G}_{\pm 1/2}, \mathcal{G}^\dagger_{\pm 1/2}\}} \, y^{\pm(J_L^3 + J_R^3)} \right], \label{eq: index D210}
	\end{equation}
	where the trace is taken over gauge-invariant states. Here $J_L^3$ and $J_R^3$ are the Cartan generators of the $\mathfrak{su}(2)_L \times \mathfrak{su}(2)_R$ R-symmetry, and $\mathcal{G}_{\pm 1/2}$  denotes the complex supercharge:
	\begin{equation}
		\mathcal{G}_{\pm 1/2} = G^{-+}_{\pm 1/2}, \qquad  \qquad (\mathcal{G}_{\pm 1/2})^\dagger = - G^{+-}_{\mp 1/2}.
	\end{equation} 
	
	The supercharge used for localization is the corresponding real combination
	\begin{equation}
		\mathscr Q_\pm
		:=
		\mathcal G_{\pm1/2}
		+
		\mathcal G_{\pm1/2}^{\dagger}.
		\label{eq: dirac operator}
	\end{equation}
	
	The chosen supercharge has unit charge under $2J_L^3$,
	\begin{equation}
		[2J_L^3,\calg_{\pm 1/2}] = \calg_{\pm 1/2} .
	\end{equation} 
	Thus $(-1)^{2J_L^3}$ provides the grading of the supersymmetric complex. The refinement generator commutes with the supercharge,
	\begin{equation}
		[\mathcal{G}_{\pm 1/2}, J_L^3+J_R^3]=0,
	\end{equation}
	and can therefore be inserted as an equivariant fugacity in the index.
	
	Thus $\mathcal{I}_\pm$ receives contributions only from states annihilated by both $\mathcal{G}_{\pm 1/2}$ and $\mathcal{G}^\dagger_{\pm 1/2}$,
	\begin{equation}
		\mathcal{G}_{\pm 1/2} \left| h, j_L, j_R \right\rangle = 0, \qquad \mathcal{G}^\dagger_{\pm 1/2} \left| h, j_L, j_R \right\rangle = 0, \label{eq: index BPS conditions}
	\end{equation}
	since the $(-1)^{2J_L^3}$ grading ensures cancellations among non-BPS states.  $\mathcal{I}_+ $ (respectively $\mathcal{I}_- $) counts chiral (respectively anti-chiral) lowest-weight multiplets of $D(2,1;0)$, as the conditions~\eqref{eq: index BPS conditions} coincide with the BPS conditions~\eqref{eq: D210 chiral primary}--\eqref{eq: D210 antichiral primary}. Thus, (\ref{eq: index D210}) is a protected quantity counting only the $D(2,1;0)$ short multiplets and is invariant under continuous symmetry-preserving deformations.
	
	The anticommutator defines an auxiliary Hamiltonian,
	\begin{equation}
		\{\mathcal{G}_{\pm 1/2}, \mathcal{G}^\dagger_{\pm 1/2} \}
		=
		L_0 \mp J_R^3
		\equiv
		\mathcal H_\pm .
		\label{eq: aux hamiltonian D210}
	\end{equation}
	
	Thus (\ref{eq: index D210}) takes the form
	\begin{equation}
		{\mathcal{I}}_\pm(\lambda) = \mathrm{Tr} \left[(-1)^{2J_L^3} e^{-\beta \breve{\mathcal{H}}_\pm} \right], \label{eq: refined index general}
	\end{equation}
	with
	\begin{equation}
		\breve{\mathcal H}_\pm =L_0\mp J_R^3 \mp\frac{i\lambda}{\beta}(J_L^3+J_R^3), \label{eq: refined hamiltonian}
	\end{equation}
	where we have written the fugacity as
	\begin{equation}
		y=e^{i\lambda}.
	\end{equation}
	
	Therefore (\ref{eq: index D210}) is represented by the Euclidean path integral over the Euclidean action $\breve S^{E}_{\pm}$ obtained from the refined auxiliary Hamiltonian $\breve{\calh}_\pm$, which includes an $\mathcal{O}\left(1/\beta\right)$ deformation of the unrefined auxiliary model $\calh_\pm$. The field configurations thus retain periodic boundary conditions.\footnote{An equivalent approach would be to compute the path integral over the Euclidean Lagrangian corresponding to $\calh_\pm$ (instead of $\breve{\calh}_\pm$) under twisted boundary conditions.}  Thus the problem is now reduced to evaluating the equivariant Witten index of the auxiliary gauged sigma model. 
	
	\subsection{Localization}
	
	The localization of superconformal quantum-mechanical indices has been formulated in \cite{Dorey:2018klg,Barns-Graham:2018xdd,Dorey:2019kaf,Raeymaekers:2024usy,Raeymaekers:2024ics}. The protected quantity is defined as a superconformal trace, but its path-integral representation is most naturally written in the auxiliary supersymmetric problem generated by the anticommutator of the chosen supercharge with its conjugate. In the present case this auxiliary Hamiltonian is \eqref{eq: aux hamiltonian D210}. It contains the special conformal generator $K$, which provides the standard confining term of conformal quantum mechanics. There is also a second effect. Passing to the auxiliary problem shifts the magnetic data by the Reeb one-form. These shifted magnetic data no longer obey the original conformal-invariance conditions of the scaling sigma model. Thus the auxiliary problem preserves the localizing supersymmetry, but it is not conformally invariant. 
	
	The refinement by $J_L^3+J_R^3$ is implemented either as twisted boundary conditions or, equivalently, as an equivariant deformation of the auxiliary Hamiltonian. We use the latter description. In this formulation the fields remain periodic, while the refined Hamiltonian contains an $\mathcal O(\beta^{-1})$ coupling to the vector field generating the $J_L^3+J_R^3$ action. This equivariant parameter lifts the transverse rotational zero modes and turns them into the oscillator determinants. In this section we first write the index as the path integral of this refined auxiliary supersymmetric problem. We then obtain the corresponding fixed-locus equations and specialize them to the quiver model.
	
	We start by deriving the corresponding Lagrangian for the refined Hamiltonian (\ref{eq: refined hamiltonian}) in the most general gauged sigma models with $D(2,1;0)$ superconformal symmetry for completeness, although we will specialize our computation to the quiver system later on. Using the canonical expressions for the conserved charges (\ref{eq:charge_MI}-\ref{eq:charge_Rtilde}), together with the identifications (\ref{eq: dictionary L0 L-},\ref{eq: JL and JR}), we obtain 
	\begin{align}
		L_0 &= \frac{1}{2\omega} \left( \frac{1}{2} \Pi_A \Pi^A + \frac{i}{2} F_{AB} \chi^A \chi^B + \frac{1}{12} \partial_{[A} C_{BCD]} \chi^A \chi^B \chi^C \chi^D + a^I M_I + 2 \omega^2 \xi_{\perp A} \xi_\perp^A \right), \\
		J_R^3 &= - \omega^{3A} \Pi_A + i (\nabla_A \omega^3_B)  \chi^A \chi^B, \\
		J_L^3 &= \frac{i}{4} (J^3)_{AB} \chi^A \chi^B.
	\end{align}
	where, to simplify the expression for $J_R^3$ we used the identity (\ref{eq: identity omega torsion}).
	
	The unrefined auxiliary Hamiltonian (\ref{eq: aux hamiltonian D210}) is then given by
	\begin{align}
		{\mathcal{H}}_\pm &= \frac{1}{2} \Pi_A \Pi^A + \frac{i}{2} F_{AB} \chi^A \chi^B + \frac{1}{12} \partial_{[A} C_{BCD]} \chi^A \chi^B \chi^C \chi^D + \frac{1}{2} \omega^{3A} \omega^3_A  + a^I M_I \nonumber \\
		&\quad \pm \left(\omega^{3A} \Pi_A - i \nabla_{[A} \omega^3_{B]} \chi^A \chi^B \right),
	\end{align}
	where we fixed the normalization by setting $2\omega = 1$.
	
	The chemical potential for the refinement enters through the refined auxiliary Hamiltonian $\breve{\calh}_\pm$ (\ref{eq: refined hamiltonian}), which gives the momentum conjugacy relation
	\begin{equation}
		\frac{\partial \breve{\mathcal{H}}_\pm}{\partial p_A} = \Pi^A + a^I k_I^A \pm \omega^{3A} \pm i \beta^{-1} \lambda \omega^{3A}. \label{eq: Hamilton EOM}
	\end{equation} 
	
The corresponding refined Lagrangian $\breve{\mathcal{L}}_\pm$ is obtained via the inverse Legendre transform as 
	\begin{equation}
		\breve{\mathcal{L}}_\pm = p_A \dot{x}^A + \frac{i}{2} G_{AB} \chi^B\dot{\chi}^A  - \breve{\mathcal{H}}_\pm,
	\end{equation} 
	which, using the canonical expression for $\Pi_A$ and after cancellations, gives
	\begin{align}
		\breve{\mathcal{L}}_\pm &= A_A \dot{x}^A + a^I v_I + \frac{1}{2} G_{AB} D_t x^A D_t x^B + \frac{i}{2} G_{AB} \chi^A \hat{D}_t \chi^B - \frac{i}{2} F_{AB} \chi^A \chi^B \nonumber \\
		&\quad - \frac{1}{12} \partial_{[A} C_{BCD]} \chi^A \chi^B \chi^C \chi^D - \frac{\lambda^2}{2\beta^2} \omega^{3A} \omega^3_A + \frac{i\lambda}{\beta} \omega^{3A} \omega^3_A \nonumber \\
		&\quad \mp \dot{x}^A \omega^3_A \mp i \beta^{-1} \lambda \dot{x}^A \omega^3_A \pm a^I k^A_I \omega^3_A \pm i \beta^{-1} \lambda a^I  k^A_I \omega^3_A  \nonumber \\
		&\quad \pm i\nabla_{[A}\omega^3_{B]}\chi^A\chi^B
		\mp\frac{\lambda}{\beta}
		\left(  \nabla_{[A}\omega^3_{B]}+\frac14(J^3)_{AB} 	\right)\chi^A\chi^B . \label{eq:call-final-simplified}
	\end{align}
	
	A more useful expression however is its expanded form
	\begin{align}
		\breve{\mathcal L}_\pm
		={}&a^Iv_I+A_A\dot x^A
		-\frac i2F_{AB}\chi^A\chi^B
		-\frac1{12}\partial_{[A}C_{BCD]}
		\chi^A\chi^B\chi^C\chi^D
		+\frac i4C_{ABC}\chi^A\dot x^B\chi^C \nonumber \\
		&+\frac12G_{AB}\dot x^A\dot x^B
		-a^Ik_{IA}\dot x^A
		+\frac i2G_{AB}\chi^A\dot\chi^B
		+\frac i2\Gamma_{ABC}\chi^A\dot x^B\chi^C \nonumber \\
		&-\frac{\lambda^2}{2\beta^2}\omega^3_A\omega^{3A}
		+\frac{i\lambda}{\beta}\omega^3_A\omega^{3A} \nonumber \\
		&\mp\omega^3_A\dot x^A
		\mp\frac{i\lambda}{\beta}\omega^3_A\dot x^A
		\pm a^Ik_{IA}\omega^{3A}
		\pm\frac{i\lambda}{\beta}a^Ik_{IA}\omega^{3A} \nonumber \\
		&\pm\frac i2
		\left(
		\omega^{3C}C_{CAB}
		+2k_{IA}V_B^{3I}
		-(J^3)_{AB}
		\right)\chi^A\chi^B \nonumber \\
		&\mp\frac{\lambda}{2\beta}
		\left(
		\nabla_A\omega^3_B
		+k_{IA}V_B^{3I}
		+\frac12\omega^{3C}C_{CAB}
		\right)\chi^A\chi^B \nonumber \\
		&+\frac12a^Ia^Jk_{IA}k_J^A
		+\frac i2a^I\nabla_Ak_{IB}\chi^A\chi^B
		+\frac i4a^Ik_I^AC_{ABC}\chi^B\chi^C .
		\label{eq: susy exact generalization minkowski}
	\end{align}
	
	At this stage, it is useful to introduce the gauge covariant frame. For this, we decompose
	the Reeb vector into the part along the gauge orbits and the part perpendicular to them,
	\begin{equation}
		\omega^{3A}
		=
		\omega_\perp^{3A}
		+
		\Omega^{3I}k_I^A,
		\qquad
		\Omega^{3I}:=G^{IJ}k_{JA}\omega^{3A},
		\qquad
		k_{IA}\omega_\perp^{3A}=0 ,
		\label{eq:main-omega-horizontal-vertical}
	\end{equation}
	where $G_{IJ}:=k_I^Ak_{JA}$.
	
	We Wick-rotate and rescale the circle according to
	\begin{equation}
		t_{\rm M}=-i\tau=-i\beta\tilde\tau,
		\qquad
		a_{\rm M}^I=\frac{i}{\beta}a^I,
		\qquad 0\leq\tilde\tau\leq1.
		\label{eq:wick-gauge-connection}
	\end{equation}
	The vertical piece $\Omega^{3I}k_I^A$ can then be combined with the gauge field by defining
	\begin{equation}
		a_{\rm eff}^I
		:=
		a^I \pm \lambda\Omega^{3I},
		\label{eq:main-effective-gauge-field}
	\end{equation}
	which transforms as a gauge connection and thus the refined action can be written in a gauge-covariant form most transparently in terms of $a_{\rm eff}^I$.
	
	We also introduce
	\begin{equation}
		A_A^{(\pm)}:=A_A\mp\omega_A^3,
		\qquad
		F^{(\pm)}:=F\mp d\omega^3,
		\qquad
		v_I^{(\pm)}:=v_I\pm k_{IA}\omega^{3A},
		\label{eq:main-shifted-magnetic-data}
	\end{equation}
	which satisfy the moment map equations
	\begin{equation}
		i_{k_I}F^{(\pm)}
		=
		dv_I^{(\pm)} .
	\end{equation}
	Then the refined Euclidean action takes the general form
	\begin{align}
		\breve S_\pm^E
		=
		\int_0^1 d\tilde\tau\,\bigg[
		&
		\frac{1}{2\beta}
		G_{AB}
		\left(
		D_{t,{\rm eff}}x^A \mp \lambda\omega_\perp^{3A}
		\right)
		\left(
		D_{t,{\rm eff}}x^B \mp \lambda\omega_\perp^{3B}
		\right)
		\nonumber
		\\
		&-i a_{\rm eff}^Iv_I^{(\pm)}
		-iA_A^{(\pm)}\dot x^A
		\pm i\lambda\Omega^{3I}v_I
		-i\lambda\omega^3_{\perp A}\omega_\perp^{3A}
		\nonumber
		\\
		&+
		\frac12G_{AB}\chi^A\hat{D}_{t,{\rm eff}}\chi^B
		\nonumber
		\\
		&\pm
		\frac{\lambda}{2}
		\left(
		\nabla_A\omega^3_{\perp B}
		+
		(\nabla_A\Omega^{3I})k_{IB}
		+
		\frac12\omega_\perp^{3C}C_{CAB}
		+
		k_{IA}V_B^{3I}
		\right)\chi^A\chi^B
		\nonumber
		\\
		&+
		\beta\bigg(
		\frac{i}{2}F_{AB}^{(\pm)}\chi^A\chi^B
		+
		\frac{1}{12}
		\partial_{[A}C_{BCD]}
		\chi^A\chi^B\chi^C\chi^D
		\bigg)
		\bigg],
		\label{eq: SE leading terms}
	\end{align}
	Here $D_{t,\rm eff}$ and $\hat{D}_{t,\rm eff}$ have the forms \eqref{eq:Dt-bosonic} and \eqref{eq:Dt-fermionic}, with $a\to a_{\rm eff}$ and derivatives taken on the unit Euclidean circle. The derivation of \eqref{eq: SE leading terms} and its gauge invariance are given in Appendix~\ref{sec:refined-gauged-sigma-model}. 
	
	The kinetic square in \eqref{eq: SE leading terms} identifies the equivariant rotational locus
	\begin{equation}
		D_{t,{\rm eff}}x^A
		=
		\pm \lambda\omega_\perp^{3A},
		\label{eq:fixed-locus-effective-form}
	\end{equation}
	where the component of the $R$-symmetry vector along the gauge orbits has been absorbed into the effective gauge connection, while the perpendicular part remains as the physical fixed-locus condition. 
	
	Equivalently, in terms of the original gauge field,
	\begin{equation}
		\dot x^A
		=
		a^Ik_I^A
		\pm\lambda\omega^{3A}.
		\label{eq: BPS locus}
	\end{equation}
	
	The same fixed locus can be obtained directly from the refined supersymmetry transformations, which we derive in Appendix~\ref{app:D21a-transformations}. For the real localizing combination $\mathscr Q_\pm$ defined in \eqref{eq: dirac operator}, we obtain
	\begin{equation}
		\delta_{\mathscr Q_\pm}x^A
		=
		i\epsilon\chi^A,
		\qquad
		\delta_{\mathscr Q_\pm}\chi^A
		=
		-\frac{i\epsilon}{\beta}
		\left(
		\dot x^A-a^Ik_I^A
		\mp
		\lambda\omega^{3A}
		\right),
		\label{eq:refined-Q-variation}
	\end{equation}
	so $\delta_{\mathscr Q_\pm}\chi^A=0$ gives \eqref{eq: BPS locus}.

	\subsection{The fixed-point formula} \label{subsec:coulomb-fixed-point-contribution}
	
	We now specialize to the quiver mechanics. The gauge generators and the Reeb vector are
	\begin{equation}
		k_b^{\mu a}=\delta_b^a\delta_4^\mu,\qquad
		\omega^{3,\mu a}=(x^{2a},-x^{1a},0,-x^{3a}).
		\label{eq: Killing kI}
	\end{equation}
	Thus $\Omega^{3a}=-x^{3a}$ and $\omega_\perp^{3,\mu a}=(x^{2a},-x^{1a},0,0)$. We remove the common translation multiplet. The relative metric is positive for a connected quiver, since for a real node vector $u^a$,
	\begin{equation}
		u^aG_{ab}u^b
		=\sum_{a<b}\frac{|\Gamma_{ab}|}{4r_{ab}^3}(u^a-u^b)^2,
		\label{eq:relative-metric-positive}
	\end{equation}
	and so its only null vector is the common translation. Inverses and primed determinants below are taken on the relative space, and the centers in each interacting pair are assumed separated.
	
	The fiber coordinates enter through the gauge-invariant combination $B^a=a^a-\dot x^{4a}$. Define its refined shift by
	\begin{equation}
		b^a=B^a-s\lambda x^{3a},
		\label{eq:quiver-shifted-auxiliary}
	\end{equation}
	where 
	\begin{equation}
		s=\pm 1. 
	\end{equation}
	
	We choose the translation gauge $x^{4a}=0$, so $B^a=a^a$. The Faddeev--Popov operator is $\delta x^{4a}/\delta\varepsilon^b =\delta^a_b$, including on the constant modes. Both this gauge fixing and the shift from $a$ to $b$ have unit Jacobian. The quotient contains three spatial bosons, four fermions and the auxiliary field $b$ per relative multiplet.
	
	The equivariant velocity and its components are
	\begin{equation}
		\mathcal L_s^A=\dot x^A-a^ck_c^A-s\lambda\omega^{3A},
		\label{eq:quiver-fixed-locus-L}
	\end{equation}
	\begin{align}
		\mathcal L_s^{1a}&=\dot x^{1a}-s\lambda x^{2a},&
		\mathcal L_s^{2a}&=\dot x^{2a}+s\lambda x^{1a},\nonumber\\
		\mathcal L_s^{3a}&=\dot x^{3a},&
		\mathcal L_s^{4a}&=-b^a.
		\label{eq:quiver-localizing-vector-components}
	\end{align}
	Dots now denote derivatives on the unit Euclidean circle. The auxiliary source and the rotation moment map are
	\begin{equation}
		U_{s,a}=v_a-sG_{ab}x^{3b},\qquad
		\mu_s=x^{3a}v_a+sG_{ab}(x^{1a}x^{1b}+x^{2a}x^{2b}),
		\label{eq:localization-U-mu}
	\end{equation}
	which follow directly from the bosonic terms of \eqref{eq: SE leading terms}:
	\begin{equation}
		-iB^av_a+si\omega_A^3\mathcal L_s^A-iA_{ia}\dot x^{ia}
		=-ib^aU_{s,a}-iA^{(s)}_{ia}\dot x^{ia}-si\lambda\mu_s,
		\label{eq:localization-bosonic-regrouping}
	\end{equation}
	where the shifted connection
	\begin{equation}
	A^{(s)}_{ia}=A_{ia}-s\omega^3_{ia}, \qquad F^{(s)}=dA^{(s)}
	\end{equation}
	obeys
	\begin{equation}
		A^{(s)}_{4a}=-U_{s,a},\qquad
		F^{(s)}_{4a,ib}=\partial_{ib}U_{s,a},\qquad
		\omega_\perp^{3,ia}F^{(s)}_{ia,jb}=\partial_{jb}\mu_s.
		\label{eq:localization-shifted-identities}
	\end{equation}
	
	The supersymmetry on the quotient is
	\begin{equation}
		Qx^{ia}=i\chi^{ia},\qquad
		Q\chi^{ia}=-\frac{i}{\beta}\mathcal L_s^{ia},\qquad
		Q\chi^{4a}=\frac{i}{\beta}b^a,\qquad
		Qb^a=-i\dot\chi^{4a},
		\label{eq:quiver-off-shell-Q}
	\end{equation}
	which follows from the parent transformations.
	
	Introducing
	\begin{align}
		\Psi_{\rm loc}&=\int_0^1d\tilde\tau\,
		\left[\frac{i}{2}G_{ab}\chi^{ia}\mathcal L_s^{ib}
		-\frac{i}{2}G_{ab}\chi^{4a}b^b-\beta U_{s,a}\chi^{4a}\right],
		\label{eq:localization-fermion}\\
		\Psi_C&=-\frac{i\beta}{12}\int_0^1d\tilde\tau\,
		C_{ABC}\chi^A\chi^B\chi^C,
		\label{eq:localization-torsion-fermion}\\
		S_{{\rm eq},s}&=\int_0^1d\tilde\tau\,
		\left[-iA^{(s)}_{ia}\dot x^{ia}
		+\frac{i\beta}{2}F^{(s)}_{ia,jb}\chi^{ia}\chi^{jb}
		-si\lambda\mu_s\right],
		\label{eq:localization-equivariant-action}
	\end{align}
	we find that applying $Q$ gives the exact decomposition
	\begin{equation}
		\breve S_s^E=Q\Psi_{\rm loc}+Q\Psi_C+S_{{\rm eq},s},
		\qquad QS_{{\rm eq},s}=0.
		\label{eq:localization-exact-decomposition}
	\end{equation}

	For a positive parameter $t$, set
	\begin{equation}
		S_s(t)=S_{{\rm eq},s}+Q\Psi_C+tQ\Psi_{\rm loc}.
		\label{eq:localization-t-family}
	\end{equation}
	At $t=1$ this is the original refined action. With the supersymmetric measure and boundary conditions for which integrated $Q$-variations vanish, the usual localization argument gives
	\begin{equation}
		\frac{d}{dt}\int D\Phi\,e^{-S_s(t)}
		=-\int D\Phi\,Q\!\left(\Psi_{\rm loc}e^{-S_s(t)}\right)=0,
		\qquad \Phi=(x^{ia},b^a,\chi^{\mu a}).
		\label{eq:localization-t-independence}
	\end{equation}
	We may therefore take $t\to\infty$ while keeping $\beta$ fixed.
	
	The bosonic part multiplied by $t$ is
	\begin{equation}
		t\int_0^1d\tilde\tau\,
		\left[\frac{G_{ab}}{2\beta}
		(\mathcal L_s^{ia}\mathcal L_s^{ib}+b^ab^b)-ib^aU_{s,a}\right].
		\label{eq:localization-bosonic-deformation}
	\end{equation}
	Completing the auxiliary square gives
	\begin{equation}
		\frac1{2\beta}b^TGb-ib^TU_s
		=\frac1{2\beta}(b-i\beta G^{-1}U_s)^TG(b-i\beta G^{-1}U_s)
		+\frac\beta2U_s^TG^{-1}U_s.
		\label{eq:localization-square}
	\end{equation}
	The real Gaussian integral over $b$ therefore leaves the nonnegative bosonic exponent
	\begin{equation}
		t\int_0^1d\tilde\tau\,
		\left[\frac{G_{ab}}{2\beta}\mathcal L_s^{ia}\mathcal L_s^{ib}
		+\frac\beta2U_s^TG^{-1}U_s\right].
		\label{eq:localization-positive-functional}
	\end{equation}
	Its zeros obey $\mathcal L_s^{ia}=0$ and $U_s=0$, and the auxiliary saddle is then $b=0$. For real $\lambda\notin2\pi\mathbb Z$, $x^{1a}+ix^{2a}$ is proportional to $e^{-is\lambda\tilde\tau}$, and so periodicity forces it to vanish. The third velocity fixes $x^3$ to be constant. Hence the saddle equations are
	\begin{equation}
		x^{1a}=x^{2a}=0,\qquad x^{3a}=z^a,\qquad B^a=s\lambda z^a,
		\qquad E_{s,a}(z):=v_a(z)-sG_{ab}(z)z^b=0.
		\label{eq:section equations}
	\end{equation}

	Using the fields and coefficient functions defined above, the refined quiver action is
	\begin{align}
		S_s(t)=\int_0^1d\tilde\tau\,\bigg[&
		\frac{t}{2\beta}G_{ab}\Big[
		(\dot x^{1a}-s\lambda x^{2a})
		(\dot x^{1b}-s\lambda x^{2b})\nonumber\\
		&\qquad +(\dot x^{2a}+s\lambda x^{1a})
		(\dot x^{2b}+s\lambda x^{1b})
		+\dot x^{3a}\dot x^{3b}+b^ab^b\Big]\nonumber\\
		&-itb^aU_{s,a}
		+\frac t2G_{ab}\chi^{\mu a}\dot\chi^{\mu b}
		-st\lambda G_{ab}\chi^{1a}\chi^{2b}
		+it\beta\chi^{4a}(\partial_{ib}U_{s,a})\chi^{ib}\nonumber\\
		&-\frac t2(\partial_{jc}G_{ab})
		\chi^{jc}\chi^{ia}\mathcal L_s^{ib}
		+\frac t2(\partial_{jc}G_{ab})\chi^{jc}\chi^{4a}b^b\nonumber\\
		&+\frac12\epsilon_{\ell ij}(\partial_{\ell c}G_{ab})
		\mathcal L_s^{ic}\chi^{ja}\chi^{4b}
		-\frac14\epsilon_{\ell ij}(\partial_{\ell c}G_{ab})
		b^c\chi^{ia}\chi^{jb}\nonumber\\
		&-\frac\beta{12}\epsilon_{\ell\nu\rho\sigma}
		(\partial_{\mu a}\partial_{\ell b}G_{cd})
		\chi^{\mu a}\chi^{\nu b}\chi^{\rho c}\chi^{\sigma d}\nonumber\\
		&-iA^{(s)}_{ia}\dot x^{ia}
		+\frac{i\beta}{2}F^{(s)}_{ia,jb}\chi^{ia}\chi^{jb}
		-si\lambda\mu_s\bigg].
		\label{eq:full-deformed-quiver-component-action}
	\end{align}

	Let $p$ be a nondegenerate solution of \eqref{eq:section equations}, and expand the fields around it as
	\begin{equation}
		x^{1,2}=t^{-1/2}X^{1,2},\qquad
		x^3=z_p+t^{-1/2}X^3,\qquad b=t^{-1/2}h,\qquad
		\chi^{\mu a}=t^{-1/2}\psi^{\mu a}.
		\label{eq:quiver-bosonic-fluctuation-expansion}
	\end{equation}
	
	The equivariant velocities and the metric expand as
	\begin{align}
		\mathcal L_s^{1a}&=t^{-1/2}(\dot X^{1a}-s\lambda X^{2a}),&
		\mathcal L_s^{2a}&=t^{-1/2}(\dot X^{2a}+s\lambda X^{1a}),\nonumber\\
		\mathcal L_s^{3a}&=t^{-1/2}\dot X^{3a},&
		\mathcal L_s^{4a}&=-t^{-1/2}h^a,
		\label{eq:localization-expanded-velocities}\\
		G_{ab}(x)&=G_{ab}(p)+t^{-1/2}(\partial_{ic}G_{ab})_pX^{ic}
		+O(t^{-1}).&&
		\label{eq:localization-expanded-metric}
	\end{align}
	Substitution into the first group gives
	\begin{align}
		\frac{t}{2\beta}G_{ab}
		(\mathcal L_s^{ia}\mathcal L_s^{ib}+b^ab^b)
		={}&\frac{G_{ab}(p)}{2\beta}\Big[
		(\dot X^{1a}-s\lambda X^{2a})(\dot X^{1b}-s\lambda X^{2b})\nonumber\\
		&+(\dot X^{2a}+s\lambda X^{1a})(\dot X^{2b}+s\lambda X^{1b})
		+\dot X^{3a}\dot X^{3b}+h^ah^b\Big]+O(t^{-1/2}).
		\label{eq:localization-group-one}
	\end{align}
	
	Since $U_s(p)=0$, the source starts at first order in the fluctuations. At the collinear background, the transverse derivatives of the pairwise distances vanish,
	\begin{equation}
	\left.\partial_{ic}r_{ab}\right|_p
	=(\delta_{ac}-\delta_{bc})
	\left.\frac{x^{ia}-x^{ib}}{r_{ab}}\right|_p
	=0,\qquad i=1,2.
	\label{eq:localization-transverse-distance-derivative}
	\end{equation}
	Both $G_{ab}$ and $v_a$ depend only on these distances, so $(\partial_{1b}U_{s,a})_p=(\partial_{2b}U_{s,a})_p=0$. Write $\partial_b=\partial/\partial z^b$ on the collinear locus. Using the symmetry of $\partial_bG_{ac}$ in its node indices and the degree $-3$ homogeneity of $G$ gives
	\begin{equation}
		\partial_b(G_{ac}z^c)
		=G_{ab}+z^c\partial_bG_{ac}
		=G_{ab}+z^c\partial_cG_{ab}=-2G_{ab}.
		\label{eq:localization-differentiate-Gz}
	\end{equation}
	Define the relative Jacobian
	\begin{equation}
		\mathcal N_{s,ab}(p):=(\partial_{3b}U_{s,a})_p
		=\left[\partial_bv_a+2sG_{ab}\right]_p.
		\label{eq:zero-mode-Jacobian-matrix}
	\end{equation}
	It is real symmetric since for $a\ne b$,
	$\partial_bv_a=\Gamma_{ab}(z^a-z^b)/(2|z^a-z^b|^3)
	=\partial_av_b$. Thus
	\begin{align}
		U_{s,a}(x)&=t^{-1/2}\mathcal N_{s,ab}(p)X^{3b}+O(t^{-1}),\nonumber\\
		-itb^aU_{s,a}(x)&=-ih^a\mathcal N_{s,ab}(p)X^{3b}+O(t^{-1/2}).
		\label{eq:localization-surviving-source}
	\end{align}
	For each separated saddle choose a local monopole gauge regular at $p$. Then $A^{(s)}_{ia}(x)\dot x^{ia}=O(t^{-1/2})$, while $\mu_s(x)=z_p^av_a(p)+O(t^{-1/2})$. Hence,
	\begin{equation}
		-itb^aU_{s,a}-iA^{(s)}_{ia}\dot x^{ia}-si\lambda\mu_s
		=-ih^a\mathcal N_{s,ab}(p)X^{3b}
		-si\lambda z_p^av_a(p)+O(t^{-1/2}).
		\label{eq:localization-group-two}
	\end{equation}
	
	Finally, the fermionic bilinears give
	\begin{align}
		&\frac t2G_{ab}\chi^{\mu a}\dot\chi^{\mu b}
		-st\lambda G_{ab}\chi^{1a}\chi^{2b}
		+it\beta\chi^{4a}(\partial_{ib}U_{s,a})\chi^{ib}\nonumber\\
		&\qquad=\frac12G_{ab}(p)\psi^{\mu a}\dot\psi^{\mu b}
		-s\lambda G_{ab}(p)\psi^{1a}\psi^{2b}
		+i\beta\psi^{4a}\mathcal N_{s,ab}(p)\psi^{3b}
		+O(t^{-1/2}).
		\label{eq:localization-group-three}
	\end{align}
	
	Every remaining term in \eqref{eq:full-deformed-quiver-component-action} is suppressed:
	\begin{align}
		-\frac t2(\partial_{jc}G_{ab})\chi^{jc}\chi^{ia}\mathcal L_s^{ib}
		+\frac t2(\partial_{jc}G_{ab})\chi^{jc}\chi^{4a}b^b
		&=O(t^{-1/2}),\nonumber\\
		\frac12\epsilon_{\ell ij}(\partial_{\ell c}G_{ab})
		\mathcal L_s^{ic}\chi^{ja}\chi^{4b}
		-\frac14\epsilon_{\ell ij}(\partial_{\ell c}G_{ab})b^c\chi^{ia}\chi^{jb}
		&=O(t^{-3/2}),\nonumber\\
		-\frac\beta{12}\epsilon_{\ell\nu\rho\sigma}
		(\partial_{\mu a}\partial_{\ell b}G_{cd})
		\chi^{\mu a}\chi^{\nu b}\chi^{\rho c}\chi^{\sigma d}
		&=O(t^{-2}),\nonumber\\
		\frac{i\beta}{2}F^{(s)}_{ia,jb}\chi^{ia}\chi^{jb}
		&=O(t^{-1}).
		\label{eq:localization-group-four}
	\end{align}
	
	Collecting the four groups gives the complete finite action
	\begin{align}
		S_s(t)={}&-si\lambda z_p^av_a(p)\nonumber\\
		&+\int_0^1d\tilde\tau\,\bigg[
		\frac{G_{ab}(p)}{2\beta}\Big(
		(\dot X^{1a}-s\lambda X^{2a})(\dot X^{1b}-s\lambda X^{2b})\nonumber\\
		&\qquad +(\dot X^{2a}+s\lambda X^{1a})(\dot X^{2b}+s\lambda X^{1b})
		+\dot X^{3a}\dot X^{3b}+h^ah^b\Big)\nonumber\\
		&\qquad-ih^a[\partial_bv_a+2sG_{ab}]_pX^{3b}
		+\frac12G_{ab}(p)\psi^{\mu a}\dot\psi^{\mu b}\nonumber\\
		&\qquad-s\lambda G_{ab}(p)\psi^{1a}\psi^{2b}
		+i\beta\psi^{4a}[\partial_bv_a+2sG_{ab}]_p\psi^{3b}
		\bigg]+O(t^{-1/2}).
		\label{eq:localization-full-quadratic-action}
	\end{align}
	We now split this as
	\begin{equation}
		S_s(t)=S_{12,s}^{(2)}+S_{34,s}^{(2)}+S_{{\rm cl},s}(p)
		+O(t^{-1/2}),\qquad S_{{\rm cl},s}(p)=-si\lambda z_p^av_a(p),
		\label{eq:localization-complete-saddle-expansion}
	\end{equation}
	where, with all terms evaluated at $p$,
	\begin{align}
		S_{12,s}^{(2)}=\int_0^1d\tilde\tau\,\bigg[&
		\frac{G_{ab}}{2\beta}\Big(
		(\dot X^{1a}-s\lambda X^{2a})(\dot X^{1b}-s\lambda X^{2b})
		+(\dot X^{2a}+s\lambda X^{1a})(\dot X^{2b}+s\lambda X^{1b})\Big)
		\nonumber\\
		&+\frac12G_{ab}(\psi^{1a}\dot\psi^{1b}+\psi^{2a}\dot\psi^{2b})
		-s\lambda G_{ab}\psi^{1a}\psi^{2b}\bigg],
		\label{eq:S12-total-definition}\\
		S_{34,s}^{(2)}=\int_0^1d\tilde\tau\,\bigg[&
		\frac{G_{ab}}{2\beta}(\dot X^{3a}\dot X^{3b}+h^ah^b)
		-ih^a\mathcal N_{s,ab}X^{3b}\nonumber\\
		&+\frac12G_{ab}(\psi^{3a}\dot\psi^{3b}+\psi^{4a}\dot\psi^{4b})
		+i\beta\psi^{4a}\mathcal N_{s,ab}\psi^{3b}\bigg].
		\label{eq:S34-total-definition}
	\end{align}
	So the local contribution factorizes as
	\begin{equation}
		\mathcal Z_s(p)=Z_{{\rm cl},s}(p)\,Z_{12,s}(p)\,Z'_{34}(p)\,Z_0(p),
		\label{eq:localization-factorization-roadmap}
	\end{equation}
	where $Z_{12,s}$ includes all transverse Fourier modes, $Z'_{34}$ contains the nonconstant longitudinal modes, and $Z_0$ contains the longitudinal constants. Hence at quadratic order the fluctuations split into a transverse $(1,2)$ sector and a longitudinal $(3,4)$ sector. The transverse sector is an equivariant oscillator in each relative two-plane. The longitudinal sector contains fluctuations of $x^{3a}$ and of the shifted fiber combination $B^a-s\lambda x^{3a}$.

	We compute the corresponding determinants in Appendix~\ref{app:nonconstant-determinants}. The longitudinal nonconstant modes give
	\begin{equation}
	Z'_{34}=1,
	\end{equation}
	while each relative transverse two-plane, including its Fourier constants, gives
	\begin{equation}
	Z_{{\rm osc},s}
	=
	\frac{s\,y^{1/2}}{1-y}.
	\end{equation}
	Thus
	\begin{equation}
	Z_{12,s}Z'_{34}
	=
	\left(
	\frac{s\,y^{1/2}}{1-y}
	\right)^{N-1}.
	\label{eq:nonconstant-determinant-main-result}
	\end{equation}
	
	It remains to integrate the longitudinal Fourier constants. Denote them by $X_0^{3a},h_0^a,\psi_0^{3a},\psi_0^{4a}$, and set $d=N-1$. Their action is read directly from the surviving quadratic action:
	\begin{equation}
		S_{0,s}
		=\frac1{2\beta}h_0^aG_{ab}h_0^b
		-ih_0^a\mathcal N_{s,ab}X_0^{3b}
		+i\beta\psi_0^{4a}\mathcal N_{s,ab}\psi_0^{3b}.
		\label{eq:zero-mode-action-E}
	\end{equation}
	With the paired Gaussian normalization, first integrate $h_0$ and then $X_0^3$:
	\begin{align}
	\int\frac{d^dX_0^3\,d^dh_0}{(2\pi\beta)^d}
		e^{-h_0^TGh_0/(2\beta)+ih_0^T\mathcal N_sX_0^3}
		&=\frac{1}{(2\pi\beta)^{d/2}\sqrt{\det{}'G}}
		\int d^dX_0^3\,
		e^{-\beta(X_0^3)^T\mathcal N_s^TG^{-1}\mathcal N_sX_0^3/2}
		\nonumber\\
		&=\frac{\beta^{-d}}
		{\sqrt{\det{}'G}\,
			\sqrt{\det{}'(\mathcal N_s^TG^{-1}\mathcal N_s)}}
		=\frac{\beta^{-d}}{|\det{}'\mathcal N_s|}.
		\label{eq:localization-constant-bosons}
	\end{align}
	Multiplying by the fermionic contribution\footnote{The grading in the defining index trace fixes the overall fermionic phase. With the transverse factor normalized as above, this fixes the phase of the longitudinal fermionic integral.} then leaves
	\begin{equation}
	Z_0(p)
	=
	\operatorname{sign}\det{}'
	[\partial_bv_a+2sG_{ab}]_p.
	\label{eq:Gauss-local-sign}
	\end{equation}
	
	Finally, using the pairwise expression for $v_a$ and $\Gamma_{ba}=-\Gamma_{ab}$,
	\begin{equation}
		z_p^av_a(p)
		=\sum_{a<b}\frac{\Gamma_{ab}(z_p^a-z_p^b)}{2|z_p^a-z_p^b|}
		=-\frac12\sum_{a<b}\Gamma_{ab}\operatorname{sign}_p(z^b-z^a),
		\end{equation}
	we obtain
	\begin{equation}
	Z_{{\rm cl},s}(p)=e^{-S_{{\rm cl},s}(p)}
	=y^{-\frac{s}{2}\sum_{a<b}\Gamma_{ab}\operatorname{sign}_p(z^b-z^a)}. \label{eq:classical-fixed-point-weight}
	\end{equation}
	Combining the factors in \eqref{eq:localization-factorization-roadmap}, we obtain
	\begin{equation}
		\mathcal Z_s(p)=
		\left(\frac{s\,y^{1/2}}{1-y}\right)^{N-1}
		y^{-\frac{s}{2}\sum_{a<b}\Gamma_{ab}\operatorname{sign}_p(z^b-z^a)}
		\operatorname{sign}\det{}'[\partial_bv_a+2sG_{ab}]_p.
		\label{eq:local-isolated-contribution}
	\end{equation}
	The superconformal quiver index is then given by summing over the contributing roots of \eqref{eq:section equations}:
	\begin{equation}
		\begin{aligned}
			\mathcal I_s(y)
			&=\left(\frac{s\,y^{1/2}}{1-y}\right)^{N-1}
			\sum_p\sigma_s(p)\,
			y^{-\frac{s}{2}\sum_{a<b}\Gamma_{ab}\operatorname{sign}_p(z^b-z^a)},\\
			\sigma_s(p)&=\operatorname{sign}\det{}'[\partial_bv_a+2sG_{ab}]_p,
			\qquad v_a(p)-sG_{ab}(p)z_p^b=0.
		\end{aligned}
		\label{eq:final-D210-quiver-index}
	\end{equation}

	\subsection{Relation to the Manschot--Pioline--Sen index}
	
	We now compare \eqref{eq:final-D210-quiver-index} with the Manschot--Pioline--Sen refined configurational index $g_{\rm ref}$ of multi-centered BPS black hole solutions \cite{Manschot:2010qz,Manschot:2011xc}.
	
	Let us first clarify what is being compared. In the MPS construction, $g_{\rm ref}$ is the refined index of the configurational degrees of freedom of an $N$-centered black-hole molecule. MPS compute this factor by quantizing the classical phase space of multi-centered supergravity solutions and localizing the $J_3$ action on its fixed points. By contrast, our starting point is the protected trace of the gauged $D(2,1;0)$ superconformal quiver quantum mechanics. As we show below, the two fixed-point expressions share their oscillator and angular momentum weights, while their saddle equations differ. 
	
	For a compact configurational phase space with isolated rotational fixed points, the MPS index is written as
	\begin{equation}
		\begin{aligned}
			g_{\rm ref}^{\rm MPS}(\{\Gamma_a\};y_{\rm MPS})
			&=
			(-1)^{\sum_{a<b}\Gamma_{ab}+N-1}
			\left(y_{\rm MPS}-y_{\rm MPS}^{-1}\right)^{1-N}\\
			&\quad\times\sum_p
			s_{\rm MPS}(p)\,
			y_{\rm MPS}^{\sum_{a<b}
				\Gamma_{ab}\operatorname{sign}_{p}(z^b-z^a)} ,
			\label{eq:MPS-coulomb-index-comparison}
		\end{aligned}
	\end{equation}
	where $s_{\rm MPS}(p)$ is the fixed-point orientation sign, and the sum is over collinear solutions obeying
	\begin{equation}
		\sum_{b=1\atop b\neq a}^{N}
		\frac{\Gamma_{ab}}{|z_a-z_b|}
		=c_a,
		\qquad
		1\leq  a \leq N-1,
		\label{eq:MPS-Denef-equations}
	\end{equation}
	where the $c_a$ are stability parameters fixed by the asymptotic moduli and satisfy $\sum_a c_a=0$. In terms of our function $v_a(z)$, the MPS saddle equations read $v_a(z)=c_a/2$, whereas our superconformal saddles obey $E_{\pm,a}(z)=v_a(z)\mp G_{ab}(z)z^b=0$.
	
	To compare the local weights in (\ref{eq:final-D210-quiver-index}), introduce
	\begin{equation}
		q:=y^{\mp1/2}, \label{eq: qfugacity defn}
	\end{equation}
	where the upper/lower choice corresponds to the two chiral choices in $\mathcal I_\pm$. Then
	\begin{equation}
		\frac{\pm y^{1/2}}{1-y}
		=
		\frac{1}{q-q^{-1}},
	\end{equation}
	and
	\begin{equation}
		y^{\mp\frac12
			\sum_{a<b}
			\Gamma_{ab}\operatorname{sign}_{p}(z^b-z^a)}
		=
		q^{\sum_{a<b}
			\Gamma_{ab}\operatorname{sign}_{p}(z^b-z^a)} .
	\end{equation}
	
	MPS use the refined trace convention
	\begin{equation}
		\operatorname{Tr}'(-y_{\rm MPS})^{2J_3}.
	\end{equation}
	Therefore the precise fugacity conversion is
	\begin{equation}
		y_{\rm MPS}=-q=-y^{\mp1/2}.
		\label{eq:MPS-fugacity-conversion}
	\end{equation}
	With this replacement, the MPS denominator becomes
	\begin{equation}
		\left(y_{\rm MPS}-y_{\rm MPS}^{-1}\right)^{1-N}
		=
		(-1)^{N-1}
		\left(q-q^{-1}\right)^{1-N},
	\end{equation}
	while the fixed-point weight becomes
	\begin{equation}
		y_{\rm MPS}^{\sum_{a<b}
			\Gamma_{ab}\operatorname{sign}_{p}(z^b-z^a)}
		=
		(-1)^{\sum_{a<b}\Gamma_{ab}}
		q^{\sum_{a<b}
			\Gamma_{ab}\operatorname{sign}_{p}(z^b-z^a)} .
	\end{equation}
	
	Hence the global MPS factor
	\begin{equation}
		(-1)^{\sum_{a<b}\Gamma_{ab}+N-1}
	\end{equation}
	is exactly accounted for by the replacement $y_{\rm MPS}=-y^{\mp1/2}$.
	
	Let us also compare the fixed-point signs. MPS define \cite{Manschot:2014fua,Manschot:2013sya} the fixed-point sign
	\begin{equation}
		s_{\rm MPS}(p)
		=
		\operatorname{sign}\det{}'M(p),
	\end{equation}
	as the sign of the reduced determinant of the Hessian matrix of the potential 
	\begin{equation}
		V(z)
		=
		-\sum_{a<b}
		\Gamma_{ab}\,
		\operatorname{sign}(z_b-z_a)\,
		\log |z_a-z_b|
		-
		\sum_a c_a z_a ,
		\label{eq:MPS-potential}
	\end{equation}
	which gives
	\begin{equation}
		M_{ab}
		=
		\frac{\partial^2 V}{\partial z^a\partial z^b}
		=
		\partial_{z^b}
		\left(
		\sum_{c\neq a}
		\frac{\Gamma_{ac}}{|z^a-z^c|}
		\right).
	\end{equation}
	On the other hand, in our normalization,
	\begin{equation}
		\partial_{z^b}v_a
		=
		\frac12\,
		\partial_{z^b}
		\left(
		\sum_{c\neq a}
		\frac{\Gamma_{ac}}{|z^a-z^c|}
		\right).
	\end{equation}
	The orientation entering our formula is instead
	\begin{equation}
		\sigma_\pm(p)
		=\operatorname{sign}\det{}'
		\left[\frac12M_{ab}\pm2G_{ab}\right]_p,
		\qquad E_\pm(p)=0 .
	\end{equation}
	
	Combining these identities, the MPS expression in our fugacity is
	\begin{equation}
		\begin{aligned}
			g_{\rm ref}^{\rm MPS}(\{\Gamma_a\};-q)
			&=
			(q-q^{-1})^{1-N}
			\sum_{p:\,v_a(p)=c_a/2}
			\operatorname{sign}\det{}'[\partial_bv_a]_p\\
			&\qquad\times
			q^{\sum_{a<b}\Gamma_{ab}
				\operatorname{sign}_p(z^b-z^a)}.
		\end{aligned}
		\label{eq:MPS-converted-fixed-point-sum}
	\end{equation}	
	Comparing this to the quiver superconformal index (\ref{eq:final-D210-quiver-index}) shows the common factors encoding the transverse rotation weights and the angular momentum determined by the DSZ pairings and collinear ordering, while the superconformal deformation changes the saddle equations and the Jacobians determining their orientation signs. 
	
	The microscopic quiver index contains scaling contributions in addition to this regular Coulomb contribution. We now compare our result directly with the calculation of those contributions.\footnote{I thank the anonymous referee for drawing my attention to this fact and suggesting the following comparison with \cite{Beaujard:2021fsk}.} 
	
	\subsection{Comparison with the microscopic scaling contribution}
	\label{subsec:BMP-scaling-comparison}
	
	Beaujard, Mondal and Pioline (BMP) \cite{Beaujard:2021fsk} compute the microscopic refined Witten index of Abelian cyclic quivers with generic superpotential and stability parameters. Their index is given by
	\begin{equation}
		\Omega_Q(y_{\rm B})
		=
		\operatorname{Tr}_{\mathcal H}(-1)^F
		y_{\rm B}^{2J_-}e^{-\beta_{\rm BMP}H_{\rm BMP}},
		\qquad y_{\rm B}=e^{i\pi z_{\rm BMP}},
		\label{eq:BMP-index-convention}
	\end{equation}
	where $H_{\rm BMP}$ is the microscopic Hamiltonian, $J_-$ is their refinement generator and $\beta_{\rm BMP}$ is the Euclidean time period. Their localization separates regular and scaling saddles:
	\begin{equation}
		\Omega_Q=g_C+I_{\rm sc}^{\rm BMP},
		\qquad
		I_{\rm sc}^{\rm BMP}
		=
		I_{\rm same}^{\rm BMP}+I_{\rm uneq}^{\rm BMP}
		=\Omega_S+H.
		\label{eq:BMP-scaling-decomposition}
	\end{equation}
	Here $g_C$ is the MPS configurational Coulomb contribution discussed above, in the corresponding conventions. The terms $\Omega_S$ and $H$ are respectively the single-centered invariant and the minimal-modification term. The split into \emph{same} and \emph{unequal} signs is a classification of BMP's scaling saddles, which we describe next.
	
	For a cyclic quiver with $N\geq3$, let $\kappa_a=\Gamma_{a,a+1}\in\mathbb Z_{>0}$, including $\kappa_N=\Gamma_{N1}$, and set $A=\sum_a\kappa_a$. BMP's real localization variables $\Sigma_a$ enter through the shifted edge separations\footnote{Here $\Sigma_a$ is the constant longitudinal scalar in BMP's vector multiplet, corresponding to the coordinate $z^a$ in our collinear background.}
	\begin{equation}
		\Sigma_{a,a+1}
		=
		\Sigma_a-\Sigma_{a+1}
		-\frac{\pi\operatorname{Im}z_{\rm BMP}}{\beta_{\rm BMP}}R_a,
		\qquad
		\sigma_a=\operatorname{sign}\Sigma_{a,a+1},
		\label{eq:BMP-shifted-separations}
	\end{equation}
	where $R_a$ is the R-charge of the chiral fields on edge $a$. The superpotential has R-charge two, so $\sum_aR_a=2$. Summing the shifted separations around the cycle therefore gives
	\begin{equation}
		\sum_a\Sigma_{a,a+1}
		=
		-\frac{2\pi\operatorname{Im}z_{\rm BMP}}{\beta_{\rm BMP}}.
		\label{eq:BMP-shifted-closure}
	\end{equation}
	Ordinary coordinate differences around a cycle sum to zero and cannot all have the same nonzero sign, e.g. for the three-centered case: $(\Sigma_1-\Sigma_2)+(\Sigma_2-\Sigma_3)+(\Sigma_3-\Sigma_1)=0$. So having them all positive would require $\Sigma_1>\Sigma_2>\Sigma_3>\Sigma_1$, which is impossible.  Instead, shifted differences satisfy $\sum_a\Sigma_{a,a+1} >0$ in the chosen convention $\operatorname{Im}z_{\rm BMP}<0$. Therefore they \textit{can} all be positive. The scaling saddles with $\sigma_a=+1$ for every edge contribute to BMP's same-sign sector $I_{\rm same}^{\rm BMP}$. The scaling saddles with both positive and negative shifted separations contribute to the unequal-sign sector $I_{\rm uneq}^{\rm BMP}$.
	
	Write $A=\sum_a\kappa_a$ and $T_\sigma=\sum_a\sigma_a\kappa_a$. For pairings with no vanishing signed sums and $\operatorname{Im}z_{\rm BMP}<0$, BMP's unequal-sign scaling contribution is \cite{Beaujard:2021fsk}
	\begin{equation}
		I_{\rm uneq}^{\rm BMP}(y_{\rm B})
		=
		\frac{(-1)^A}{(y_{\rm B}-y_{\rm B}^{-1})^{N-1}}
		\sum_{\substack{\sigma_a=\pm1\\T_\sigma>0}}^{\prime}
		\left(\prod_a\sigma_a\right)y_{\rm B}^{-T_\sigma}.
		\label{eq:BMP-unequal-original}
	\end{equation}
	Here $\sigma_a$ is the sign of the shifted edge separation defined above, and the prime excludes the all-equal sign choices. The condition $T_\sigma>0$ follows from BMP's scaling-saddle solution
	\begin{equation}
		|\Sigma_{a,a+1}|
		=
		-\frac{2\pi\operatorname{Im}z_{\rm BMP}}{\beta_{\rm BMP}}
		\frac{\kappa_a}{T_\sigma}.
		\label{eq:BMP-scaling-saddle}
	\end{equation}

	We first compare the general fixed-point expressions. Choose our minus index, put $y=q^2$, and set
	\begin{equation}
		y_{\rm B}=-q^{-1},\qquad
		\epsilon_a=-\sigma_a,\qquad
		S_\epsilon=\sum_a\epsilon_a\kappa_a=-T_\sigma.
		\label{eq:BMP-convention-dictionary}
	\end{equation}
	For this chirality $y_{\rm MPS}=-q$, so $y_{\rm B}=y_{\rm MPS}^{-1}$.\footnote{The regular Coulomb index is invariant under this inversion.} The conversion of BMP's expression follows from
	\begin{equation}
		\begin{aligned}
			y_{\rm B}-y_{\rm B}^{-1}&=q-q^{-1},&
			\prod_a\sigma_a&=(-1)^N\prod_a\epsilon_a,\\
			(-1)^Ay_{\rm B}^{-T_\sigma}
			&=(-1)^{A-T_\sigma}q^{-S_\epsilon}=q^{-S_\epsilon},&
			A-T_\sigma&=2\sum_{\sigma_a=-1}\kappa_a.
		\end{aligned}
		\label{eq:BMP-fugacity-conversion}
	\end{equation}
	Thus
	\begin{equation}
		I_{\rm uneq}^{\rm BMP}(-q^{-1})
		=
		(q-q^{-1})^{1-N}
		\sum_{\substack{\epsilon_a=\pm1\\S_\epsilon<0}}^{\prime}
		(-1)^N\left(\prod_a\epsilon_a\right)q^{-S_\epsilon},
		\label{eq:BMP-unequal-converted}
	\end{equation}
	where the domain $|y_{\rm B}|>1$ becomes $|q|<1$.
	
	For a collinear saddle $p$ in our calculation, define its edge signs by $\epsilon_a(p)=\operatorname{sign}(z_p^a-z_p^{a+1})$, with cyclic node labels. Since $\sum_a(z_p^a-z_p^{a+1})=0$, these signs are necessarily mixed. Its angular momentum obeys
	\begin{equation}
		2J_3(p)=\sum_a\kappa_a\epsilon_a(p)=S_{\epsilon(p)}.
	\end{equation}
	Our general formula therefore takes the form
	\begin{equation}
		\begin{aligned}
			\mathcal I_-(q^2)
			&=(q-q^{-1})^{1-N}
			\sum_{p:\,E_-(p)=0}
			\sigma_-(p)\,q^{-S_{\epsilon(p)}},\\
			E_{-,a}(p)&=v_a(p)+G_{ab}(p)z_p^b,\\
			\sigma_-(p)&=\operatorname{sign}\det{}'
			[\partial_bv_a-2G_{ab}]_p.
		\end{aligned}
		\label{eq:BMP-superconformal-sum-comparison}
	\end{equation}
	Equations \eqref{eq:BMP-unequal-converted} and \eqref{eq:BMP-superconformal-sum-comparison} have the same universal denominator and angular momentum weights. However, their sums are prescribed differently: BMP sum over the allowed mixed-sign vectors with sign $(-1)^N\prod_a\epsilon_a$, while our formula sums the actual roots of $E_-=0$, weighted by their Jacobian signs. We now evaluate both sums for the symmetric three-center family, which gives an exact match for every positive integer $k$.
	
	Take $\kappa_1=\kappa_2=\kappa_3=k>0$. The unequal BMP sign vectors satisfying $T_\sigma>0$ are precisely
	\begin{equation}
		(\sigma_1,\sigma_2,\sigma_3)
		=
		(+,+,-),\quad (+,-,+),\quad (-,+,+).
	\end{equation}
	Each has $T_\sigma=k$ and $\prod_a\sigma_a=-1$, while $A=3k$. The all-positive vector has $T_\sigma=3k$, but is excluded because it belongs to the same-sign sector. Thus
	\begin{equation}
		I_{{\rm uneq},(k,k,k)}^{\rm BMP}(y_{\rm B})
		=
		-\frac{3(-1)^{3k}y_{\rm B}^{-k}}
		{(y_{\rm B}-y_{\rm B}^{-1})^2},
		\label{eq:BMP-symmetric-three-evaluation}
	\end{equation}
	or,
	\begin{equation}
		I_{{\rm uneq},(k,k,k)}^{\rm BMP}(-q^{-1})
		=
		-\frac{3q^k}{(q-q^{-1})^2}.
		\label{eq:BMP-symmetric-three-converted}
	\end{equation}
	
	Our calculation in Appendix~\ref{app:two-three-center-checks} gives three minus-index saddles for the same pairings. At each saddle,
	\begin{equation}
		J_3=-\frac{k}{2},
		\qquad
		\operatorname{sign}\det{}'[\partial_bv_a-2G_{ab}]=-1.
	\end{equation}
	The transverse determinant is $(q-q^{-1})^{-2}$ and the classical weight is $q^{-2J_3}=q^k$. Each saddle therefore contributes $-q^k/(q-q^{-1})^2$. Summing the three yields
	\begin{equation}
		\mathcal I_{-,\,3{\rm c}}(q^2)
		=
		-\frac{3q^k}{(q-q^{-1})^2}
		=
		I_{{\rm uneq},(k,k,k)}^{\rm BMP}(-q^{-1}).
		\label{eq:BMP-symmetric-three-match}
	\end{equation}
	Thus, for the symmetric three-center family, our superconformal index reproduces exactly BMP's unequal-sign contribution to the microscopic scaling term $\Omega_S+H$. However, we note that the equality in the general cases need not hold since it requires the resulting signed multiplicities at each angular momentum weight to agree, which the common form of the local weights alone does not establish. 
	
	\section{Discussion}
	\label{sect: discussion}

	In this paper, we formulated a $D(2,1;0)$ superconformal index for scaling quiver quantum mechanics and evaluated it by supersymmetric localization. The result is the fixed-point formula \eqref{eq:final-D210-quiver-index}. Its three factors have a direct origin in the gauged sigma model: the transverse modes give the universal equivariant oscillator factor, the classical action gives the DSZ ordering phase, and the constant-mode integral gives the orientation sign for each term. A natural next step would be to determine the normalizable BPS states underlying this index. The match with BMP's unequal-sign contribution makes this investigation particularly interesting. This is because although BMP related their same-sign contribution to stacky invariants at trivial stability, the mathematical interpretation of the unequal-sign contribution has remained open \cite{Beaujard:2021fsk}. As this turns out to be exactly equal to the protected trace over superconformal short multiplets in the symmetric three-center case, it suggests that we may understand this directly through a direct computation of the BPS states in the quantum mechanics of the centers. The gauged formulation would be especially useful for this purpose as then the BPS states admit a description in terms of twisted Dolbeault cohomology, subject to the gauge constraints. Solving these equations for normalizable states would determine their multiplicities and $R$-symmetry quantum numbers, including information lost in the graded trace. For the symmetric three-center family, this could then provide a cohomological explanation of the unequal-sign contribution. It would also be interesting to understand when the agreement extends to other cyclic quivers and whether the same-sign contribution admits a related superconformal description.
	
	Let us now comment on some further directions. First, it would be natural to extend the localization analysis to general $D(2,1;\alpha)$ superconformal mechanics. The family $D(2,1;\alpha)$ is the most general exceptional $\mathcal N=4$ superconformal algebra in one dimension, and many mechanical realizations are known \cite{Ivanov:2003tm,Fedoruk:2009xf,Krivonos:2010zy,Galajinsky:2014hza}. Moreover, $D(2,1;\alpha)$ appears naturally in the superisometry algebra of $AdS_3\times S^3\times S^3\times S^1$ string backgrounds. In this case the parameter $\alpha$ is fixed by the relative radii of the two three-spheres \cite{Babichenko:2009dk,Borsato:2015mma}. It would therefore be interesting to understand which protected traces, if any, admit similarly concrete equivariant localization formulae for gauged quantum-mechanical models with generic $D(2,1;\alpha)$ symmetry.
		
	A second direction is to relax the scaling limit. Away from the scaling region, the quiver mechanics contains additional terms controlled by the asymptotic moduli. These terms break the exact conformal symmetry but control the Denef chamber structure and wall-crossing. It would be useful to understand whether they can be incorporated as deformations of the localization problem.
	
	Finally, the role of the $SU(2)_L$ symmetry deserves a more geometric interpretation. In the present construction $J_L^3$ enters the grading of the index and is realized in the gauged sigma model as a Lefschetz-type action on the HKT target. Since the scaling quiver mechanics describes a near $AdS_2\times S^2$ region of multi-centered BPS black-hole solutions, it would be interesting to understand how this $SU(2)_L$ action is encoded in the bulk geometry. In particular, the near $AdS_2$ limit of multicenter scaling solutions can retain nontrivial $S^2$-fibration data, even when the total angular momentum vanishes \cite{Mirfendereski:2018tob}. This suggests that the bulk interpretation of the two $SU(2)$ actions in the $D(2,1;0)$ quantum mechanics may be more subtle than in the empty single-center $AdS_2\times S^2$ throat. It would also be interesting to relate the superconformal quiver index to the $AdS_2$ microstate-geometry perspective, where deep scaling solutions define asymptotically $AdS_2$ configurations with nontrivial IR caps \cite{Bena:2018bbd}.
	
	\subsection*{Acknowledgements}
	
	I would like to thank Delaram Mirfendereski, Joris Raeymaekers, and Dieter Van den Bleeken for a great collaboration \cite{Mirfendereski:2022omg, Raeymaekers:2024usy}, which led to this project. I also thank Nick Dorey, Tom\'a\v{s} Proch\'azka, Paolo Rossi, and Andy Zhao for valuable discussions. I am grateful to Joris Raeymaekers for helpful comments on an earlier draft. I was supported by the European Union's Horizon Europe programme under grant agreement No. 101109743, project Quivers, by the Czech Academy of Sciences under the grant number LQ100102101, and by European Structural and Investment Funds and the Czech Ministry of Education, Youth and Sports (Project FORTE CZ.02.01.01/00/22 008/0004632).

	\appendix

	\section{Gauged superconformal transformations}
	\label{app:D21a-transformations}

	In this appendix we collect the local transformations of the gauged sigma model \eqref{eq:gauged_model_general} which realize the $D(2,1;\alpha)$ superconformal symmetry.

	\paragraph{Gauge transformations:}
	\begin{align}
	\delta_\lambda x^A &= \lambda^I k_I^A, \\
	\delta_\lambda a^I &= \dot{\lambda}^I + f_{JK}{}^I a^J \lambda^K, \\
	\delta_\lambda \chi^A &= \lambda^I \partial_B k_I^A \chi^B.
	\end{align}

	\paragraph{Conformal transformations:}
	The bosonic conformal transformations are generated by a time-dependent function
	\begin{equation}
	P(t) = u + v t + w t^2,
	\label{eq:infinitesimal_P}
	\end{equation}
	where the parameters $u$, $v$, $w$ correspond to time translations, dilatations, and special conformal transformations, respectively. These generate the following infinitesimal variations
	\begin{align}
	\delta_P x^A &= P \dot{x}^A + \dot{P} \xi^A, \\
	\delta_P \chi^A &= P \dot{\chi}^A + \dot{P} \left( \frac{1}{2} \delta^A_B + \partial_B \xi^A \right) \chi^B, \\
	\delta_P a^I &= P \dot{a}^I + \dot{P} (\delta^I_J + \gamma^I{}_J) a^J + \ddot{P} h^I,
	\end{align}
	where $\gamma^I{}_J = \delta^I_J$ for the $D(2,1;0)$ case.

	\paragraph{Fermionic transformations:}
	Supersymmetry and conformal supersymmetry transformations are packaged into a single fermionic variation parameter
	\begin{equation}
	\Sigma_\rho(t) = \epsilon_\rho + \eta_\rho t,
	\end{equation}
	where $\epsilon_\rho$ and $\eta_\rho$ are constant spinors, and $\rho = 1,\dots,4$ labels the supercharges. The combined variations are given by
	\begin{align}
	\delta_\Sigma x^A &= -i (J^\rho)^A{}_B \Sigma_\rho \chi^B, \label{eq: delta Sigma x} \\
	\delta_\Sigma \chi^A &= (\bar{J}^\rho)^A{}_B \left( \Sigma_\rho D_t x^B + 2 \dot{\Sigma}_\rho \xi_\perp^B \right) + i \partial_C (J^\rho)^A{}_B \Sigma_\rho \chi^C \chi^B, \label{eq: delta Sigma chi} \\
	\delta_\Sigma a^I &= -2i (\alpha+1)^{-1} V^{\rho I}_A \dot{\Sigma}_\rho \chi^A, \label{eq:delta-Sigma-a}
	\end{align} where $J^\rho=(J^i,\mathds{1}),\; \bar J^\rho=(-J^i,\mathds{1})$, and $V^{\rho I}_A$ is a shorthand for
	\begin{equation}
	V^{\rho I}_A = -(\alpha + 1) \, \partial_B h^I \, (J^\rho)^B{}_{A},
	\end{equation} introduced for convenience.

	\paragraph{R-symmetry transformations:}
	The total R-symmetry group is $SU(2)_L \times SU(2)_R$, realized by particular combinations (see~\eqref{eq: JL and JR}) of the two non-commuting actions:

	\textit{Factor 1:}
	\begin{align}
	\delta_r x^A &= r^i \omega^i{}^A, \label{eq: reeb transfo x} \\
	\delta_r \chi^A &= r^i \left( \partial_B \omega_i^A - V^{iI}_B k_I^A \right) \chi^B, \label{eq: reeb transfo chi}\\
	\delta_r a^I &= r^i V^{iI}_A D_t x^A. \label{eq: reeb transfo a}
	\end{align}

	\textit{Factor 2:}
	\begin{align}
	\delta_{\tilde{r}} x^A = 0, \qquad
	\delta_{\tilde{r}} a^I = 0, \qquad
	\delta_{\tilde{r}} \chi^A = \frac{1}{2} \tilde{r}^i (J^i)^A{}_B \chi^B.
	\end{align}

	The second factor acts only on the fermions and can be interpreted as a fermion-number-like symmetry. 

	\paragraph{Refined localizing transformations.}

	For the index computation we use the complex supercharge
	\begin{equation}
	\mathcal G_{\pm1/2}=G^{-+}_{\pm1/2},
	\qquad
	\mathcal G_{\pm1/2}^{\dagger}
	=
	-G^{+-}_{\mp1/2}.
	\end{equation}
	The real localizing supercharge is the combination $\mathscr Q_\pm$ defined in \eqref{eq: dirac operator}.
	\begin{equation}
	\mathscr Q_\pm
	=
	\mathcal G_{\pm1/2}
	+
	\mathcal G_{\pm1/2}^{\dagger}.
	\end{equation}

	Using the dictionary between the spinor-index basis and the sigma-model charges in Appendix~\ref{app:D21a-dictionary}, the action of $\mathcal G_{\pm1/2}$ on the refined sigma-model fields is
	\begin{align}
	\delta_{\mathcal G_{\pm1/2}}x^A
	&=
	\frac{\epsilon}{2}
	\left(
	(J^3)^A{}_{B}
	+
	i\delta^A_B
	\right)\chi^B,
	\label{eq:Gpm-first-x}
	\\
	\delta_{\mathcal G_{\pm1/2}}\chi^A
	&=
	-\frac{i\epsilon}{2}
	\left(
	(J^3)^A{}_{B}
	-
	i\delta^A_B
	\right)
	\left(
	D_t x^B
	\mp
	i\beta^{-1}\lambda\omega^{3B}
	\right)
	-
	\frac{\epsilon}{2}
	\partial_C(J^3)^A{}_{B}\chi^C\chi^B.
	\label{eq:Gpm-first-chi}
	\end{align}

	The conjugate supercharge acts as
	\begin{align}
	\delta_{\mathcal G_{\pm1/2}^{\dagger}}x^A
	&=
	\frac{\epsilon}{2}
	\left(
	-(J^3)^A{}_{B}
	+
	i\delta^A_B
	\right)\chi^B,
	\label{eq:Gpmdagger-first-x}
	\\
	\delta_{\mathcal G_{\pm1/2}^{\dagger}}\chi^A
	&=
	\frac{i\epsilon}{2}
	\left(
	(J^3)^A{}_{B}
	+
	i\delta^A_B
	\right)
	\left(
	D_t x^B
	\mp
	i\beta^{-1}\lambda\omega^{3B}
	\right)
	+
	\frac{\epsilon}{2}
	\partial_C(J^3)^A{}_{B}\chi^C\chi^B.
	\label{eq:Gpmdagger-first-chi}
	\end{align}
	Adding these gives
	\begin{equation}
	\delta_{\mathscr Q_\pm}x^A
	=
	i\epsilon\chi^A,
	\qquad
	\delta_{\mathscr Q_\pm}\chi^A
	=
	-\epsilon
	\left(
	D_t x^A
	\mp
	i\beta^{-1}\lambda\omega^{3A}
	\right).
	\label{eq:Gpm-Gpmdagger-Hamiltonian}
	\end{equation}
	For the gauge field, we have
	\begin{equation}
	\delta_{\mathscr Q_\pm}a^I
	=
	\pm
	\epsilon\beta^{-1}\lambda
	V_A^{3I}\chi^A .
	\end{equation}

	After Wick rotation and the rescaling $\tau=\beta\tilde\tau$, the induced Euclidean transformations are
	\begin{equation}
	\delta_{\mathscr Q_\pm}x^A
	=
	i\epsilon\chi^A,
	\qquad
	\delta_{\mathscr Q_\pm}\chi^A
	=
	-\frac{i\epsilon}{\beta}
	\left(
	\dot x^A-a^Ik_I^A
	\mp
	\lambda\omega^{3A}
	\right),
	\qquad
	\delta_{\mathscr Q_\pm}a^I
	=
	\mp
	i\epsilon\lambda
	V_A^{3I}\chi^A  ,
	\label{eq:Gpm-Gpmdagger-Euclidean}
	\end{equation}
	where now $\dot x^A=\frac{dx^A}{d\tilde\tau}$.
	
	\section{$D(2,1;\alpha)$} \label{app:D21a-algebra}

	In this appendix, we collect the (anti-)commutation relations of the $D(2,1;\alpha)$ superalgebra as realized in the gauged superconformal sigma model. The bosonic subalgebra is
	\begin{equation}
	\mathfrak{sl}(2,\mathbb{R}) \oplus \mathfrak{su}(2)_L \oplus \mathfrak{su}(2)_R,
	\end{equation}
	with generators $H$, $D$, $K$ for time translations, dilations, and special conformal transformations, and $R^i$, $\tilde{R}^i$ for the two $\mathfrak{su}(2)$ R-symmetry factors.

	The nonvanishing bosonic commutators are
	\begin{align}
	[H, K] &= 2i D, & [H, D] &= i H, & [D, K] &= i K. \label{eq:bosonic_comm}
	\end{align}

	The supercharges $Q^\mu$ and $S^\mu$ satisfy
	\begin{align}
	[Q^\mu, D] &= \frac{i}{2} Q^\mu, &
	[S^\mu, D] &= -\frac{i}{2} S^\mu, \\
	[S^\mu, H] &= i Q^\mu, &
	[Q^\mu, K] &= -i S^\mu, \\
	\{ Q^\mu, Q^\nu \} &= 2 \delta^{\mu\nu} H, &
	\{ S^\mu, S^\nu \} &= 2 \delta^{\mu\nu} K.
	\end{align}

	The mixed anticommutator reads
	\begin{equation}
	\{ Q^\mu, S^\nu \} = -2 \delta^{\mu\nu} D - \frac{2}{\alpha+1} (j_-^i)_{\mu\nu} R^-_i - \frac{2\alpha}{\alpha+1} (j_+^i)_{\mu\nu} R^+_i. \label{eq:QS_comm}
	\end{equation}

	The R-symmetry generators act as
	\begin{align}
	[Q^\mu, R^\pm_i] &= \frac{i}{2} (j_\pm^i)_{\mu\nu} Q^\nu, &
	[S^\mu, R^\pm_i] &= \frac{i}{2} (j_\pm^i)_{\mu\nu} S^\nu, \\
	[R^\pm_i, R^\pm_j] &= i \epsilon^{ijk} R^\pm_k,
	\end{align}
	where $j_\pm$ are the (anti-)selfdual 't Hooft symbols
	\begin{equation}
	(j^i_\pm)_{\mu \nu} = \mp \left(\delta_{\mu i} \delta_{\nu 4} - \delta_{\mu 4} \delta_{\nu i} \right) - \epsilon_{i \mu \nu 4}.
	\end{equation}

	The commuting $SU(2)_L \times SU(2)_R$ structure is recovered from the linear combinations
	\begin{equation}
	R^-_i = -(R_i + \tilde{R}_i), \qquad R^+_i = \tilde{R}_i. \label{eq:Rplusminus}
	\end{equation}

	In order for the transformations given in the main text to close on the full $D(2,1;\alpha)$ superalgebra, the target space geometry must satisfy several structural conditions. Most notably, the complex structures $J^i$ must be integrable: 
	\begin{equation}
	\mathcal{N}(J^i, J^j)^A{}_{BC} = (J^{(i})^D{}_{[B} \partial_{|D|} (J^{j)})^A{}_{C]} - (J^{(i})^A{}_{D} \partial_{[B} (J^{j)})^D{}_{C]} = 0,
	\end{equation}
	and the conformal Killing vector $\xi^A$ must preserve them:
	\begin{equation}
	\mathcal{L}_\xi (J^i)^A{}_B = 0.
	\end{equation}

	We refer to \cite{Mirfendereski:2022omg} for a comprehensive treatment of the geometric and algebraic constraints required for full $D(2,1;\alpha)$ invariance.

	An alternative and widely used realization of $D(2,1;\alpha)$, particularly useful for studying unitary representations and BPS bounds, expresses the fermionic generators as $G^{\alpha\alpha'}_p$, where $\alpha,\alpha' = \pm$ label spinor indices of the two $\mathfrak{su}(2)$ R-symmetry factors, and $p = \pm \frac{1}{2}$ denotes conformal weight. We mainly used \cite{VanDerJeugt:1985hq, Gunaydin:1986fe, Frappat:1996pb, deBoer:1999gea, Gaiotto:2004pc} as references in the following.

	In this basis, the bosonic conformal algebra is realized through three generators $L_{-1}$, $L_0$, and $L_{+1}$ satisfying:
	\begin{equation}
	[L_m, L_n] = (m - n) L_{m+n}, \qquad m, n = 0, \pm 1.
	\end{equation}

	The eight supercharges $G^{\alpha\alpha'}_p$ satisfy the anticommutation relations
	\begin{align}
	\{G_{1/2}^{\alpha\alpha'}, G_{-1/2}^{\beta\beta'}\} &= \epsilon^{\alpha\beta} \epsilon^{\alpha'\beta'} L_0 + \gamma \epsilon^{\alpha\beta} T_L^{\alpha'\beta'} + (1 - \gamma) \epsilon^{\alpha'\beta'} T_R^{\alpha\beta}, \qquad \gamma = \frac{\alpha}{1 + \alpha}, \label{eq: D21a BPS conditions} \\[4pt]
	\{G_{\pm 1/2}^{\alpha\alpha'}, G_{\pm 1/2}^{\beta\beta'}\} &= \epsilon^{\alpha\beta} \epsilon^{\alpha'\beta'} L_{\pm 1}. \label{eq: D21a BPS conditions 2}
	\end{align}

	The R-symmetry structure is encoded in the matrices
	\begin{equation}
	T^{\alpha\beta}_{L,R} = J^i_{L,R} (\sigma_i)_\gamma{}^\beta \epsilon^{\gamma\alpha},
	\end{equation}
	where $\sigma^i$ are the Pauli matrices. This gives the component expressions
	\begin{equation}
	T^{++}_{L,R} = -J^+_{L,R}, \qquad
	T^{--}_{L,R} = J^-_{L,R}, \qquad
	T^{+-}_{L,R} = T^{-+}_{L,R} = J^3_{L,R}, \label{eq: Talphabeta components}
	\end{equation}
	with ladder operators defined by
	\begin{equation}
	J^\pm_{L,R} = J^1_{L,R} \pm i J^2_{L,R}, \qquad
	[J^+_{L,R}, J^-_{L,R}] = 2 J^3_{L,R}. \label{eq: JLR ladder}
	\end{equation}

	The R-symmetry generators act on the fermionic charges as
	\begin{equation}
	[J^i_R, G^{\alpha\beta}_p] = \frac{1}{2} (\sigma^i)_\gamma{}^\alpha G^{\gamma\beta}_p, \qquad
	[J^i_L, G^{\alpha\beta}_p] = \frac{1}{2} (\sigma^i)_\gamma{}^\beta G^{\alpha\gamma}_p. \label{eq: D21a JR commutators}
	\end{equation}

	The conjugation property of the supercharges is
	\begin{equation}
	(G^{\alpha\alpha'}_{\pm 1/2})^\dagger = \epsilon_{\alpha\beta} \epsilon_{\alpha'\beta'} G^{\beta\beta'}_{\mp 1/2}. \label{eq: D21a conjugation property}
	\end{equation}

	This spinor-index formulation makes manifest the $SU(2)_L \times SU(2)_R$ symmetry and facilitates the analysis of shortening conditions and BPS bounds, to which we now turn.

	Unitary irreducible representations of $D(2,1;\alpha)$ are classified by their lowest weight primary states, defined by annihilation conditions
	\begin{align}
	L_0 |h, j_L, j_R\rangle &= h |h, j_L, j_R\rangle, \quad L_{+1} |h, j_L, j_R\rangle = 0, \quad
	G_{+1/2}^{\alpha\alpha'} |h, j_L, j_R\rangle = 0\\
	J_L^3 |h, j_L, j_R\rangle &= j_L |h, j_L, j_R\rangle, \quad
	J_R^3 |h, j_L, j_R\rangle = j_R |h, j_L, j_R\rangle,
	\end{align}
	where $h$ is the conformal weight (lowest eigenvalue of $L_0$), while $j_L$ and $j_R$ denote the $J_L^3$ and $J_R^3$ weights.
	
	The BPS shortening conditions arise from the anticommutators in Eq.~\eqref{eq: D21a BPS conditions}. For lowest-weight representations, shortening occurs if
	\begin{equation}
	G^{++}_{-1/2} |h, j_L, j_R\rangle = 0 \quad \text{or} \quad G^{--}_{-1/2} |h, j_L, j_R\rangle = 0.
	\end{equation}

	Applying the conjugation condition \eqref{eq: D21a conjugation property}, one finds
	\begin{equation}
	|G^{++}_{-1/2} |h, j_0, r_0\rangle|^2 = h - \gamma r_0 - (1 - \gamma) j_0, \qquad
	|G^{--}_{-1/2} |h, j_0, r_0\rangle|^2 = h + \gamma r_0 + (1 - \gamma) j_0, \label{d21abpslw}
	\end{equation}
	giving the unitarity bound
	\begin{equation}
	h \geq |(1 - \gamma) j_0 + \gamma r_0|, \label{boundd21alw}
	\end{equation}
	which is saturated when the state is BPS. In particular:
	\begin{align}
	h = (1 - \gamma) j_0 + \gamma r_0 &\quad \Leftrightarrow \quad G^{++}_{-1/2} |h, j_0, r_0\rangle = 0 \quad \text{(chiral primary)}, \\
	h = -(1 - \gamma) j_0 - \gamma r_0 &\quad \Leftrightarrow \quad G^{--}_{-1/2} |h, j_0, r_0\rangle = 0 \quad \text{(anti-chiral primary)}.
	\end{align}

	For generic spins, the corresponding shortened lowest-weight multiplet contains
	\begin{align}
	(j_L, j_R)_S \longrightarrow &\ 
	(j_L, j_R)_h + (j_L - \frac{1}{2}, j_R - \frac{1}{2})_{h + \frac{1}{2}} + (j_L - \frac{1}{2}, j_R + \frac{1}{2})_{h + \frac{1}{2}} \nonumber \\
	&+ (j_L + \frac{1}{2}, j_R - \frac{1}{2})_{h + \frac{1}{2}} + (j_L, j_R - 1)_{h + 1} + (j_L - 1, j_R)_{h + 1} \nonumber \\
	&+ (j_L, j_R)_{h + 1} + (j_L - \frac{1}{2}, j_R - \frac{1}{2})_{h + \frac{3}{2}}. \label{shortd21a}
	\end{align}

	\paragraph{Specialization to $D(2,1;0)$.}

	In the special case $\alpha = 0$, the superconformal algebra simplifies significantly and takes the form
	\begin{equation}
	D(2,1;0) = \mathfrak{psu}(1,1|2) \ltimes \mathfrak{su}(2).
	\end{equation}
	In this limit, the anticommutation relations reduce to
	\begin{align}
	\{G_{1/2}^{\alpha\alpha'}, G_{-1/2}^{\beta\beta'}\} &= \epsilon^{\alpha\beta} \epsilon^{\alpha'\beta'} L_0 + \epsilon^{\alpha'\beta'} T_R^{\alpha\beta}, \\[6pt]
	\{G_{\pm 1/2}^{\alpha\alpha'}, G_{\pm 1/2}^{\beta\beta'}\} &= \epsilon^{\alpha\beta} \epsilon^{\alpha'\beta'} L_{\pm 1},
	\end{align}
	where we see that the left R-symmetry $\mathfrak{su}(2)_L$ decouples from these relations.

	Another key property of $D(2,1;0)$ is that there is a single unitarity bound
	\begin{equation}
	h \geq |j_0|, \label{boundd210}
	\end{equation}
	which in this case is also a sufficient condition for the unitarity of all states in the multiplet, and is saturated precisely for chiral or anti-chiral primaries:
	\begin{align}
	G^{++}_{-1/2} |h,h,r_0\rangle = G^{+-}_{-1/2} |h,h,r_0\rangle = 0 &\quad \Leftrightarrow \quad |h,h,r_0\rangle\ \text{is a chiral primary,} \label{eq: D210 chiral primary} \\
	G^{--}_{-1/2} |h,-h,r_0\rangle = G^{-+}_{-1/2} |h,-h,r_0\rangle = 0 &\quad \Leftrightarrow \quad |h,-h,r_0\rangle\ \text{is an anti-chiral primary.} \label{eq: D210 antichiral primary}
	\end{align}

	The chiral primary lowest-weight state $|lw\rangle \equiv |h,h,r_0\rangle$ is annihilated by the fermionic generators
	\begin{equation}
	G^{\alpha\alpha'}_{1/2}, \quad G^{++}_{-1/2}, \quad G^{+-}_{-1/2},
	\end{equation}
	and the corresponding short multiplet decomposes as
	\begin{equation}
	(j_L,h)_S \longrightarrow (j_L,h)_h + (j_L - \frac{1}{2}, h - \frac{1}{2})_{h + \frac{1}{2}} + (j_L + \frac{1}{2}, h - \frac{1}{2})_{h + \frac{1}{2}} + (j_L, h - 1)_{h + 1}, \label{chiralshortd210}
	\end{equation}
	where the subscript $S$ denotes a short BPS multiplet.

	Compared with the $\mathfrak{psu}(1,1|2)$ multiplet alone, the $D(2,1;0)$ multiplet carries an additional spectator $SU(2)_L$ quantum number. This refines the state counting by $j_L$, while the BPS bound depends only on the $SU(2)_R$ spin.

	For special values of the charges, the short multiplet becomes even smaller. In particular, when the conformal weight vanishes ($h = 0$), we obtain the conformal singlet multiplets 
	\begin{equation}
	(j_L, 0)_{\text{singlet}} \longrightarrow (j_L, 0)_{h=0}, \label{eq:singletD210}
	\end{equation}
	which may carry non-zero $j_L$ charge, but are uncharged under $J_R$. This means that no $J_R$-charged states can appear as conformal singlets.

	\subsection{Dictionary of $D(2,1;\alpha)$ conventions} \label{app:D21a-dictionary}

	It is quite useful to relate two commonly used conventions for the $D(2,1;\alpha)$ algebra:
	\begin{itemize}
	\item The off-shell realization in sigma models, involving real supercharges $Q^\mu$ and $S^\mu$, as used in equations \eqref{eq:bosonic_comm}--\eqref{eq:QS_comm}.
	\item The representation-theoretic spinor-index basis involving $G^{\alpha\alpha'}_{\pm 1/2}$, as used in equations \eqref{eq: D21a BPS conditions}--\eqref{eq: D21a JR commutators}.
	\end{itemize}

	Although we define and analyze the superconformal index in terms of the latter (spinor) basis, explicit computations via localization require translation to the sigma model supercharges. Below we provide the relevant change of basis.

	We first define complex combinations of the real supercharges:
	\begin{align}
	\mathcal{Q}_- &= \frac{1}{2} (Q^3 + i Q^4), & \mathcal{Q}_+ &= \frac{1}{2} (Q^1 + i Q^2), \label{eq: calq definitions} \\
	\mathcal{S}_- &= \frac{1}{2} (S^3 + i S^4), & \mathcal{S}_+ &= \frac{1}{2} (S^1 + i S^2). \label{eq: cals definitions}
	\end{align}

	We also introduce the notation for complex conjugates, for example:
	\begin{equation}
	\bar{\mathcal{Q}}_- = \frac{1}{2} (Q^3 - i Q^4),
	\end{equation}
	and similarly for $\bar{\mathcal{Q}}_+$, $\bar{\mathcal{S}}_\pm$.

	Now define the spinor-index supercharges $G^{\alpha\beta}_{\pm 1/2}$ in terms of the complex combinations as
	\begin{align}
	G^{+-}_{\pm 1/2} &= \frac{i}{\sqrt{2\omega}} \left(\bar{\mathcal{Q}}_- \mp i \omega \bar{\mathcal{S}}_- \right), &
	G^{-+}_{\pm 1/2} &= \frac{i}{\sqrt{2\omega}} \left(\mathcal{Q}_- \mp i \omega \mathcal{S}_- \right), \label{eq: Gpm Gmp defns} \\
	G^{++}_{\pm 1/2} &= \frac{-i}{\sqrt{2\omega}} \left(\mathcal{Q}_+ \mp i \omega \mathcal{S}_+ \right), &
	G^{--}_{\pm 1/2} &= \frac{i}{\sqrt{2\omega}} \left(\bar{\mathcal{Q}}_+ \mp i \omega \bar{\mathcal{S}}_+ \right). \label{eq: Gpp Gmm defns}
	\end{align}

	This structure is equivalent to the standard $SU(2) \times SU(2)$ doublet notation, as expressed by
	\begin{align}
	G^{\alpha\beta}_{+1/2} &= \frac{i}{2\sqrt{2\omega}} (Q^\mu - i \omega S^\mu) (\bar{\sigma}_\mu)^{\beta\alpha}, \label{eq: Galpha beta su2 form1} \\
	G^{\alpha\beta}_{-1/2} &= \frac{i}{2\sqrt{2\omega}} (Q^\mu + i \omega S^\mu) ({\sigma}_\mu)^{\alpha\beta}, \label{eq: Galpha beta su2 form}
	\end{align}
	with the extended sigma matrix definitions
	\begin{equation}
	(\sigma_\mu)_\alpha{}^{\beta} = (\sigma_i, i), \qquad
	(\bar{\sigma}_\mu)_\alpha{}^{\beta} = (\sigma_i, -i).
	\end{equation}

	To work out index contractions in these conventions, it is useful to write them in components
	\begin{align}
	(\sigma_\mu)^{\alpha \tilde{\alpha}} &= (\sigma_\mu)_\beta{}^{\tilde{\alpha}} \, \epsilon^{\beta\alpha}, \\
	(\sigma_\mu)^{++} &= -(\sigma_\mu)_-{}^+, &
	(\sigma_\mu)^{--} &= (\sigma_\mu)_+{}^-, \\
	(\sigma_\mu)^{+-} &= -(\sigma_\mu)_-{}^-, &
	(\sigma_\mu)^{-+} &= (\sigma_\mu)_+{}^+.
	\end{align}

	From these relations, we can also define real combinations of supercharges $\Upsilon^\mu_{\pm 1/2}$, which are linearly related to the spinor charges
	\begin{align}
	\Upsilon^2_{\pm 1/2} &\equiv G^{++}_{\pm 1/2} + G^{--}_{\mp 1/2}, &
	\Upsilon^4_{\pm 1/2} &\equiv G^{+-}_{\pm 1/2} - G^{-+}_{\mp 1/2}, \label{eq: definition Upsilon1,2} \\
	\Upsilon^1_{\pm 1/2} &\equiv i(G^{++}_{\pm 1/2} - G^{--}_{\mp 1/2}), &
	\Upsilon^3_{\pm 1/2} &\equiv -i(G^{+-}_{\pm 1/2} + G^{-+}_{\mp 1/2}). \label{eq: definition Upsilon3,4}
	\end{align}

	These satisfy the following relations when expressed in terms of the original sigma model supercharges:
	\begin{align}
	\Upsilon^2_{\pm 1/2} &= \frac{1}{\sqrt{2\omega}} (Q^2 \mp \omega S^1), &
	\Upsilon^4_{\pm 1/2} &= \frac{1}{\sqrt{2\omega}} (Q^4 \pm \omega S^3), \label{eq: Upsilon1,2 Qmu} \\
	\Upsilon^1_{\pm 1/2} &= \frac{1}{\sqrt{2\omega}} (Q^1 \pm \omega S^2), &
	\Upsilon^3_{\pm 1/2} &= \frac{1}{\sqrt{2\omega}} (Q^3 \mp \omega S^4). \label{eq: Upsilon3,4 Qmu}
	\end{align}

	Finally, we collect the identifications for the bosonic generators. The commuting $\mathfrak{su}(2)$ $R$ symmetry factors are
	\begin{equation}
	J^i_R = -(R^i + \tilde{R}^i), \qquad J^i_L = \tilde{R}^i, \label{eq: noncommuting R sym dictionary}
	\end{equation}
	and the conformal generators in terms of $H$, $D$, $K$ are
	\begin{equation}
	L_0 = \frac{1}{2\omega}(H + \omega^2 K), \qquad\quad
	L_{\pm} = \frac{1}{2\omega}(H - \omega^2 K \pm 2i \omega D).
	\end{equation}

	\section{Adapted coordinates and metric decomposition} \label{app:metric-decomposition}

	\paragraph{Ungauged Models} We start by reviewing the geometric origin of the complex coordinate definition for ungauged $(4,4,0)$ sigma models \cite{Michelson:1999zf}, which will be useful for a generalization to the gauged case. We consider a $4N$ dimensional target space of an ungauged sigma model, parameterized by the flat Cartesian coordinates:
	\begin{equation}
	(x^{\mu a}) \in \mathbb{R}^{4N}, \qquad \mu = 1,\dots,4,\quad a = 1,\dots,N,
	\end{equation} with fermionic superpartners $\chi^{\mu a}$.

	We recall that the flat space $\mathbb{R}^4$ admits a family of complex structures generated by the self-dual 't Hooft symbols $j^i_+$. In particular, we have
	\begin{equation}
	(J^3)^{\mu a}{}_{\nu b}
	=
	\delta^a_b (j^3_+)^\mu{}_\nu,
	\qquad
	(j^3_+)^\mu{}_\nu
	=
	\begin{pmatrix}
		0 & -1 & 0 & 0 \\
		1 & 0 & 0 & 0 \\
		0 & 0 & 0 & -1 \\
		0 & 0 & 1 & 0
	\end{pmatrix}.
	\end{equation}

	We can drop the node index now since each node carries the same almost complex structure acting on differential forms
	\begin{equation}
	J^3: dx^\mu \mapsto (J^3)^\mu_{\ \nu} dx^\nu.
	\end{equation}

	The complex 1-forms
	\begin{equation}
	dz^{1} = \frac{1}{\sqrt{2}}(dx^{1} + i dx^{2}), \qquad
	dz^{2} = \frac{1}{\sqrt{2}}(dx^{3} + i dx^{4})
	\end{equation}
	are then holomorphic 1-forms adapted to $J^3$, i.e.
	\begin{equation}
	J^3 \cdot dz^{1} = idz^{1} \qquad\qquad J^3 \cdot dz^{2} = idz^{2},
	\end{equation}
	and integrating them gives the standard complex coordinates.

	\paragraph{Gauged Models}

	In the gauged model, the {physical} bosonic degrees of freedom are
	\begin{equation}
	(x^{ia}) \in \mathbb{R}^{3N}, \qquad i = 1,2,3, \quad a = 1,\dots,N
	\end{equation}
	while the fourth coordinate $x^{4a}$ corresponds to the gauged isometries. We can similarly introduce a radial-angular coordinate system on the physical configuration space $\mathbb{R}^{3N}$, using a complex stereographic coordinate on the angular $S^2$ associated with each node.

	Each bosonic vector $\vec{x}^a \in \mathbb{R}^{3}$ (for each $a = 1, \dots, N$) can be written in spherical coordinates
	\begin{equation}
	\vec{x}^a = r^a  \hat{r}^a (\theta^a,\phi^a),
	\end{equation}
	where $r^a \in \mathbb{R}_{\ge 0}$ is the radial distance and $\hat{r}^a \in S^2$ is the unit direction vector, with the standard parametrization
	\begin{equation}
	\hat{r}^a =
	\left(
	\sin\theta^a \cos\phi^a,\;
	\sin\theta^a \sin\phi^a,\;
	\cos\theta^a
	\right).
	\end{equation}

	Rather than using standard spherical angles $(\theta^a, \phi^a)$, we use the complex stereographic coordinates\footnote{In this appendix only, $z^a$ denotes the stereographic angular coordinate of the $a$-th node and should not be confused with the collinear position variables used in the localization computation.} $z^a, \bar{z}^a \in \mathbb{C}$ at each node $a$
	\begin{equation}
	z^a = \tan\left( \frac{\theta^a}{2} \right) e^{i \phi^a},
	\end{equation}
	which gives a holomorphic parametrization via
	\begin{equation}
	\hat{r}^a =
	\frac{1}{1 + |z^a|^2}
	\left(
	z^a + \bar{z}^a,\;
	\frac{1}{i}(z^a - \bar{z}^a),\;
	1 - |z^a|^2
	\right). \label{eq: hat r^a}
	\end{equation}

	Thus, the bosonic coordinates of node $a$ become
	\begin{equation}
	x^{ia} = r^a  \hat{r}^{ia}(z^a, \bar{z}^a), \label{eq: stero full xia}
	\end{equation}
	which associates a complex frame for each node at a fixed radial distance. 

	The $\mathcal{N}=4$ gauged quantum mechanics target space has a local quaternionic structure, described node-by-node by:
	\begin{equation}
	(J^\rho)^{\mu a}{}_{\nu b}
	=
	\delta^a_b (j^\rho)^\mu{}_\nu,
	\qquad
	j^\rho=(j^1_+,j^2_+,j^3_+,\mathbb I  )
	\end{equation}
	where $(j^i_+)_{\mu\nu}$ are the self-dual 't Hooft symbols.

	At each node, the complex structures $J^i$ act on the full four-dimensional $(x^{ia},x^{4a})$ space. After separating the gauge-fiber direction, their spatial components induce the standard rotations on the physical $\mathbb R^3$ coordinates. In particular, $j^3_+$ acts on the spatial components as
	\begin{equation}
	(j^3_+)_{ij} =
	\begin{pmatrix}
		0 & -1 & 0 \\
		1 & 0 & 0 \\
		0 & 0 & 0
	\end{pmatrix},
	\end{equation}
	and defines a complex structure on the tangent plane orthogonal to the $x^3$-axis, that is locally compatible with the stereographic frame. Thus, the stereographic coordinates $(z^a, \bar{z}^a)$ of each node capture the angular orientation of each quiver node purely via $\hat{r}^a$, while $r^a$ captures its radial information.

	From (\ref{eq: stero full xia}), we have 
	\begin{equation}
	dx^{ia} = \hat{r}^{ia} \, dr^a + r^a \, \frac{\partial \hat{r}^{ia}}{\partial z^a} dz^a + r^a \, \frac{\partial \hat{r}^{ia}}{\partial \bar{z}^a} d\bar{z}^a,
	\end{equation}
	which gives 
	\begin{align}
	dx^{ia} dx^{ib} &= \hat{r}^{ia} \hat{r}^{ib}  dr^a dr^b
	+ r^b\, \hat{r}^{ia}  \partial_{z^b} \hat{r}^{ib}  dr^a dz^b
	+ r^b\, \hat{r}^{ia}  \partial_{\bar{z}^b} \hat{r}^{ib}  dr^a d\bar{z}^b \notag \\
	&\quad
	+ r^a\, \hat{r}^{ib}  \partial_{z^a} \hat{r}^{ia}  dr^b dz^a
	+ r^a\, \hat{r}^{ib}  \partial_{\bar{z}^a} \hat{r}^{ia}  dr^b d\bar{z}^a \notag \\
	&\quad
	+ r^a r^b  \partial_z \hat{r}^{ia} \partial_z \hat{r}^{ib}  dz^a dz^b
	+ r^a r^b  \partial_{\bar{z}} \hat{r}^{ia} \partial_{\bar{z}} \hat{r}^{ib}  d\bar{z}^a d\bar{z}^b \notag \\
	&\quad
	+ r^a r^b  \partial_z \hat{r}^{ia} \partial_{\bar{z}} \hat{r}^{ib}  dz^a d\bar{z}^b
	+ r^a r^b  \partial_{\bar{z}} \hat{r}^{ia} \partial_z \hat{r}^{ib}  d\bar{z}^a dz^b. \label{eq: metric dx^ia dx^ib full}
	\end{align}

	From (\ref{eq: hat r^a}), we obtain the \textit{angular frame vectors}:
	\begin{equation}
	\partial_z \hat{r}^a  =
	\frac{1}{(1 + |z^a |^2)^2}
	\begin{pmatrix}
		1 -  (\bar{z}^a)^2 \\
		-i(1 + (\bar{z}^a)^2) \\
		-2\bar{z}^a 
	\end{pmatrix}  \qquad\qquad
	\partial_{\bar{z}} \hat{r}^a  =
	\frac{1}{(1 + |z^a|^2)^2}
	\begin{pmatrix}
		1 - (z^a)^2 \\
		i(1 + (z^a)^2) \\
		-2z^a 
	\end{pmatrix}, \label{eq: angular frame vectors}
	\end{equation} which satisfy the crucial identities
	\begin{align}
	\hat{r}^a \cdot \partial_z \hat{r}^a  &= 0, \quad \hat{r}^a  \cdot \partial_{\bar{z}} \hat{r}^a  = 0 \label{eq: vec ids ra 1} \\
	\partial_z \hat{r}^a  \cdot \partial_z \hat{r}^a  &= 0, \quad \partial_{\bar{z}} \hat{r}^a  \cdot \partial_{\bar{z}} \hat{r}^a  = 0 \label{eq: vec ids ra 2}  \\
	\partial_z \hat{r}^a  \cdot \partial_{\bar{z}} \hat{r}^a  &= \dfrac{2}{(1 + |z^a |^2)^2}, \label{eq: vec ids ra 3} 
	\end{align}
	and the conjugation property
	\begin{equation}
	\overline{\partial_z \hat{r}^a}  = \partial_{\bar z} \hat{r}^a . \label{eq: frame fields conjugation}
	\end{equation}
	As expected, (\ref{eq: metric dx^ia dx^ib full}) thus simplifies when evaluated on the same node ($a=b$) to 
	\begin{equation}
	dx^{ia} dx^{ia} = 
	dr^a dr^a
	+ r^a r^a \Omega^{(a)}_{S^2}
	, \label{eq: flat gauged general metric}
	\end{equation} 
	where 
	\begin{equation}
	\Omega^{(a)}_{S^2}(z^a, \bar{z}^a) =  \frac{4}{(1 + |z^a|^2)^2} \, dz^a d\bar{z}^a.
	\end{equation}
	In particular, (\ref{eq: flat gauged general metric}) corresponds to the reduced metric for the $\textit{flat}$ gauged sigma models with $(3,4,1)$ origin, which has $G_{ab}= \delta_{ab}$. However, the identities  (\ref{eq: vec ids ra 1}-\ref{eq: vec ids ra 3}) hold only at a fixed node, and for general non-flat metrics all terms in (\ref{eq: metric dx^ia dx^ib full}) are generically non-vanishing.  

	We define the following functions of $(z^{a},\bar{z}^{a})$:
	\begin{align}
	&\chi_{ab} = 2\partial_z \hat{r}^{a} \cdot \partial_{\bar{z}} \hat{r}^{b}, \qquad
	\xi_{ab} = \partial_z \hat{r}^{a} \cdot \partial_z \hat{r}^{b}, \qquad
	\bar{\xi}_{ab} = \partial_{\bar{z}} \hat{r}^{a} \cdot \partial_{\bar{z}} \hat{r}^{b}, \label{eq: angular metric couplings}
	\end{align}
	which capture the generalized angular metric couplings. The full angular sector of the metric then takes the form:
	\begin{align}
	ds^2_{\text{angular}} = G_{ab} \left[ r^a r^b (\chi_{ab} dz^a d\bar{z}^b + \bar{\xi}_{ab} d\bar{z}^a d\bar{z}^b + \xi_{ab} dz^a dz^b) \right], \label{eq: pure angular metric}
	\end{align}
	with each coupling depending nontrivially on both $z^a$ and $z^b$. This gives the angular part of the metric for gauged sigma models with $(3,4,1)$ origin for any given node metric $G_{ab}$ and a particular configuration in the physical target space which fix the coupling functions (\ref{eq: angular metric couplings}). Geometrically, the couplings $\chi_{ab}$ and $\xi_{ab}$ measure the metric overlap between angular variations at different nodes. Such angular deformations generalize the standard $S^2$ geometry and remain non-trivial, as we will show explicitly, for symmetric or scaling quiver configurations.

	We also have cross-terms mixing radial and angular derivatives,  which appear through 
	\begin{align}
	\mu_{ab} = \hat{r}^{a} \cdot \partial_{{z}} \hat{r}^{b}, \qquad
	\bar{\mu}_{ab} = \hat{r}^{a} \cdot \partial_{\bar{z}} \hat{r}^{b}, 
	\label{eq: mu couplings}
	\end{align}
	describing the coupling between the radial motion of node $a$ and the angular motion of node $b$. $\mu_{ab}$ and $\bar{\mu}_{ab}$ are complex conjugates due to (\ref{eq: frame fields conjugation}). Similarly to the angular couplings, though these coefficients are also in general complex, they contribute to the metric (\ref{eq: metric dx^ia dx^ib full}) only in Hermitian combinations.

	Unlike $\chi_{ab}$ and $\xi_{ab}$, which describe purely angular deformations, the $\mu_{ab}$ couplings describe inter-node correlations including radial displacements. Similarly to the angular deformations, these effects trivialize in the flat case (or in highly symmetric cases) but are generically nonzero.

	Thus, the full sigma model metric now includes the complete set of radial, angular, gauge-fiber, and crucially mixed couplings derived earlier and explicitly computable once the configuration (and the node-metric $G_{ab}$) is fixed. It decomposes into node-covariant blocks:
	\begin{equation}
	\begin{aligned}
		ds^2 = G_{ab}(x)\big[&
		\gamma_{ab}\,dr^a dr^b
		+ dx^{4a}dx^{4b}
		+ r^a r^b
		\left(
		\chi_{ab}\,dz^a d\bar z^b
		+\xi_{ab}\,dz^a dz^b
		+\bar\xi_{ab}\,d\bar z^a d\bar z^b
		\right)
		\\
		&\quad
		+2r^b\mu_{ab}\,dr^a dz^b
		+2r^b\bar\mu_{ab}\,dr^a d\bar z^b
		\big].
	\end{aligned} \label{eq: metric decomposition main}
	\end{equation}
	To summarize, here $\gamma_{ab}$ is the cosine of the angular difference between nodes (located at $z^a$ and $z^b$):
	\begin{equation}
	\gamma_{ab} = \hat{r}^a \cdot \hat{r}^b , \label{eq: gamma_ab defn}
	\end{equation}
	$\chi_{ab}$ (\ref{eq: angular metric couplings}) captures the generalized angular overlap, $\xi_{ab}$ and $\bar{\xi}_{ab}$ (\ref{eq: angular metric couplings}) encode holomorphic and anti-holomorphic angular interference between nodes, $\mu_{ab}$ and $\bar{\mu}_{ab}$ (\ref{eq: mu couplings}) describe radial-angular mixing across different nodes, $G_{ab}(x)$ mixes the node contributions in all sectors:
	\begin{equation}
	G_{ab}[r^c,z^c,\bar{z}^c] = \frac{1}{2} \partial_{ia} \partial_{ib}\calh(r^c, \hat{r}^{ic}(z^c, \bar{z}^c)).
	\end{equation}
	All couplings $\chi_{ab},\xi_{ab},\bar\xi_{ab},\mu_{ab},\bar \mu_{ab}$ depend nontrivially on $z^a, z^b$, and encode the detailed geometry of the $S^2$ fiber structure per node. We note that the decomposition (\ref{eq: metric decomposition main}) is general for the gauged sigma models with $(3,4,1)$ origin.

	\subsection{Explicit quiver metrics} 

	We now evaluate the decomposition \eqref{eq: metric decomposition main} for simple quiver configurations.  First, we consider a rotationally aligned configuration in which each node lies along the $x^3$-axis
	\begin{equation}
	z^1 = z^2 = 0 \qquad  \qquad \hat{r}^{ia}(z^a = 0) = \delta^{i3} \qquad  \qquad x^{ia} = r^a \delta^{i3}, \qquad\qquad a=1,2,
	\end{equation}
	which describes an arbitrary 2-node quiver configuration due to rotational symmetry. 

	With our coordinate frame fixed, angular frame vectors (\ref{eq: angular frame vectors}) become
	\begin{equation}
	\partial_z \hat{r}^a = \begin{pmatrix}
		1 \\ -i \\ 0
	\end{pmatrix} \qquad\qquad \partial_{\bar z} \hat{r}^a = \begin{pmatrix}
		1 \\ i \\ 0
	\end{pmatrix}, \qquad\qquad a=1,2,
	\end{equation}
	which give the angular metric couplings (\ref{eq: angular metric couplings}): 
	\begin{equation}
	\chi_{ab} = 4 \qquad\qquad \xi_{ab}=\bar{\xi}_{ab}=0,\qquad\qquad a=1,2,
	\end{equation} and also calculating the radial metric couplings (\ref{eq: mu couplings},\ref{eq: gamma_ab defn}): 
	\begin{equation}
	\gamma_{ab}=1 \qquad\qquad \mu_{ab}=\bar{\mu}_{ab}=0,\qquad\qquad a=1,2,
	\end{equation} 
	we find that the sigma model metric in spherical-complex coordinates becomes
	\begin{equation}
	ds^2 = G_{ab} \left( dr^a dr^b + r^a r^b \, \Omega_{S^2} \right) + G_{ab}(x) dx^{4a} dx^{4b},
	\end{equation}
	where the generic angular metric (\ref{eq: pure angular metric}) simplifies in this case to the round metric 
	\begin{equation}
	\Omega_{S^2}^{(ab)} = 4 dz^a d\bar{z}^b=4dz d\bar{z}=\Omega_{S^2},
	\end{equation}
	where we denoted $dz := dz^1 = dz^2$.

	Finally the node space metric (\ref{eq: scaling Gab}) becomes
	\begin{equation}
	G_{ab}(x) = \frac{|\Gamma|}{4 r_{12}^3}
	\begin{pmatrix}
		1 & -1 \\
		-1 & 1
	\end{pmatrix},
	\end{equation}
	and depends only on the relative distance $r_{12}= |r^1 - r^2|$, with $\Gamma=\Gamma_{12}$ denoting the DSZ product:
	\begin{equation}
	\Gamma_{ab} = \left<\Gamma_a,\Gamma_b\right> =-  \left<\Gamma_b,\Gamma_a\right>.
	\end{equation}	

	We thus obtain
	\begin{align}
	ds^2 &= G_{ab} dr^a dr^b + G_{ab} r^a r^b \Omega_{S^2}^{(a)} + G_{ab} dx^{4a} dx^{4b} \notag \\
	&= \frac{|\Gamma|}{4 r_{12}^3} \Big(\left[ (dr^1)^2 + (dr^2)^2 - 2 dr^1 dr^2 \right] +
	\left[(r^1)^2 + (r^2)^2 - 2 r^1 r^2 \right] \Omega_{S^2} \notag \\
	&\quad + \left[ (dx^{41})^2 + (dx^{42})^2 - 2 dx^{41} dx^{42} \right]\Big) \notag \\
	&= \frac{|\Gamma|}{4 {\tilde r}^3} \left( d\tilde r^2 + \tilde r^2 \Omega_{S^2} + (d\tilde x^4)^2 \right), \label{eq: 2node metric}
	\end{align}
	where in the last step we work in an ordered patch and define $\tilde r=r^1-r^2>0$, $\tilde x^4=x^{41}-x^{42}$.

	We next consider a symmetric three-node configuration with
	\begin{equation}
	r_{12}=r_{13}=r_{23}=\rho,
	\qquad
	r^1=r^2=r^3=r_0 ,
	\end{equation}
	which strikes a nice balance between providing analytic control and accommodating nontrivial physical features, such as the angular coupling terms appearing in (\ref{eq: metric decomposition main}), which were completely absent in the 2-node case.

	We realize this configuration by placing the three nodes at equally spaced angular positions on the unit circle by choosing the frame vectors as
	\begin{equation}
	\hat{r}^{1} =
	\begin{pmatrix}
		1 \\
		0 \\
		0
	\end{pmatrix} \qquad\qquad  		\hat{r}^{2} =
	\frac{1}{2}
	\begin{pmatrix}
		-1  \\
		\sqrt{3} \\
		0
	\end{pmatrix} \qquad  	\qquad	\hat{r}^{3} =
	\frac{1}{2}
	\begin{pmatrix}
		-1  \\
		-\sqrt{3} \\
		0
	\end{pmatrix}, \label{eq: 3node frame vectors}
	\end{equation}
	which precisely describe the node locations in the Cartesian plane, as expected. These determine the purely radial metric couplings 
	\begin{equation}
	\gamma_{ab} = \begin{pmatrix}
		1 & -1/2 & -1/2 \\ 
		-1/2 & 1 & -1/2 \\ 
		-1/2 & -1/2 & 1
	\end{pmatrix}. \label{eq: 3node gamma_ab explicit}
	\end{equation}

	Key objects are the angular frame vectors 
	\begin{equation}
	\partial_z \hat{r}^1  =
	\begin{pmatrix}
		0 \\ -\frac{i}{2} \\ -\frac12
	\end{pmatrix},
	\qquad
	\partial_z \hat{r}^2  =
	\frac{1}{4}
	\begin{pmatrix}
		1-e^{-4\pi i/3} \\
		-i\left( 1+ e^{-4\pi i/3}\right) \\
		-2e^{-2\pi i/3}
	\end{pmatrix},
	\qquad
	\partial_z \hat{r}^3  =
	\frac{1}{4}
	\begin{pmatrix}
		1-e^{-2\pi i/3} \\
		-i\left( 1+ e^{-2\pi i/3}\right) \\
		-2e^{2\pi i/3}
	\end{pmatrix}.
	\label{eq: angular frame vector 3node}
	\end{equation} 
	which determine the angular couplings $\chi_{ab},\xi_{ab},\bar\xi_{ab}$ and the radial-angular couplings $\mu_{ab},\bar\mu_{ab}$ via (\ref{eq: angular metric couplings}) and (\ref{eq: mu couplings}).

	The general Coulomb branch metric (\ref{eq: scaling Gab})  $G_{ab}$ for the 3-node quiver in the special case where all pairwise distances are equal: $r_{12} = r_{13} = r_{23} = \rho$ becomes
	\begin{equation}
	G_{ab} = \frac{1}{4 \rho^3} \begin{pmatrix}
		|\Gamma_{12}| + |\Gamma_{13}| & -|\Gamma_{12}| & -|\Gamma_{13}| \\
		-|\Gamma_{12}| & |\Gamma_{12}| + |\Gamma_{23}| & -|\Gamma_{23}| \\
		-|\Gamma_{13}| & -|\Gamma_{23}| & |\Gamma_{13}| + |\Gamma_{23}|
	\end{pmatrix}, \label{eq: 3node Gab}
	\end{equation}
	where $\Gamma_{ab}=-\Gamma_{ba}$ are the DSZ pairings. This expression encodes the pairwise interactions between nodes, with the overall $\rho^{-3}$ scaling fixed by the symmetric configuration.

	For the scaling charge assignment \eqref{eq: 3node scaling DSZ assignment}, the equal-distance configuration satisfies \eqref{eq: sum gamma zero}. The node-space metric then reduces to
	\begin{equation}
	G_{ab}\big|_{(\Gamma,\Gamma,-\Gamma)}
	=
	\frac{|\Gamma|}{4\rho^3}
	\widetilde G_{ab},
	\qquad
	\widetilde G_{ab}
	=
	\begin{pmatrix}
		2 & -1 & -1 \\
		-1 & 2 & -1 \\
		-1 & -1 & 2
	\end{pmatrix}.
	\label{eq: scaling metric simpler}
	\end{equation}
	Together with the angular and radial-angular couplings computed from \eqref{eq: angular frame vector 3node}, this determines the full metric \eqref{eq: metric decomposition main} for this symmetric scaling three-node example.

	\paragraph{Regular cyclic $N$-node examples.}
	The same metric decomposition applies to the regular cyclic scaling configurations introduced in the main text. Consider a regular $N$-gon with equal nearest-neighbor separations $r_{a,a+1}=\rho$, and nearest-neighbor DSZ pairings
	\begin{equation}
	\Gamma_{a,a+1}=k,
	\qquad
	\Gamma_{a+1,a}=-k,
	\qquad
	k>0,
	\end{equation}
	with cyclic identification of the node labels. Since only nearest-neighbor pairings are nonzero, the node-space metric \eqref{eq: scaling Gab} becomes
	\begin{align}
	G_{ab}(x)
	&=
	\frac{k}{4\rho^3}
	\begin{cases}
		2, & a=b, \\
		-1, & a \text{ and } b \text{ are nearest neighbors}, \\
		0, & \text{otherwise},
	\end{cases}
	\nonumber\\
	&=
	\frac{k}{4\rho^3}
	\left(
	2\delta_{ab}-A^{(N)}_{ab}
	\right),
	\label{eq: regular N node metric}
	\end{align}
	where $A^{(N)}_{ab}$ is the adjacency matrix of the cycle graph $C_N$. The full adapted-coordinate metric is then obtained from \eqref{eq: metric decomposition main} by evaluating the universal couplings $\gamma_{ab},\chi_{ab},\xi_{ab},\bar\xi_{ab},\mu_{ab},\bar\mu_{ab}$ for the chosen regular $N$-gon embedding.
	
	\section{Refined gauged sigma model}
	\label{sec:refined-gauged-sigma-model}
	
	In this appendix, we write the refined Euclidean action in a form adapted to the gauge symmetry and identify the effective gauged sigma-model data which appear after the refined localizing vector is decomposed into its horizontal and vertical parts.
	
	Our starting point is \eqref{eq: susy exact generalization minkowski}. We Wick-rotate and rescale according to
	\begin{equation}
		t_{\rm M}=-i\beta\tilde\tau,
		\qquad a_{\rm M}^I=\frac{i}{\beta}a^I,
		\qquad 0\leq\tilde\tau\leq1.
	\end{equation}
	With $s=\pm1$, $W_A^{(s)}=-s\lambda\omega_A^3-a^Ik_{IA}$ and $\mathcal L_s^A=\dot x^A+W^{(s)A}$, the full Euclidean action is
	\begin{align}
		\breve S_s^E=\int_0^1d\tilde\tau\,\bigg[&
		-ia^Iv_I-iA_A\dot x^A+si\omega_A^3\mathcal L_s^A
		+\frac{1}{2\beta}G_{AB}\mathcal L_s^A\mathcal L_s^B
		\nonumber\\
		&-\frac12\nabla_AW_B^{(s)}\chi^A\chi^B
		-\frac14W^{(s)C}C_{CAB}\chi^A\chi^B
		+\frac{s\lambda}{2}k_{IA}V_B^{3I}\chi^A\chi^B
		\nonumber\\
		&+\frac12G_{AB}\chi^A\dot\chi^B
		+\frac12\hat\Gamma_{ABC}\chi^A\dot x^B\chi^C
		\nonumber\\
		&+\frac{i\beta}{2}(F-s\,d\omega^3)_{AB}\chi^A\chi^B
		+\frac{\beta}{12}\partial_{[A}C_{BCD]}
		\chi^A\chi^B\chi^C\chi^D\bigg].
		\label{eq: SE leading terms raw}
	\end{align}
	
	Let us first recall the properties we need. The gauge generators obey
	\begin{equation}
		[k_I,k_J]^A=f_{IJ}{}^Kk_K^A .
		\label{eq:rgsm-gauge-algebra}
	\end{equation}
	They preserve the sigma-model tensors:
	\begin{equation}
		\mathcal L_{k_I}G_{AB}=0,
		\qquad
		\mathcal L_{k_I}C_{ABC}=0 .
		\label{eq:rgsm-gauge-preserves-GC}
	\end{equation}
	The original magnetic data obey
	\begin{equation}
		i_{k_I}F=dv_I,
		\label{eq:rgsm-original-moment-map}
	\end{equation}
	and the moment maps are equivariant:
	\begin{equation}
		\mathcal L_{k_K}v_I
		=
		-f_{IK}{}^Jv_J .
		\label{eq:rgsm-v-equivariance}
	\end{equation}
	The $R$-symmetry vector $\omega^3$ is gauge invariant:
	\begin{equation}
		\mathcal L_{k_I}\omega^3=0 .
		\label{eq:rgsm-omega3-gauge-invariant}
	\end{equation}
	We also have the gauged $D(2,1;0)$ compatibility condition
	\begin{equation}
		\omega^{3C}C_{CAB}
		=
		(J^3)_{AB}
		+
		2\nabla_{[A}\omega^3_{B]}
		-
		2k_{I[A}V^{3I}_{B]} .
		\label{eq:rgsm-D210-identity}
	\end{equation}
	The one-forms $V^{3I}$ are adjoint-valued:
	\begin{equation}
		\mathcal L_{k_K}V^{3I}
		=
		f_{JK}{}^I V^{3J}.
		\label{eq:rgsm-V-adjoint}
	\end{equation}
	
	We decompose $\omega^3$ into its vertical and horizontal parts. Define\footnote{We assume that the gauge action is locally free on the locus of interest, so
		that $G_{IJ}=k_I^Ak_{JA}$ is invertible.}
	\begin{equation}
		G_{IJ}:=k_I^Ak_{JA},
		\qquad
		\Omega^{3I}:=G^{IJ}k_{JA}\omega^{3A},
		\qquad
		\mu_I^3:=k_{IA}\omega^{3A}=G_{IJ}\Omega^{3J}.
		\label{eq:rgsm-Omega-definition}
	\end{equation}
	Then
	\begin{equation}
		\omega^{3A}
		=
		\omega_\perp^{3A}
		+
		\Omega^{3I}k_I^A,
		\qquad
		k_{IA}\omega_\perp^{3A}=0 .
		\label{eq:rgsm-omega-split}
	\end{equation}
	For $s=\pm1$, define
	\begin{equation}
		a_{\rm eff}^I:=a^I+s\lambda\Omega^{3I},
		\label{eq:rgsm-aeff}
	\end{equation}
	
	\begin{equation}
		D_{t,{\rm eff}}x^A:=\dot x^A-a_{\rm eff}^Ik_I^A,
		\label{eq:rgsm-Dteff}
	\end{equation}
	The refined localizing vector is therefore
	\begin{equation}
		\mathcal L_s^A
		=
		\dot x^A-a^Ik_I^A-s\lambda\omega^{3A}
		=
		D_{t,{\rm eff}}x^A-s\lambda\omega_\perp^{3A} .
		\label{eq:rgsm-Ls-effective}
	\end{equation}
	
	The magnetic data are shifted by
	\begin{equation}
		A_A^{(s)}:=A_A-s\omega_A^3,
		\qquad
		F^{(s)}:=dA^{(s)}=F-s\,d\omega^3,
		\label{eq:rgsm-AF-shift}
	\end{equation}
	and
	\begin{equation}
		v_I^{(s)}:=v_I+s\mu_I^3 .
		\label{eq:rgsm-v-shift}
	\end{equation}
	These shifted data obey the same moment-map equation as the original data.
	Indeed,
	\begin{equation}
		i_{k_I}d\omega^3
		=
		\mathcal L_{k_I}\omega^3-d(i_{k_I}\omega^3)
		=
		-d\mu_I^3,
		\label{eq:rgsm-i-k-domega}
	\end{equation}
	and hence
	\begin{equation}
			i_{k_I}F^{(s)}
			=
			i_{k_I}F-s\,i_{k_I}d\omega^3 =
			dv_I+s\,d\mu_I^3 =
			dv_I^{(s)} .
		\label{eq:rgsm-effective-moment-map}
	\end{equation}
	Moreover $v_I^{(s)}$ is equivariant:
	\begin{equation}
		\mathcal L_{k_K}v_I^{(s)}
		=
		-f_{IK}{}^Jv_J^{(s)} .
		\label{eq:rgsm-vs-equivariance}
	\end{equation}
	To see this, use $\mu_I^3=i_{k_I}\omega^3$ and \eqref{eq:rgsm-omega3-gauge-invariant}:
	\begin{equation}
			\mathcal L_{k_K}\mu_I^3
			=
			\mathcal L_{k_K}(i_{k_I}\omega^3)
			=
			i_{[k_K,k_I]}\omega^3
			=
			f_{KI}{}^J\mu_J^3
			=
			-f_{IK}{}^J\mu_J^3 .
		\label{eq:rgsm-mu-equivariance}
	\end{equation}
	Together with \eqref{eq:rgsm-v-equivariance}, this gives \eqref{eq:rgsm-vs-equivariance}.
	
	The effective fermion covariant derivative is
	\begin{equation}
		\begin{aligned}
			\hat{D}_{t,{\rm eff}}\chi^A
			:=
			&\dot\chi^A
			+
			\left(
			\Gamma^A{}_{BC}
			+
			\frac12 C^A{}_{BC}
			\right)\dot x^B\chi^C
			\\
			&+
			a_{\rm eff}^I
			\left(
			\nabla^Ak_{IB}
			+
			\frac12 C^A{}_{BC}k_I^C
			\right)\chi^B .
		\end{aligned}
		\label{eq:rgsm-Dchi-eff}
	\end{equation}
	We use
	\begin{equation}
		\hat\Gamma^A{}_{BC}
		:=
		\Gamma^A{}_{BC}
		+
		\frac12C^A{}_{BC}.
		\label{eq:rgsm-hat-Gamma}
	\end{equation}
	Equivalently,
	\begin{equation}
		\begin{aligned}
			\frac12G_{AB}\chi^A\hat D_{t,{\rm eff}}\chi^B
			&=
			\frac12G_{AB}\chi^A\dot\chi^B
			+
			\frac12 \hat\Gamma_{ABC}\chi^A\dot x^B\chi^C
			\\
			&\quad
			+
			\frac12a_{\rm eff}^I\nabla_Ak_{IB}\chi^A\chi^B
			+
			\frac14a_{\rm eff}^Ik_I^CC_{CAB}\chi^A\chi^B .
		\end{aligned}
		\label{eq:rgsm-Dchi-expanded}
	\end{equation}
	
	We now rewrite the refined action. First, the first-order terms satisfy
	\begin{equation}
		\begin{aligned}
			&-i a^Iv_I
			-iA_A\dot x^A
			+si\,\omega_A^3\mathcal L_s^A
			\\
			&\qquad =
			-i a_{\rm eff}^Iv_I^{(s)}
			-iA_A^{(s)}\dot x^A
			+i s\lambda\Omega^{3I}v_I
			-i\lambda\omega^3_{\perp A}\omega_\perp^{3A}.
		\end{aligned}
		\label{eq:rgsm-first-order-rewrite}
	\end{equation}

	Hence
	\begin{equation}
		\begin{aligned}
			&-\frac12(\nabla_AW_B^{(s)})\chi^A\chi^B
			-\frac14W^{(s)C}C_{CAB}\chi^A\chi^B
			+
			\frac{s\lambda}{2}k_{IA}V_B^{3I}\chi^A\chi^B
			\\
			&\qquad =
			\frac12a_{\rm eff}^I\nabla_Ak_{IB}\chi^A\chi^B
			+
			\frac14a_{\rm eff}^Ik_I^CC_{CAB}\chi^A\chi^B
			\\
			&\qquad\quad
			+
			\frac{s\lambda}{2}
			\left(
			\nabla_A\omega^3_{\perp B}
			+
			(\nabla_A\Omega^{3I})k_{IB}
			+
			\frac12\omega_\perp^{3C}C_{CAB}
			+
			k_{IA}V_B^{3I}
			\right)\chi^A\chi^B .
		\end{aligned}
		\label{eq:rgsm-fermion-rewrite}
	\end{equation}
	Only the antisymmetric part in $A,B$ contributes.
	
	Finally,
	\begin{equation}
		\frac{i}{2}F_{AB}\chi^A\chi^B
		-si\nabla_A\omega_B^3\chi^A\chi^B
		=\frac{i}{2}F_{AB}^{(s)}\chi^A\chi^B .
		\label{eq:rgsm-beta-rewrite}
	\end{equation}
	
	Putting these identities together, the refined action is
	\begin{equation}
		\begin{aligned}
			\breve S_s^E
			=
			\int_0^1d\tilde\tau\,\bigg[
			&
			\frac{1}{2\beta}
			G_{AB}
			\left(
			D_{t,{\rm eff}}x^A-s\lambda\omega_\perp^{3A}
			\right)
			\left(
			D_{t,{\rm eff}}x^B-s\lambda\omega_\perp^{3B}
			\right)
			\\
			&-i a_{\rm eff}^I v_I^{(s)}
			-iA_A^{(s)}\dot x^A
			+i s\lambda\Omega^{3I}v_I
			-i\lambda\omega^3_{\perp A}\omega_\perp^{3A}
			\\
			&+
			\frac12G_{AB}\chi^A \hat D_{t,{\rm eff}}\chi^B
			\\
			&+
			\frac{s\lambda}{2}
			\left(
			\nabla_A\omega^3_{\perp B}
			+
			(\nabla_A\Omega^{3I})k_{IB}
			+
			\frac12\omega_\perp^{3C}C_{CAB}
			+
			k_{IA}V_B^{3I}
			\right)\chi^A\chi^B
			\\
			&+
			\beta\bigg(
			\frac{i}{2}F_{AB}^{(s)}\chi^A\chi^B
			+
			\frac{1}{12}
			\partial_{[A}C_{BCD]}
			\chi^A\chi^B\chi^C\chi^D
			\bigg)
			\bigg].
		\end{aligned}
		\label{eq:rgsm-refined-effective-action}
	\end{equation}
	
	We recall that gauge transformations are
	\begin{equation}
		\delta_\Lambda x^A=\Lambda^Ik_I^A,
		\qquad
		\delta_\Lambda\chi^A=\Lambda^I\partial_Bk_I^A\chi^B,
		\qquad
		\delta_\Lambda a^I
		=
		\dot\Lambda^I+f_{JK}{}^Ia^J\Lambda^K .
		\label{eq:rgsm-gauge-transformations}
	\end{equation}
	
	Since $\mathcal L_{k_K}\omega^3=0$, the vertical component of $\omega^3$ transforms in the adjoint representation. Indeed,
	\begin{equation}
		0
		=
		\mathcal L_{k_K}(\Omega^{3I}k_I)
		=
		(\mathcal L_{k_K}\Omega^{3I})k_I
		+
		\Omega^{3I}[k_K,k_I].
		\label{eq:rgsm-Omega-adjoint-derivation}
	\end{equation}
	Using \eqref{eq:rgsm-gauge-algebra}, we obtain
	\begin{equation}
		\mathcal L_{k_K}\Omega^{3I}
		=
		f_{JK}{}^I\Omega^{3J}.
		\label{eq:rgsm-Omega-adjoint}
	\end{equation}
	Therefore
	\begin{equation}
		\begin{aligned}
			\delta_\Lambda a_{\rm eff}^I
			&=
			\delta_\Lambda a^I
			+
			s\lambda\,\delta_\Lambda\Omega^{3I}
			\\
			&=
			\dot\Lambda^I
			+
			f_{JK}{}^Ia^J\Lambda^K
			+
			s\lambda f_{JK}{}^I\Omega^{3J}\Lambda^K
			\\
			&=
			\dot\Lambda^I
			+
			f_{JK}{}^Ia_{\rm eff}^J\Lambda^K .
		\end{aligned}
		\label{eq:rgsm-aeff-gauge-transform}
	\end{equation}
	Thus $a_{\rm eff}^I$ transforms as a worldline gauge connection.
	
	We now check the gauge invariance of the full refined action. The bosonic gauge covariant term transforms as
	\begin{equation}
		\delta_\Lambda(D_{t,{\rm eff}}x^A)
		=
		\Lambda^I\partial_Bk_I^A\,D_{t,{\rm eff}}x^B .
		\label{eq:rgsm-Dteff-gauge-cov}
	\end{equation}
	Moreover,
	\begin{equation}
		\omega_\perp^{3A}
		=
		\omega^{3A}-\Omega^{3I}k_I^A
		\label{eq:rgsm-omega-perp-definition-repeat}
	\end{equation}
	is gauge invariant as a vector field:
	\begin{equation}
		\mathcal L_{k_I}\omega_\perp^3=0 .
		\label{eq:rgsm-omega-perp-gauge-invariant}
	\end{equation}
	Thus
	\begin{equation}
		\delta_\Lambda\omega_\perp^{3A}
		=
		\Lambda^I\partial_Bk_I^A\,\omega_\perp^{3B} ,
		\label{eq:rgsm-Us-gauge-cov}
	\end{equation}
	which gives
	\begin{equation}
		\delta_\Lambda
		\left(
		D_{t,{\rm eff}}x^A-s\lambda\omega_\perp^{3A}
		\right)
		=
		\Lambda^I\partial_Bk_I^A
		\left(
		D_{t,{\rm eff}}x^B-s\lambda\omega_\perp^{3B}
		\right).
		\label{eq:rgsm-effective-vector-covariant}
	\end{equation}
	So the bosonic square in \eqref{eq:rgsm-refined-effective-action} is gauge invariant.
	
	Using \eqref{eq:rgsm-effective-moment-map} and \eqref{eq:rgsm-vs-equivariance}, the first-order terms are gauge invariant up to a total derivative:
	\begin{equation}
		\delta_\Lambda
		\left(
		-iA_A^{(s)}\dot x^A
		-i a_{\rm eff}^Iv_I^{(s)}
		\right)
		=
		-i\frac{d}{d\tilde\tau}
		\left[
		\Lambda^I
		\left(
		A_A^{(s)}k_I^A+v_I^{(s)}
		\right)
		\right].
		\label{eq:rgsm-first-order-gauge}
	\end{equation}
	
	The scalar remainder
	\begin{equation}
		R_s
		:=
		i s\lambda\Omega^{3I}v_I
		-
		i\lambda\omega_{\perp A}^3\omega_\perp^{3A}
		\label{eq:rgsm-scalar-remainder}
	\end{equation}
	is gauge invariant. The first term is invariant by \eqref{eq:rgsm-Omega-adjoint} and \eqref{eq:rgsm-v-equivariance}. The second term is invariant by \eqref{eq:rgsm-gauge-preserves-GC} and \eqref{eq:rgsm-omega-perp-gauge-invariant}.
	
	The effective fermion derivative is gauge covariant:
	\begin{equation}
		\delta_\Lambda(\hat D_{t,{\rm eff}}\chi^A)
		=
		\Lambda^I\partial_Bk_I^A\,\hat D_{t,{\rm eff}}\chi^B ,
		\label{eq:rgsm-Dchi-gauge-cov}
	\end{equation}
	which follows as in the ordinary gauged sigma model, with $a_{\rm eff}^I$ playing the role of the gauge connection. Hence
	\begin{equation}
		\delta_\Lambda
		\left(
		G_{AB}\chi^A\hat D_{t,{\rm eff}}\chi^B
		\right)
		=
		0.
		\label{eq:rgsm-Dchi-bilinear-gauge-invariant}
	\end{equation}
	
	For the remaining fermion bilinear, define
	\begin{equation}
		\mathcal M^{(s)}_{AB}
		:=
		\left[
		\nabla_A\omega^3_{\perp B}
		+
		(\nabla_A\Omega^{3I})k_{IB}
		+
		\frac12\omega_\perp^{3C}C_{CAB}
		+
		k_{IA}V_B^{3I}
		\right]_{[AB]}.
		\label{eq:rgsm-M-two-form}
	\end{equation}
	Using
	\begin{equation}
		\mathcal L_{k_I}\omega_\perp^3=0,
		\qquad
		\mathcal L_{k_I}G=0,
		\qquad
		\mathcal L_{k_I}C=0,
		\qquad
		\mathcal L_{k_K}V^{3I}=f_{JK}{}^I V^{3J},
		\label{eq:rgsm-M-invariance-inputs}
	\end{equation}
	this two-form is gauge invariant. More explicitly, \eqref{eq:rgsm-Omega-adjoint} implies
	\begin{equation}
		\mathcal L_{k_K}d\Omega^{3I}
		=
		f_{JK}{}^I d\Omega^{3J}.
		\label{eq:rgsm-dOmega-adjoint}
	\end{equation}
	Also, since $k_I^\flat:=k_{IA}dx^A$ and $\mathcal L_{k_K}G=0$, \eqref{eq:rgsm-gauge-algebra} implies
	\begin{equation}
		\mathcal L_{k_K}k_I^\flat
		=
		f_{KI}{}^J k_J^\flat
		=
		-f_{IK}{}^J k_J^\flat .
		\label{eq:rgsm-kflat-coadjoint}
	\end{equation}
	Together with \eqref{eq:rgsm-V-adjoint}, this gives
	\begin{equation}
		\mathcal L_{k_K}
		\left(
		d\Omega^{3I}\wedge k_I^\flat
		+
		k_I^\flat\wedge V^{3I}
		\right)
		=
		0.
		\label{eq:rgsm-M-mixed-invariant-combinations}
	\end{equation}
	Therefore
	\begin{equation}
		\delta_\Lambda
		\left(
		\mathcal M^{(s)}_{AB}\chi^A\chi^B
		\right)
		=
		0.
		\label{eq:rgsm-M-fermion-bilinear}
	\end{equation}
	
	Finally,
	\begin{equation}
		\mathcal L_{k_I}F^{(s)}=0
		\label{eq:rgsm-Fs-gauge-invariant}
	\end{equation}
	follows from \eqref{eq:rgsm-effective-moment-map}, and
	\begin{equation}
		\mathcal L_{k_I}dC=0
		\label{eq:rgsm-dC-gauge-invariant}
	\end{equation}
	follows from \eqref{eq:rgsm-gauge-preserves-GC}. Thus the terms proportional to $\beta$ are gauge invariant.
	
	\section{Nonconstant-mode determinants}
	\label{app:nonconstant-determinants}
	
	In this appendix we compute the Gaussian determinants of the fluctuations around the collinear localization saddle. The computation splits into a longitudinal $(3,4)$ sector and a transverse $(1,2)$ sector. The longitudinal nonconstant modes cancel between bosons and fermions, while the transverse sector gives the universal equivariant oscillator factor appearing in \eqref{eq:nonconstant-determinant-main-result}.
	
	The node metric $G_{ab}(p)$ has the common translation direction as a null direction, so all node matrices are restricted to the relative subspace of dimension $N-1$. We choose an orthonormal frame $e^A{}_a(p)$ on this relative subspace such that
	\begin{equation}
		G_{ab}(p)=e^A{}_a(p)e^A{}_b(p).
	\end{equation}
	
	We start from the quadratic actions \eqref{eq:S12-total-definition} and \eqref{eq:S34-total-definition}, and first consider the longitudinal sector. Define
	\begin{equation}
		Z^A=\beta^{-1/2}e^A{}_aX^{3a},
		\qquad
		T^A=\beta^{-1/2}e^A{}_ah^a,
		\qquad
		\xi^A=e^A{}_a\psi^{3a},
		\qquad
		\eta^A=e^A{}_a\psi^{4a},
		\label{eq: frame nonconstant modes}
	\end{equation}
	and
	\begin{equation}
		M=e^{-T}\mathcal N_s e^{-1},
		\qquad
		\mathcal N_s=[\partial_bv_a+2sG_{ab}]_p.
	\end{equation}
	Since $M$ is real symmetric\footnote{Note that $\caln_{s,ab}=\caln_{s,ba}$.}, a common orthogonal transformation of $Z,T,\xi,\eta$ diagonalizes it. Denote its eigenvalues by $m_A$. For one relative direction the bosonic action becomes
	\begin{equation}
		S^{(2)}_{B,34,A}
		=
		\frac12
		\int_0^1d\tilde\tau\,
		\left[
		\dot Z^2+T^2-2i\beta m_A TZ
		\right].
		\label{eq:S34-boson-one-copy-no-shift}
	\end{equation}
	
	Expanding in nonzero Fourier modes,
	\begin{equation}
		Z(\tilde\tau)=\sum_{n\neq0}Z_n e^{2\pi in\tilde\tau},
		\qquad
		T(\tilde\tau)=\sum_{n\neq0}T_n e^{2\pi in\tilde\tau},
	\end{equation}
	the quadratic form for each mode is
	\begin{equation}
		S^{(2)}_{B,34,A,n}
		=
		\frac12
		\begin{pmatrix}
			Z_{-n} & T_{-n}
		\end{pmatrix}
		\begin{pmatrix}
			(2\pi n)^2 & -i\beta m_A\\
			-i\beta m_A & 1
		\end{pmatrix}
		\begin{pmatrix}
			Z_n\\
			T_n
		\end{pmatrix}.
		\label{eq:S34-boson-mode-matrix}
	\end{equation}
	The determinant of this $2\times2$ matrix is
	\begin{equation}
		\det
		\begin{pmatrix}
			(2\pi n)^2 & -i\beta m_A\\
			-i\beta m_A & 1
		\end{pmatrix}
		=
		(2\pi n)^2+\beta^2m_A^2.
		\label{eq:S34-boson-mode-det}
	\end{equation}
	Thus
	\begin{equation}
		Z_{ZT,A}
		=
		\prod_{n\neq0}
		\left[(2\pi n)^2+\beta^2m_A^2\right]^{-1/2}
		=
		\left[
		\det{}'(-\partial_{\tilde\tau}^2+\beta^2m_A^2)
		\right]^{-1/2}.
		\label{eq:ZZB-no-shift}
	\end{equation}
	
	The two real longitudinal fermions have action
	\begin{equation}
		S^{(2)}_{F,34,A}
		=
		\frac12\int_0^1d\tilde\tau\,
		\begin{pmatrix}\xi&\eta\end{pmatrix}
		\begin{pmatrix}
			\partial_{\tilde\tau}&-i\beta m_A\\
			i\beta m_A&\partial_{\tilde\tau}
		\end{pmatrix}
		\begin{pmatrix}\xi\\\eta\end{pmatrix}.
	\end{equation}
	Pairing the positive and negative Fourier modes gives
	\begin{align}
		Z_{\xi\eta,A}
		&=
		\operatorname{Pf}'\!
		\begin{pmatrix}
			\partial_{\tilde\tau}&-i\beta m_A\\
			i\beta m_A&\partial_{\tilde\tau}
		\end{pmatrix}
		\nonumber\\
		&=
		\prod_{n>0}\left[(2\pi n)^2+\beta^2m_A^2\right]
		=
		\left[
		\det{}'(-\partial_{\tilde\tau}^2+\beta^2m_A^2)
		\right]^{1/2}.
		\label{eq:Zpsieta-no-shift}
	\end{align}
	where the primes exclude the Fourier constants. Therefore one relative longitudinal copy contributes
	\begin{equation}
		Z'_{34,A}
		=
		Z_{ZT,A}Z_{\xi\eta,A}
		=
		1,
	\end{equation}
	and hence
	\begin{equation}
		Z'_{34}=1.
		\label{eq:Z34-normal-frame-one}
	\end{equation}
	
	We next turn to the transverse $(1,2)$ sector. Using the same frame,
	\begin{equation}
		Y^{iA}=\beta^{-1/2}e^A{}_aX^{ia},
		\qquad
		\zeta^{iA}=e^A{}_a\psi^{ia},
		\qquad
		i=1,2,
	\end{equation}
	the transverse action \eqref{eq:S12-total-definition} factorizes over $A=1,\ldots,N-1$. For one relative two-plane,
	\begin{align}
		S_{12,A}^{(2)}
		=
		&
		\frac12
		\int_0^1d\tilde\tau\,
		\left[
		(\dot Y^1-s\lambda Y^2)^2
		+
		(\dot Y^2+s\lambda Y^1)^2
		\right]
		\nonumber\\
		&+
		\frac12
		\int_0^1d\tilde\tau\,
		\zeta^i
		\left(
		\delta_{ij}\partial_{\tilde\tau}
		+
		s\lambda J_{ij}
		\right)
		\zeta^j,
		\qquad
		J=\begin{pmatrix}0&-1\\1&0\end{pmatrix}.
		\label{eq:S12-one-copy}
	\end{align}
	
	Define the first-order operator
	\begin{equation}
		D_s
		=
		\partial_{\tilde\tau}+s\lambda J.
		\label{eq:Ds-definition}
	\end{equation}
	Since $J^T=-J$ and
	$\partial_{\tilde\tau}^T=-\partial_{\tilde\tau}$
	on periodic fields,
	\begin{equation}
		D_s^T=-D_s.
	\end{equation}
	The bosonic part of \eqref{eq:S12-one-copy} is
	\begin{equation}
		S^{(2)}_{B,12,A}
		=
		\frac12
		\int_0^1d\tilde\tau\,
		Y^i(D_s^TD_s)_{ij}Y^j,
	\end{equation}
	where
	\begin{equation}
		D_s^TD_s
		=
		-D_s^2
		=
		-\partial_{\tilde\tau}^2
		-2s\lambda J\partial_{\tilde\tau}
		+\lambda^2.
		\label{eq:OB-from-Ds}
	\end{equation}
	The fermionic operator is $D_s$. Thus
	\begin{equation}
		S_{12,A}^{(2)}
		=
		\frac12
		\int_0^1d\tilde\tau\,
		\left[
		Y^i(D_s^TD_s)_{ij}Y^j
		+
		\zeta^i(D_s)_{ij}\zeta^j
		\right].
	\end{equation}
	
	With the Gaussian measures normalized as before, one relative two-plane gives
	\begin{equation}
		Z_{{\rm osc},s}
		=
		\frac{\operatorname{Pf}(D_s)}
		{\det(D_s^TD_s)^{1/2}}.
		\label{eq:transverse-real-ratio}
	\end{equation}
	The determinants here include the transverse Fourier constants.
	
	Complexifying the two-plane,
	\begin{equation}
		Y=Y^1+iY^2,
		\qquad
		\bar Y=Y^1-iY^2,
	\end{equation}
	we have
	\begin{equation}
		JY=iY,
		\qquad
		J\bar Y=-i\bar Y.
	\end{equation}
	On the Fourier mode $e^{2\pi in\tilde\tau}$,
	\begin{equation}
		D_s\big|_{Y,n}
		=
		i(2\pi n+s\lambda),
		\qquad
		D_s\big|_{\bar Y,n}
		=
		i(2\pi n-s\lambda).
		\label{eq:Ds-holo-antiholo-eigs}
	\end{equation}
	The real operator $D_s$ is antisymmetric, so the fermion integral gives a Pfaffian. Using the complex polarization defined by $J$, we take
	\begin{equation}
		\operatorname{Pf}_{J}(D_s)
		=
		\det\nolimits_{\mathbb C}(D_s|_{J=+i})
		=
		\prod_{n\in\mathbb Z}i(2\pi n+s\lambda),
		\label{eq:Pfaffian-J-orientation}
	\end{equation}
	which gives
	\begin{equation}
		Z_{{\rm osc},s}
		=
		\frac{\operatorname{Pf}_{J}(D_s)}
		{\det(D_s^TD_s)^{1/2}}
		=
		\frac{1}{
			\overline{
				\det\nolimits_{\mathbb C}(D_s|_{J=+i})
			}
		}
		=
		\prod_{n\in\mathbb Z}
		\frac{i}{2\pi n+s\lambda}.
		\label{eq:Zosc-inverse-conjugate-det}
	\end{equation}
	Using the zeta-regularized circle product
	\begin{equation}
	\prod_{n\in\mathbb Z}^{\zeta}
	(2\pi n+s\lambda)
	=
	2\sin\frac{s\lambda}{2},
	\label{eq:zeta-circle-product}
	\end{equation}
	we obtain
	\begin{equation}
	Z_{{\rm osc},s}
	=
	\frac{i}{2\sin(s\lambda/2)}
	=
	\frac{is}{2\sin(\lambda/2)}
	=
	\frac{s\,y^{1/2}}{1-y}.
	\label{eq:single-transverse-oscillator}
	\end{equation}
	Since there are $N-1$ relative two-planes,
	\begin{equation}
	Z_{12,s}
	=
	\left(
	\frac{s\,y^{1/2}}{1-y}
	\right)^{N-1}.
	\label{eq:Z12-final}
	\end{equation}	
	
	\section{Two- and three-center checks}
	\label{app:two-three-center-checks}
	
	In this appendix we evaluate the fixed-point formula \eqref{eq:final-D210-quiver-index} in the first two nontrivial cases.
	
	\paragraph{Two centers.}
	Take $\Gamma_{12}=\kappa\ne0$ and use the relative coordinate
	$z=z^1-z^2$. The relative data are
	\begin{equation}
		G=\frac{|\kappa|}{4|z|^3},\qquad
		v=\frac{\kappa}{2|z|},\qquad
		E_s=v-sGz.
		\label{eq:two-center-section}
	\end{equation}
	The equation $E_s=0$ gives
	\begin{equation}
		2\kappa|z|=s|\kappa|\operatorname{sign}z.
	\end{equation}
	There is therefore one root for each chirality,
	\begin{equation}
		|z|=\frac12,\qquad
		\operatorname{sign}z=s\operatorname{sign}\kappa.
		\label{eq:two-center-distance}
	\end{equation}
	At this root the Jacobian and classical exponent are
	\begin{equation}
		\mathcal N_s
		=\left.\frac{dE_s}{dz}\right|_p
		=2s|\kappa|,
		\qquad
		zv=\frac{s|\kappa|}{2}.
		\label{eq:two-center-reduced-Jacobian}
	\end{equation}
	Thus the orientation sign is $s$, while the classical weight is
	$y^{szv}=y^{|\kappa|/2}$, and hence the fixed point formula gives
	\begin{equation}
		\mathcal I_{s,\,2{\rm c}}(y)
		=
		s\,\frac{s\,y^{1/2}}{1-y}\,y^{|\kappa|/2}
		=
		\frac{y^{(|\kappa|+1)/2}}{1-y}.
		\label{eq:two-center-full-sum}
	\end{equation}
	Let us note however that the two-center system does not admit a scaling black-hole solution. So this simply provides an elementary check of the local orientation conventions in this case.
	
	\paragraph{Three centers.}
	Take equal cyclic pairings
	\begin{equation}
		\Gamma_{12}=\Gamma_{23}=\Gamma_{31}=k>0
	\end{equation}
	and $s=-1$. Define the cyclic differences
	\begin{equation}
		Z_1=z^1-z^2,\qquad
		Z_2=z^2-z^3,\qquad
		Z_3=z^3-z^1,
		\qquad
		Z_1+Z_2+Z_3=0.
	\end{equation}
	The section equations become
	\begin{equation}
		E_{-,a}
		=
		\frac{k}{2}\bigl[f(Z_a)-f(Z_{a-1})\bigr]=0,
		\qquad
		f(Z)=\frac1{|Z|}+\frac{Z}{2|Z|^3},
		\label{eq:three-center-edge-section}
	\end{equation}
	where the edge labels are cyclic.
	
	For separated configurations the cyclic differences are nonzero. Their sum vanishes, so their signs must be mixed. First suppose that two differences are negative. Write them as $-r_1,-r_2$, with $r_1,r_2>0$, and the positive difference as $p=r_1+r_2$. Equation \eqref{eq:three-center-edge-section} requires
	\begin{equation}
		\frac1{r_1}-\frac1{2r_1^2}
		=
		\frac1{r_2}-\frac1{2r_2^2}
		=
		\frac1p+\frac1{2p^2}.
	\end{equation}
	Equality of the first two expressions gives
	\begin{equation}
		(r_2-r_1)\bigl[2r_1r_2-(r_1+r_2)\bigr]=0.
	\end{equation}
	If $r_1r_2=(r_1+r_2)/2=p/2$, their common value is $1/p$, which cannot equal $1/p+1/(2p^2)$. Thus $r_1=r_2=r$ and $p=2r$. The remaining equation is
	\begin{equation}
		\frac1r-\frac1{2r^2}
		=
		\frac1{2r}+\frac1{8r^2},
		\qquad\Longrightarrow\qquad
		r=\frac54.
	\end{equation}
	
	If instead two cyclic differences are positive, say $r_1,\,r_2>0$, $f(r_1)=f(r_2)$ requires their lengths to agree, $r_1=r_2=r$, so the negative difference is $-2r$. These cannot solve the section equations because
	\begin{equation}
		f(r)-f(-2r)
		=
		\frac1{2r}+\frac5{8r^2}>0.
	\end{equation}
	Therefore the three separated roots are precisely the cyclic permutations of
	\begin{equation}
		(Z_1,Z_2,Z_3)
		=
		\left(-\frac54,-\frac54,\frac52\right).
		\label{eq:three-center-cyclic-roots}
	\end{equation}
	
Choose $z=(-5/2,-5/4,0)$. Using $z^1,z^2$ as relative coordinates, the Jacobian is
	\begin{equation}
		\mathcal N_{-}
		=
		\left[\partial_bv_a-2G_{ab}\right]_{a,b=1,2}
		=
		\frac{k}{125}
		\begin{pmatrix}
			-6&-8\\
			-8&16
		\end{pmatrix},
		\qquad
		\det\mathcal N_{-}
		=
		-\frac{32k^2}{3125}.
		\label{eq:three-center-cyclic-Jacobian}
	\end{equation}
	At the same root,
	\begin{equation}
		(v_1,v_2,v_3)
		=
		\left(\frac{k}{5},0,-\frac{k}{5}\right),
		\qquad
		z^av_a=-\frac{k}{2}.
	\end{equation}
	Cyclic symmetry gives the same orientation sign and classical weight at the other two roots. With $y=q^2$, each root therefore contributes
	\begin{equation}
		\mathcal Z_-(p)
		=
		-\left(\frac{-y^{1/2}}{1-y}\right)^2y^{k/2}
		=
		-\frac{q^k}{(q-q^{-1})^2}.
	\end{equation}
	Summing the three contributions gives
	\begin{equation}
		\mathcal I_{-,\,3{\rm c}}(q^2)
		=
		-\frac{3q^k}{(q-q^{-1})^2}.
		\label{eq:three-center-full-isolated-sum}
	\end{equation}

	\bibliographystyle{ytphys}
	\bibliography{IndexQuivers}

@article{Babichenko:2009dk,
	author = "Babichenko, A. and Stefanski, Jr., B. and Zarembo, K.",
	title = "{Integrability and the AdS(3)/CFT(2) correspondence}",
	eprint = "0912.1723",
	archivePrefix = "arXiv",
	primaryClass = "hep-th",
	reportNumber = "ITEP-TH-59-09, LPTENS-09-36, UUITP-25-09",
	doi = "10.1007/JHEP03(2010)058",
	journal = "JHEP",
	volume = "03",
	pages = "058",
	year = "2010"
}

@article{Borsato:2015mma,
	author = "Borsato, Riccardo and Ohlsson Sax, Olof and Sfondrini, Alessandro and Stefa{\'n}ski, Bogdan",
	title = "{The $\mathrm{AdS}_3\times \mathrm{S}^3\times \mathrm{S}^3\times\mathrm{S}^1$ worldsheet S matrix}",
	eprint = "1506.00218",
	archivePrefix = "arXiv",
	primaryClass = "hep-th",
	reportNumber = "ITP-UU-15-08, HU-EP-15-26, HU-MATHEMATIK-P-2015-06, IMPERIAL-TP-OOS-2015-01",
	doi = "10.1088/1751-8113/48/41/415401",
	journal = "J. Phys. A",
	volume = "48",
	number = "41",
	pages = "415401",
	year = "2015"
}

@article{Bena:2012hf,
	author = "Bena, Iosif and Berkooz, Micha and de Boer, Jan and El-Showk, Sheer and Van den Bleeken, Dieter",
	title = "{Scaling BPS Solutions and pure-Higgs States}",
	eprint = "1205.5023",
	archivePrefix = "arXiv",
	primaryClass = "hep-th",
	reportNumber = "IPHT-T12-041",
	doi = "10.1007/JHEP11(2012)171",
	journal = "JHEP",
	volume = "11",
	pages = "171",
	year = "2012"
}

@article{Manschot:2010qz,
	author = "Manschot, Jan and Pioline, Boris and Sen, Ashoke",
	title = "{Wall Crossing from Boltzmann Black Hole Halos}",
	eprint = "1011.1258",
	archivePrefix = "arXiv",
	primaryClass = "hep-th",
	doi = "10.1007/JHEP07(2011)059",
	journal = "JHEP",
	volume = "07",
	pages = "059",
	year = "2011"
}

@article{Manschot:2011xc,
	author = "Manschot, Jan and Pioline, Boris and Sen, Ashoke",
	title = "{A Fixed point formula for the index of multi-centered N=2 black holes}",
	eprint = "1103.1887",
	archivePrefix = "arXiv",
	primaryClass = "hep-th",
	doi = "10.1007/JHEP05(2011)057",
	journal = "JHEP",
	volume = "05",
	pages = "057",
	year = "2011"
}

@article{Strominger:1986uh,
	author = "Strominger, Andrew",
	title = "{Superstrings with Torsion}",
	reportNumber = "NSF-ITP-86-16",
	doi = "10.1016/0550-3213(86)90286-5",
	journal = "Nucl. Phys. B",
	volume = "274",
	pages = "253",
	year = "1986"
}

@article{Gates:1984nk,
	author = "Gates, Jr., S. J. and Hull, C. M. and Rocek, M.",
	title = "{Twisted Multiplets and New Supersymmetric Nonlinear Sigma Models}",
	reportNumber = "ITP-SB-84-53",
	doi = "10.1016/0550-3213(84)90592-3",
	journal = "Nucl. Phys. B",
	volume = "248",
	pages = "157--186",
	year = "1984"
}

@article{Gunaydin:1986fe,
	author = "Gunaydin, M. and Sierra, G. and Townsend, P. K.",
	title = "{The Unitary Supermultiplets of $d=3$ Anti-de Sitter and $d=2$ Conformal Superalgebras}",
	reportNumber = "UCRL-93923",
	doi = "10.1016/0550-3213(86)90293-2",
	journal = "Nucl. Phys. B",
	volume = "274",
	pages = "429--447",
	year = "1986"
}

@article{Galajinsky:2014hza,
	author = "Galajinsky, Anton",
	title = "{$\mathcal N =4$ superconformal mechanics from the $su(2)$ perspective}",
	eprint = "1412.4467",
	archivePrefix = "arXiv",
	primaryClass = "hep-th",
	doi = "10.1007/JHEP02(2015)091",
	journal = "JHEP",
	volume = "02",
	pages = "091",
	year = "2015"
}

@article{VanDerJeugt:1985hq,
	author = "Van Der Jeugt, J.",
	title = "{Irreducible Representations of the Exceptional Lie Superalgebras $D(2,1,\alpha)$}", 
	doi = "10.1063/1.526547",
	journal = "J. Math. Phys.",
	volume = "26",
	pages = "913--924",
	year = "1985"
}

@article{Frappat:1996pb,
	author = "Frappat, L. and Sorba, P. and Sciarrino, A.",
	title = "{Dictionary on Lie superalgebras}",
	eprint = "hep-th/9607161",
	archivePrefix = "arXiv",
	reportNumber = "ENSLAPP-AL-600-96, DSF-T-30-96",
	month = "7",
	year = "1996"
}

@article{Gaiotto:2004pc,
	author = "Gaiotto, Davide and Simons, Aaron and Strominger, Andrew and Yin, Xi",
	title = "{D0-branes in black hole attractors}",
	eprint = "hep-th/0412179",
	archivePrefix = "arXiv",
	doi = "10.1088/1126-6708/2006/03/019",
	journal = "JHEP",
	volume = "03",
	pages = "019",
	year = "2006"
}

@article{deBoer:1999gea,
	author = "de Boer, Jan and Pasquinucci, Andrea and Skenderis, Kostas",
	title = "{AdS / CFT dualities involving large 2-D N=4 superconformal symmetry}",
	eprint = "hep-th/9904073",
	archivePrefix = "arXiv",
	reportNumber = "KUL-TF-99-11, SPIN-1999-06, INLO-PUB-7-99",
	doi = "10.4310/ATMP.1999.v3.n3.a5",
	journal = "Adv. Theor. Math. Phys.",
	volume = "3",
	pages = "577--614",
	year = "1999"
}

@article{Mirfendereski:2022omg,
	author = "Mirfendereski, Delaram and Raeymaekers, Joris and \c{S}anl\i{}, Canberk and Van den Bleeken, Dieter",
	title = "{The geometry of gauged (super)conformal mechanics}",
	eprint = "2203.10167",
	archivePrefix = "arXiv",
	primaryClass = "hep-th",
	doi = "10.1007/JHEP08(2022)081",
	journal = "JHEP",
	volume = "08",
	pages = "081",
	year = "2022"
}

@article{Manschot:2013sya,
	author = "Manschot, Jan and Pioline, Boris and Sen, Ashoke",
	title = "{On the Coulomb and Higgs branch formulae for multi-centered black holes and quiver invariants}",
	eprint = "1302.5498",
	archivePrefix = "arXiv",
	primaryClass = "hep-th",
	reportNumber = "BONN-TH-2013-03, CERN-PH-TH-2013-030, HRI-ST-1302",
	doi = "10.1007/JHEP05(2013)166",
	journal = "JHEP",
	volume = "05",
	pages = "166",
	year = "2013"
}

@article{Manschot:2014fua,
	author = "Manschot, Jan and Pioline, Boris and Sen, Ashoke",
	title = "{The Coulomb Branch Formula for Quiver Moduli Spaces}",
	eprint = "1404.7154",
	archivePrefix = "arXiv",
	primaryClass = "hep-th",
	reportNumber = "CERN-PH-TH-2014-074",
	doi = "10.5802/cml.41",
	month = "4",
	year = "2014"
}

@article{Douglas:1996sw,
	author = "Douglas, Michael R. and Moore, Gregory W.",
	title = "{D-branes, quivers, and ALE instantons}",
	eprint = "hep-th/9603167",
	archivePrefix = "arXiv",
	reportNumber = "RU-96-15, YCTP-P5-96",
	month = "3",
	year = "1996"
}

@article{Barns-Graham:2018xdd,
	author = "Barns-Graham, Alec E. and Dorey, Nick",
	title = {{A Superconformal Index for HyperK\"ahler Cones}},
	eprint = "1812.04565",
	archivePrefix = "arXiv",
	primaryClass = "hep-th",
	month = "12",
	year = "2018"
}

@article{Krivonos:2010zy,
	author = "Krivonos, Sergey and Lechtenfeld, Olaf",
	title = "{Many-particle mechanics with D(2,1:alpha) superconformal symmetry}",
	eprint = "1012.4639",
	archivePrefix = "arXiv",
	primaryClass = "hep-th",
	doi = "10.1007/JHEP02(2011)042",
	journal = "JHEP",
	volume = "02",
	pages = "042",
	year = "2011"
}

@article{Beaujard:2021fsk,
	author = "Beaujard, Guillaume and Mondal, Swapnamay and Pioline, Boris",
	title = "{Multi-centered black holes, scaling solutions and pure-Higgs indices from localization}",
	eprint = "2103.03205",
	archivePrefix = "arXiv",
	primaryClass = "hep-th",
	doi = "10.21468/SciPostPhys.11.2.023",
	journal = "SciPost Phys.",
	volume = "11",
	number = "2",
	pages = "023",
	year = "2021"
}

@article{Raeymaekers:2024usy,
	author = "Raeymaekers, Joris and Sanli, Canberk and Van den Bleeken, Dieter",
	title = "{Superconformal indices and localization in N = 2B quantum mechanics}",
	eprint = "2403.07665",
	archivePrefix = "arXiv",
	primaryClass = "hep-th",
	doi = "10.1007/JHEP05(2024)275",
	journal = "JHEP",
	volume = "05",
	pages = "275",
	year = "2024"
}

@article{Raeymaekers:2024ics,
	author = "Raeymaekers, Joris and Rossi, Paolo and Sanli, Canberk",
	title = "{Index and localization for type B superconformal mechanics on singular spaces}",
	eprint = "2412.04390",
	archivePrefix = "arXiv",
	primaryClass = "hep-th",
	doi = "10.1007/JHEP04(2025)199",
	journal = "JHEP",
	volume = "04",
	pages = "199",
	year = "2025"
}

@article{DeMarco:2026jje,
	author = {De Marco, Andrea and Mullahasanoglu, Mustafa and Raeymaekers, Joris and Rossi, Paolo and Seng{\"o}r, Gizem},
	title = "{On conformal symmetry in large-$N$ quiver mechanics}",
	eprint = "2607.15348",
	archivePrefix = "arXiv",
	primaryClass = "hep-th",
	month = "7",
	year = "2026"
}

@article{Dorey:2018klg,
	author = "Dorey, Nick and Singleton, Andrew",
	title = "{An Index for Superconformal Quantum Mechanics}",
	eprint = "1812.11816",
	archivePrefix = "arXiv",
	primaryClass = "hep-th",
	month = "12",
	year = "2018"
}

@article{deAlfaro:1976vlx,
	author = "de Alfaro, Vittorio and Fubini, S. and Furlan, G.",
	title = "{Conformal Invariance in Quantum Mechanics}",
	reportNumber = "CERN-TH-2115",
	doi = "10.1007/BF02785666",
	journal = "Nuovo Cim. A",
	volume = "34",
	pages = "569",
	year = "1976"
}

@article{Dorey:2019kaf,
	author = "Dorey, Nick and Zhang, Daniel",
	title = {{Superconformal quantum mechanics on K\"ahler cones}},
	eprint = "1911.06787",
	archivePrefix = "arXiv",
	primaryClass = "hep-th",
	doi = "10.1007/JHEP05(2020)115",
	journal = "JHEP",
	volume = "05",
	pages = "115",
	year = "2020"
}

@article{Mirfendereski:2020rrk,
	author = "Mirfendereski, Delaram and Raeymaekers, Joris and Van Den Bleeken, Dieter",
	title = "{Superconformal mechanics of AdS$_2$ D-brane boundstates}",
	eprint = "2009.07107",
	archivePrefix = "arXiv",
	primaryClass = "hep-th",
	doi = "10.1007/JHEP12(2020)176",
	month = "9",
	year = "2020"
}

@article{Bena:2018bbd,
	author         = "Bena, Iosif and Heidmann, Pierre and Turton, David",
	title          = "{AdS$_{2}$ holography: mind the cap}",
	journal        = "JHEP",
	volume         = "12",
	year           = "2018",
	pages          = "028",
	doi            = "10.1007/JHEP12(2018)028",
	eprint         = "1806.02834",
	archivePrefix  = "arXiv",
	primaryClass   = "hep-th",
	SLACcitation   = "%%CITATION = ARXIV:1806.02834;%%"
}

@article{Smilga:1986rb,
	author         = "Smilga, Andrei V.",
	title          = "{Perturbative Corrections to Effective Zero Mode
	Hamiltonian in Supersymmetric {QED}}",
	journal        = "Nucl. Phys.",
	volume         = "B291",
	year           = "1987",
	pages          = "241-255",
	doi            = "10.1016/0550-3213(87)90473-1",
	reportNumber   = "ITEP-86-171",
	SLACcitation   = "%%CITATION = NUPHA,B291,241;%%"
}

@article{Denef:2002ru,
	author         = "Denef, Frederik",
	title          = "{Quantum quivers and Hall / hole halos}",
	journal        = "JHEP",
	volume         = "10",
	year           = "2002",
	pages          = "023",
	doi            = "10.1088/1126-6708/2002/10/023",
	eprint         = "hep-th/0206072",
	archivePrefix  = "arXiv",
	primaryClass   = "hep-th",
	SLACcitation   = "%%CITATION = HEP-TH/0206072;%%"
}

@article{Anninos:2013nra,
	author         = "Anninos, Dionysios and Anous, Tarek and de Lange, Paul
	and Konstantinidis, George",
	title          = "{Conformal quivers and melting molecules}",
	journal        = "JHEP",
	volume         = "03",
	year           = "2015",
	pages          = "066",
	doi            = "10.1007/JHEP03(2015)066",
	eprint         = "1310.7929",
	archivePrefix  = "arXiv",
	primaryClass   = "hep-th",
	SLACcitation   = "%%CITATION = ARXIV:1310.7929;%%"
}

@article{Ivanov:2003tm,
	author = "Ivanov, Evgeny and Krivonos, Sergey and Lechtenfeld, Olaf",
	title = "{N=4, d = 1 supermultiplets from nonlinear realizations of D(2,1: alpha)}",
	eprint = "hep-th/0310299",
	archivePrefix = "arXiv",
	reportNumber = "ITP-UH-09-03",
	doi = "10.1088/0264-9381/21/4/021",
	journal = "Class. Quant. Grav.",
	volume = "21",
	pages = "1031--1050",
	year = "2004"
}

@article{Denef:2007vg,
	author = "Denef, Frederik and Moore, Gregory W.",
	title = "{Split states, entropy enigmas, holes and halos}",
	eprint = "hep-th/0702146",
	archivePrefix = "arXiv",
	doi = "10.1007/JHEP11(2011)129",
	journal = "JHEP",
	volume = "11",
	pages = "129",
	year = "2011"
}

@article{Mirfendereski:2018tob,
	author = "Mirfendereski, Delaram and Van Den Bleeken, Dieter",
	title = "{Twist on multicenter Ad$S_2$ solutions}",
	eprint = "1807.01879",
	archivePrefix = "arXiv",
	primaryClass = "hep-th",
	doi = "10.1103/PhysRevD.98.106001",
	journal = "Phys. Rev. D",
	volume = "98",
	number = "10",
	pages = "106001",
	year = "2018"
}

@article{Delduc:2006yp,
	author = "Delduc, F. and Ivanov, E.",
	title = "{Gauging N=4 Supersymmetric Mechanics}",
	eprint = "hep-th/0605211",
	archivePrefix = "arXiv",
	doi = "10.1016/j.nuclphysb.2006.06.031",
	journal = "Nucl. Phys. B",
	volume = "753",
	pages = "211--241",
	year = "2006"
}

@article{Fedoruk:2009xf,
	author = "Fedoruk, Sergey and Ivanov, Evgeny and Lechtenfeld, Olaf",
	title = "{New D(2,1,alpha) Mechanics with Spin Variables}",
	eprint = "0912.3508",
	archivePrefix = "arXiv",
	primaryClass = "hep-th",
	doi = "10.1007/JHEP04(2010)129",
	journal = "JHEP",
	volume = "04",
	pages = "129",
	year = "2010"
}

@article{Michelson:1999zf,
	author = "Michelson, Jeremy and Strominger, Andrew",
	title = "{The Geometry of (super)conformal quantum mechanics}",
	eprint = "hep-th/9907191",
	archivePrefix = "arXiv",
	reportNumber = "HUTP-99-A045",
	doi = "10.1007/PL00005528",
	journal = "Commun. Math. Phys.",
	volume = "213",
	pages = "1--17",
	year = "2000"
}

@article{Manschot:2012rx,
	author = "Manschot, Jan and Pioline, Boris and Sen, Ashoke",
	title = "{From Black Holes to Quivers}",
	eprint = "1207.2230",
	archivePrefix = "arXiv",
	primaryClass = "hep-th",
	reportNumber = "BONN-TH-2012-14, CERN-PH-TH-2012-191, HRI-ST-1204",
	doi = "10.1007/JHEP11(2012)023",
	journal = "JHEP",
	volume = "11",
	pages = "023",
	year = "2012"
}

@article{Maloney:1999dv,
	author = "Maloney, Alexander and Spradlin, Marcus and Strominger, Andrew",
	title = "{Superconformal multiblack hole moduli spaces in four-dimensions}",
	eprint = "hep-th/9911001",
	archivePrefix = "arXiv",
	reportNumber = "HUTP-99-A055",
	doi = "10.1088/1126-6708/2002/04/003",
	journal = "JHEP",
	volume = "04",
	pages = "003",
	year = "2002"
}

@article{BrittoPacumio:1999ax,
	author = "Britto-Pacumio, Ruth and Michelson, Jeremy and Strominger, Andrew and Volovich, Anastasia",
	editor = "Thorlacius, Lárus and Jonsson, Thordur",
	title = "{Lectures on Superconformal Quantum Mechanics and Multi-Black Hole Moduli Spaces}",
	eprint = "hep-th/9911066",
	archivePrefix = "arXiv",
	reportNumber = "HUTP-99-A060, RUNHETC-99-41",
	doi = "10.1007/978-94-011-4303-5\_6",
	journal = "NATO Sci. Ser. C",
	volume = "556",
	pages = "255--284",
	year = "2000"
}

@article{Fedoruk:2011aa,
	author = "Fedoruk, Sergey and Ivanov, Evgeny and Lechtenfeld, Olaf",
	title = "{Superconformal Mechanics}",
	eprint = "1112.1947",
	archivePrefix = "arXiv",
	primaryClass = "hep-th",
	doi = "10.1088/1751-8113/45/17/173001",
	journal = "J. Phys. A",
	volume = "45",
	pages = "173001",
	year = "2012"
}

@article{Denef:2000nb,
	author         = "Denef, Frederik",
	title          = "{Supergravity flows and D-brane stability}",
	journal        = "JHEP",
	volume         = "08",
	year           = "2000",
	pages          = "050",
	doi            = "10.1088/1126-6708/2000/08/050",
	eprint         = "hep-th/0005049",
	archivePrefix  = "arXiv",
	primaryClass   = "hep-th",
	SLACcitation   = "%%CITATION = HEP-TH/0005049;%%"
}
	
\end{document}